\documentclass[12pt]{article}
\usepackage[dvipdfmx]{graphicx}
\usepackage{bmpsize}
\usepackage{bbding}
\usepackage{orcidlink}

\usepackage[utf8]{inputenc}

\usepackage{newtxtext}

\usepackage[utf8]{inputenc}

\usepackage[square,comma,numbers,sort&compress]{natbib}
\usepackage[left=20mm,right=20mm,top=20mm,bottom=20mm]{geometry}
\usepackage{ascmac}
\usepackage{bm}

\usepackage{epsf}
\usepackage{color}
\usepackage{amsmath}
\usepackage{amssymb}
\usepackage{latexsym}
\usepackage{slashed}
\usepackage{float}
\usepackage{cases}
\usepackage{bm}
\usepackage{arydshln}
\usepackage{hyperref}
\usepackage{subcaption} 
\hypersetup{ 
 setpagesize=false,
 bookmarksnumbered=true,%
 bookmarksopen=true,%
 colorlinks=true,%
 linkcolor=blue,
 citecolor=red,
}
\usepackage{comment}

\usepackage{listings}
\usepackage{xcolor}

\definecolor{codegreen}{rgb}{0,0.6,0}
\definecolor{codegray}{rgb}{0.5,0.5,0.5}
\definecolor{codepurple}{rgb}{0.58,0,0.82}
\definecolor{backcolour}{rgb}{0.95,0.95,0.92}

\lstdefinestyle{mystyle}{
    backgroundcolor=\color{backcolour},   
    commentstyle=\color{codegreen},
    keywordstyle=\color{magenta},
    numberstyle=\tiny\color{codegray},
    stringstyle=\color{codepurple},
    basicstyle=\ttfamily\footnotesize,
    breakatwhitespace=false,         
    breaklines=true,                 
    captionpos=b,                    
    keepspaces=true,                 
    numbers=left,                    
    numbersep=5pt,                  
    showspaces=false,                
    showstringspaces=false,
    showtabs=false,                  
    tabsize=2
}

\usepackage{mathtools}

\DeclarePairedDelimiterX\braket[2]{\langle}{\rangle}{#1 \delimsize\vert #2}

\usepackage{multicol}
\usepackage{lipsum}

\usepackage{braket}

\newcommand{\eps}{\epsilon}

\newcommand{\wt}{\widetilde}

\newcommand{\al}[1]{\begin{align}#1\end{align}}
\newcommand{\als}[1]{\begin{align*}#1\end{align*}}
\newcommand{\bp}{\begin{pmatrix}}
\newcommand{\ep}{\end{pmatrix}}
\newcommand{\bb}{\begin{bmatrix}}
\newcommand{\eb}{\end{bmatrix}}
\newcommand{\nn}{\nonumber\\}

\newcommand{\ol}{\overline}

\newcommand{\paren}[1]{\left(#1\right)}
\newcommand{\sqbr}[1]{\left[#1\right]}

\newcommand{\fn}[1]{\!\left(#1\right)}
\newcommand{\br}[1]{\left\{#1\right\}}

\newcommand{\tx}[1]{\text{#1}}
\newcommand{\wh}[1]{\widehat{#1}}
\newcommand{\ov}{\over}

\newcommand{\beq}{\begin{equation}}
\newcommand{\eeq}{\end{equation}}
\newcommand{\bea}{\begin{eqnarray}}
\newcommand{\eea}{\end{eqnarray}}

\newcommand{\pal}{\partial}

\newcommand{\fulltoday}{\number\day\space \ifcase\month\or
    January\or February\or March\or April\or May\or June\or
    July\or August\or September\or October\or November\or December\fi
    \space\number\year}

 \makeatletter
    
    \@addtoreset{equation}{section}
  \makeatother

\usepackage{calc}
\newcounter{hours}\newcounter{minutes}
\makeatletter                   
\renewcommand*{\thehours}{\two@digits\c@hours}
\renewcommand*{\theminutes}{\two@digits\c@minutes}
\makeatother

\begin{document}

\allowdisplaybreaks[2]
\renewcommand{\thefootnote}{*}

\newlength{\mylength}

\title{
Muon Beam Dump Experiments {probe} \\ five-dimensional nature of $U(1)_{L_{\mu} - L_{\tau}}$ \\
}

\author{
Dibyendu Chakraborty\thanks{\orcidlink{0009-0007-0558-2782}, E-mail: \tt dc282@snu.edu.in}, \mbox{}
Arindam Chatterjee\thanks{\orcidlink{0000-0002-7256-8813}, E-mail: \tt arindam.chatterjee@snu.edu.in}, \\[5pt]
Ayushi Kaushik\thanks{\orcidlink{0009-0007-3453-6988}, E-mail: \tt ak356@snu.edu.in}, \mbox{} and
Kenji Nishiwaki\thanks{\orcidlink{0000-0001-6526-6799}, E-mail: \tt kenji.nishiwaki@snu.edu.in}
\bigskip\\
Department of Physics, School of Natural Sciences, \\ 
Shiv Nadar Institution of Eminence (Deemed to be University), \\
Tehsil Dadri, Gautam Buddha Nagar, Uttar Pradesh,
201314, India
}
\maketitle
\begin{abstract}
\noindent
{We have investigated the prospects of probing} the {five-dimensional} $U(1)_{L_\mu - L_\tau}$ interactions in present and future muon dump experiments, namely, NA64$_\mu$, M$^3$, MuSIC, and a future muon {beam} dump experiment.
These experiments are classified into two categories: the first two can probe processes where feebly interacting massive particles go into invisible channels, while the latter two can probe processes where these states decay into muon pairs.
These two types of experiments are complementary in that they allow exploration of different parameter regions of a model.
In our scenario, the presence of multiple massive gauge bosons as Kaluza-Klein~(KK) particles leads to an enhancement in the signal {events} compared to the corresponding four-dimensional scenario.
In particular, the decay process into muon pairs enables mass reconstruction of the parent particle, making it possible to directly demonstrate the existence of multiple KK particles in at least some parameter regions.
This can provide clear evidence that the origin of the $U(1)_{L_\mu - L_\tau}$ interaction lies in five dimensions.
Furthermore, the muon $(g-2)$ value, which is now consistent with the SM, can be used to exclude specific parameter regions for new particles interacting with muons.
{We also carefully discuss the non-trivial effects arising from nonzero kinetic mixing.}
\end{abstract}

\newpage
\renewcommand\thefootnote{\arabic{footnote}}
\setcounter{footnote}{0}

\section{Introduction
\label{sec:Introduction}}

In recent years, significant attention has been paid to light particles that couple with the Standard Model~(SM) sector feebly,
motivated by addressing the issues that cannot be solved within the SM, e.g., dark matter, neutrino profiles and others; a recent review is available in Ref.~\cite{Antel:2023hkf}.
The search for such particles is considered one of the most intriguing topics in particle phenomenology today.

When the mass of such particles is in the MeV range or GeV range but not particularly heavy, fixed-target beam-dump experiments are highly effective as a search method.
Examples of such experiments conducted in the past include the following:
CHARM~\cite{CHARM:1985nku}, KEK~\cite{Konaka:1986cb}, E141~\cite{Riordan:1987aw}, E137~\cite{Bjorken:1988as}, Orsay~\cite{Davier:1989wz}, E774~\cite{Bross:1989mp}, LSND~\cite{LSND:1997vqj}, NOMAD~\cite{NOMAD:2001eyx}, and NA64~\cite{NA64:2018lsq}, which are classified according to the type of incident beam:
protons~\cite{CHARM:1985nku,LSND:1997vqj,NOMAD:2001eyx} and electrons~\cite{Konaka:1986cb,Riordan:1987aw,Bjorken:1988as,Davier:1989wz,Bross:1989mp,NA64:2018lsq}.

{Generally proton or electron beams are used in such experiments. However, recently,} new types of beam dump experiments have been discussed, where the incident beam consists of particles other than protons and electrons.\footnote{
At the CERN Large Hadron Collider~(LHC), a next-generation proton beam dump experiment, SHiP, is scheduled~\cite{SHiP:2015vad,Alekhin:2015byh}.
Also, the FASER experiment has been investigating new feebly interacting light particles with a displaced detector, as well as studying high-energy neutrinos~\cite{FASER:2018bac,FASER:2018eoc}.
See~\cite{Feng:2022inv} for the Forward Physics Facility at the High-Luminosity LHC.
Note that the bounds have been reported by the ATLAS~\cite{ATLAS:2014fzk,ATLAS:2015itk,ATLAS:2022bll,ATLAS:2022izj,ATLAS:2023cjw,ATLAS:2024zxk} and LHCb~\cite{LHCb:2019vmc,LHCb:2020ysn} collaborations.
}
Among these, one particularly interesting option is to use muons as the incident beam.\footnote{
Other such options are positrons~\cite{Kanemura:2015cxa,Marsicano:2018krp,Nardi:2018cxi,Berryman:2019dme,Berlin:2020uwy,Celentano:2020vtu,Asai:2021xtg,Asai:2021ehn,Moroi:2022qwz,Asai:2022zxw,Battaglieri:2022dcy,Asai:2023dzs,Blinov:2024pza},
charged pions~\cite{Curtin:2023bcf}, and photons~\cite{Blinov:2024pza,Schulthess:2025tct}.
}
Various theoretical papers have already been published~\cite{Chen:2017awl,Kahn:2018cqs,Gninenko:2018ter,Gninenko:2020hbd,Acosta:2021qpx,Kirpichnikov:2021jev,Cesarotti:2022ttv,Acosta:2022ejc,Zhevlakov:2022vio,Voronchikhin:2022rwc,Forbes:2022bvo,Radics:2023tkn,Zhevlakov:2023jzt,Cesarotti:2023sje,Batell:2024cdl,Voronchikhin:2024ygo,Li:2025yzb}, and an experiment, NA64${}_{\mu}$, is already underway~\cite{Sieber:2021fue,NA64:2024klw}.
Naturally, these experiments are very useful to probe theoretical frameworks consisting of new feebly interacting light particles which couple to muons.

One such interesting model incorporates an additional \( U(1)_{L_\mu - L_\tau} \) symmetry within the gauge sector~\cite{Foot:1990mn, He:1990pn, He:1991qd, Foot:1994vd}.
This gauge extension is especially attractive as it is anomaly-free without introducing the right-handed neutrinos.
{
{This gauge extension has been widely employed to mediate light Dark Matter interactions~\cite{Baek:2008nz,Baek:2015fea,Patra:2016shz,Biswas:2016yan,Biswas:2016yjr,Asai:2017ryy,Arcadi:2018tly,Kamada:2018zxi,Foldenauer:2018zrz,Asai:2019ciz,Okada:2019sbb,Asai:2020qlp,Holst:2021lzm,Tapadar:2021kgw,Heeck:2022znj,Nagao:2022osm,KA:2023dyz,Figueroa:2024tmn}, and to realise various leptogenesis mechanisms~\cite{Asai:2020qax,Borah:2021mri,Eijima:2023yiw,Granelli:2023egb,Wada:2024cbe}.}
%
Furthermore, recent precise measurements of neutrino mass and mixing angles have enabled constraints to be placed on the \( U(1)_{L_\mu - L_\tau} \) interaction through information from the active neutrino mass matrix (in particular, the minimal texture of it)~\cite{Asai:2017ryy,Asai:2018ocx,Asai:2020qax,Ibe:2025rwk}.
Moreover, the information from recent precise measurements of $N_\tx{eff}$ will provide further information about very-light and feebly-interacting gauge bosons (see previous works~\cite{Escudero:2019gzq,Araki:2021xdk,Carpio:2021jhu,Asai:2023ajh}).
Therefore, irrespective of the evolving status of the muon anomalous magnetic moment,\footnote{
The latest SM value is $a_\mu^\tx{exp} - a_\mu^\tx{SM} = 38(63) \times 10^{-11}$, which is consistent within $\sim \! 0.6\,\sigma$~\cite{Aliberti:2025beg}.
}
particularly in light of recent QCD lattice determinations of the hadronic vacuum polarisation contribution, which indicate no significant tension between the SM prediction and experimental measurements, the $U(1)_{L_\mu - L_\tau}$ gauged framework remains theoretically well motivated and phenomenologically testable across multiple experimental frontiers. It is important to emphasise that the $(g-2)_\mu$ observable constitutes only one among several experimental and cosmological probes of this symmetry.
}


In our previous study~\cite{Chakraborty:2024xxc}, we looked into a five-dimensional (5D) version of the basic $U(1)_{L_\mu - L_\tau}$ framework that added a single compact spatial dimension. The $U(1)_{L_\mu - L_\tau}$ gauge symmetry spreads out in the 5D bulk, introducing an infinite number of gauge bosons that are seen as Kaluza-Klein (KK) modes~\cite{Ponton:2012bi} from a four-dimensional~(4D) point of view.
In this method, the source of the MeV to GeV mass scale can be directly connected to the inverse size of the compact extra dimension under a twisted boundary condition, without the introduction of a new scalar boson.

{
This is one concrete example of the bigger framework for physics beyond the Standard Model, hidden spatial direction (or directions) in our universe, commonly known as extra dimension(s).
By utilising the spatial structure of extra dimensions, it becomes possible to construct phenomenologically interesting models that are impossible to realise in four dimensions (see e.g., the seminal works~\cite{Arkani-Hamed:1998jmv,Randall:1999ee}).
{Of course, by combining it with mechanisms that are also viable in four dimensions, a wider variety of discussions becomes possible.}
Furthermore, (though beyond the scope of this paper's discussion), if matter coordination extends into the extra dimensions, there is also the possibility that experimentally verifiable differences in cosmic expansion may be observed compared to the case of four-dimensional spacetime in early cosmology {(see e.g.,~\cite{Anchordoqui:2024ajk})}.

On the other hand, phenomenologically, what deserves attention is the search for a method to directly distinguish between four-dimensional and five-dimensional physics.
When all SM particles are propagating in the extra dimension(s), KK particles appear for all SM constituents (including those with strong interactions), dubbed the universal extra dimensions, making verification particularly straightforward in the experiments at the LHC.
However, such investigations have already been experimentally done, and no positive results have been obtained to date~\cite{Antoniadis:1990ew,Appelquist:2000nn}, (see also~\cite{Kakuda:2013kba,Deutschmann:2017bth,Flores:2021xwx}).
On the other hand, if only new particles beyond the SM extend into the extra dimension(s), experimental constraints generally become weaker, making it possible to consider relatively light KK particles with sizes below the electroweak scale.
{In this work, we propose and systematically validate a strategy to search for the gauge boson(s) mediating such interactions, focusing on cases where these are light, couple predominantly to muons, and extend into extra dimensions. The analysis further elucidates the differences between four-dimensional and five-dimensional scenarios in this context.}
Actual calculations are performed using the aforementioned $U(1)_{L_\mu - L_\tau}$ interaction as an example, but the methodology is adaptable to other cases satisfying the aforementioned conditions.
}

In~\cite{Chakraborty:2024xxc}, as the first survey, we employed electron-neutrino elastic scattering (E$\nu$ES), specifically the interaction $e \, \nu_{X} \rightarrow e \, \nu_{X}$ $\left( X: e, \bar{e}, \mu, \bar{\mu} \right)$.
To illustrate our methodology, we examined data from studies Borexino~\cite{Saldanha:2012vba}, TEXONO~\cite{Wong:2015kgl}, and CHARM-II~\cite{CHARM-II:1993phx,CHARM-II:1994dzw} and  {considered} the future prospects in DUNE~\cite{DUNE:2016hlj,DUNE:2020fgq}.
Note that Section one of our previous work~\cite{Chakraborty:2024xxc} contains a list of other kinds of constraints on the 4D $U(1)_{L_\mu - L_\tau}$ scenario.

However, {in the context of} $U(1)_{L_\mu - L_\tau}$ {gauge extension}, {the} massive gauge bosons interact with electrons only through kinetic mixing terms~\cite{Holdom:1985ag}.
So, through E$\nu$ES, direct searches for muon interactions are not possible.
{The ongoing and proposed} muon beam dump experiments, such as NA64$_\mu$~\cite{NA64:2024nwj}, M$^3$~\cite{Kahn:2018cqs}, MuSIC~\cite{Acosta:2021qpx,Acosta:2022ejc}, and a future muon dump experiment~\cite{Cesarotti:2022ttv},  provide an excellent avenue to directly probe such interactions for muons, even in scenarios with negligible kinetic mixing.
These four experiments provide complementary results focusing on two decay processes of the massive gauge boson $Z'$, produced via Bremsstrahlung from a muon beam: $Z' \to \tx{invisible}$ (NA64$_\mu$ and M$^3$) and $Z' \to \mu^- \mu^+$ (MuSIC and a future muon {beam} dump experiment).
{In our framework, the existence of multiple $Z^{'}$ candidates leads to enhanced experimental sensitivity 
and a correspondingly broader scope for examining constraints and future prospects than in the 4D counterpart.}
In particular, the decay of $Z'$ into muon pairs is interesting because it allows for signal reconstruction, enabling direct confirmation of the existence of multiple vector bosons.

This paper is organised as follows.
In Section~\ref{sec:Model-Infro}, we introduce our 5D $U(1)_{L_\mu - L_\tau}$ scenario.
In Section~\ref{sec:cross-section}, we provide the formulas for the signal process via massive-vector Bremsstrahlung in 4D and 5D.
In Section~\ref{sec:Experiments}, the details of the four muon beam dump experiments and how to estimate the number of signal events are explained.
Section~\ref{sec:Results} provides numerical results and current status and prospects of our scenario in the future experiments, {covering both cases where the kinetic mixing effect is zero and non-zero.}
Section~\ref{sec:Summary} is devoted to providing a summary and discussion.
In Appendix~\ref{sec:Additional-Plots}, complementary plots are shown for the validity of our analysis.
{In Appendix~\ref{sec:for-nmax-choice}, we provide the justification of our application as a truncation of the summations of KK modes.}

\section{Model Specifications
\label{sec:Model-Infro}}

This part gives a short summary of the five-dimensional $U(1)_{L_\mu - L_\tau}$ framework with an extra small spatial dimension.
In this setup, all SM particles are stuck on a three-dimensional brane at $y = y_{\text{SM}}$.
The $U(1)_{L_\mu - L_\tau}$ gauge boson and its KK excitations, on the other hand, can move through the five-dimensional bulk.

We use the coordinate $y \in [0, \pi R]$ to describe the extra spatial dimension, where $R$ is the radius of compactification.
It is assumed that the SM brane can be anywhere in the extra dimension, including the ends, and that its position is a free parameter.
This generalisation is given so that a more thorough phenomenological study can be done.
This study focuses merely on the flat extra-dimensional case.
We use twisted boundary conditions on the five-dimensional $U(1)_{L_{\mu} - L_{\tau}}$ gauge field: Neumann boundary conditions at $y = 0$ and Dirichlet boundary conditions at $y = \pi R$. As a result, there is no massless mode in the KK spectrum.
Without using the Higgs mechanism or adding an extra singlet scalar field, the massless mode is taken away. 
The 5D $U(1)_{L_\mu-L_\tau}$ gauge coupling $g'_\tx{5D}$ is not dimensionless and has the mass dimension $-1/2$.
In 4D effective theories, it is always accompanied by $\sqrt{\pi R}$ as the combination
$g' = {g'_\tx{5D} / \sqrt{\pi R}}$, and we can consider $g'$ as a fundamental parameter.

\subsection{Effective four-dimensional Lagrangian via KK decomposition
\label{sec:effective-Lagrangian}}

In this subsection, the theoretical framework for phenomenological calculations is introduced in terms of effective field theory to provide a common tool for treating the flat and warped backgrounds together.\footnote{
Details of the KK decompositions are available in Appendix A of Ref.~\cite{Chakraborty:2024xxc}.
{For a string-theoritic realisation of a bulk leptonic $U(1)$, see~\cite{Antoniadis:2021mqz}.}
}
{This paper focuses on the phenomenology of the flat case.}
The {Lagrangian describing the} effective 4D vectors' free part reads
\al{
{\cal L}_\tx{eff}^{\paren{V\tx{-free}}}
	&=
		\sum_n
		\br{
			-{1\ov 4} \wh{Z'}_{\mu\nu}^{\paren{n}} \wh{Z'}^{\paren{n}\mu\nu}
			+{\epsilon_n\ov 2c_W} \wh{Z'}_{\mu\nu}^{\paren{n}} \wh{B}^{\mu\nu}
			+{1\ov 2} M_n^2 \wh{Z'}_\mu^{\paren{n}} \wh{Z'}^{\paren{n}\mu}
		} \nn
	&\quad
		-\frac{1}{4} \wh{B}_{\mu\nu} \wh{B}^{\mu\nu} -\frac{1}{4} {W}^3_{\mu\nu} {W}^{3\mu\nu},
	\label{eq:effectiveLagrangian-Vfree}
}
where $n = 1,2,3,\cdots$ discriminate massive KK states;
{ $\wh{Z'}^{\paren{n}}_\mu$ denotes the 4D gauge eigenstate field
of the $n$-th KK excitation of the $U\fn{1}_{L_\mu - L_\tau}$ gauge boson, $\wh{B}_\mu$ 
and $W^3_\mu$ are gauge-eigenstate 4D gauge fields of the 
$U(1)_Y$}, and the 3rd component of $SU(2)_W$, respectively,
where each field strength is defined as usual, $V_{\mu\nu} \leftrightarrow \pal_\mu V_\nu - \pal_\nu V_\mu$. 
{Thus, for $\wh{Z'}^{\paren{n}}_\mu$, $\wh{B}_\mu$ and $W^3_\mu$, the respective field strengths are defined} as follows,
\al{
\wh{Z'}_{\mu\nu}^{\paren{n}} &:= \pal_\mu \wh{Z'}^{\paren{n}}_\nu - \pal_\nu \wh{Z'}^{\paren{n}}_\mu,&
\wh{B}_{\mu\nu} &:= \pal_\mu \wh{B}_\nu - \pal_\nu \wh{B}_\mu,&
{W}^3_{\mu\nu} &:= \pal_\mu {W}^3_\nu - \pal_\nu {W}^3_\mu.&
}
$\epsilon_n$ is the effective kinetic mixing factor between $\wh{Z'}^{\paren{n}}_\mu$ and $\wh{B}_\mu$, which is defined as
\al{
\epsilon_n 
	&=
		\epsilon_4 f_n,&
f_n
	&:=
		f_V^{\paren{n}}\fn{y_\tx{SM}}
		= {\sqrt{2} \cos\left[ \fn{n - \frac{1}{2}} \frac{y_\tx{SM}}{R} \right]},&
}
where $f_V^{\paren{n}}\fn{y}$ is the $n$-th KK mode wavefunction toward the {flat} $y$ direction, and the effective dimensionless parameter $\epsilon_4$ is defined as
\al{
\epsilon_4 := {\epsilon_\tx{5D}/\sqrt{\pi R} }, \quad
(\epsilon_\tx{5D} \tx{ is a parameter in the 5D original action~\cite{Chakraborty:2024xxc}.})
}
$M_n$ is the $n$-th KK mass in the flat space {and is given by},
\al{
M_n := (2n-1) m_\tx{KK}, \ (n = 1,2,3,\cdots); \quad m_\tx{KK} := {1 \ov 2R}.
}
Note that $\epsilon_n$ and $M_n$ take different forms in the cases of the flat {and warped} backgrounds,
$v \simeq 246\,\tx{GeV}$ is the Higgs vacuum expectation value,
$c_W$, $s_W$, $g_2$ and $g_1$ are the cosine and the sine of the Weinberg angle, the $SU(2)_W$ 
gauge coupling, and the $U(1)_Y$ gauge coupling{, respectively}. {Introducing  
the secondary basis},
\al{
\wh{B}_\mu 
	&=
		B_\mu + \sum_n {\epsilon_n \ov c_W} \wt{Z'}^{\paren{n}}_\mu, \\
\wh{Z'}^{\paren{n}}_\mu
	&=
		\wt{Z'}^{\paren{n}}_\mu, \\
\bp \wt{Z}_\mu \\ A_\mu \ep
	&=
		\bp c_W & s_W \\ -s_W & c_W \ep
		\bp W^3_\mu \\ B_\mu \ep,
}
{the Lagrangian takes the following form,}
\al{
{ {\cal L}}_\tx{eff}^{\paren{V\tx{-free}}}
	&=
		-{1\ov 4} \sum_n \wt{Z'}_{\mu\nu}^{\paren{n}} \wt{Z'}^{\paren{n}\mu\nu}
		-{1\ov 4} \wt{Z}_{\mu\nu} \wt{Z}^{\mu\nu}
		-{1\ov 4} A_{\mu\nu} A^{\mu\nu} 
		+{1\ov 2} \sum_n M_n^2 \wt{Z'}_\mu^{\paren{n}} \wt{Z'}^{\paren{n}\mu}\nn
    &\quad    
		+{1\ov 2} m_{Z,0}^2 \wt{Z}_\mu \wt{Z}^\mu
		+ m_{Z,0}^2  t_W \sum_n\epsilon_n \wt{Z}_\mu \wt{Z'}^{\paren{n}\mu}
		+{1\ov 2} m_{Z,0}^2 t_W^2 \sum_{n,n'} \epsilon_n \epsilon_{n'} \wt{Z'}^{\paren{n}}_\mu \wt{Z'}^{\paren{n'}\mu} {,}
	\label{eq:S_eff-Vfree-1}
}
where $A_\mu$ is the photon field, {$A_{\mu \nu} := \partial_{\mu} A_{\nu}- \partial_{\nu} A_{\mu}$}, $t_W = s_W/c_W$, and $m_{Z,0}^2 := \paren{g_1^2+g_2^2}v^2/{4}$.\footnote{
If we take into account the 2nd-order perturbation for the eigenvalues, the unpurbed mass eigenvalues ($m_{Z,0}^2$ and $M_n^2$) are slightly shifted as follows:
\al{
m_{Z,0}^2
	&\to
		m_{Z,0}^2 \sqbr{ 1 - t_W^2 m_{Z,0}^2 \sum_n {\epsilon_n^2 \ov M_n^2 - m_{Z,0}^2} },&
M_n^2
	&\to
		M_n^2 \sqbr{ 1 + t_W^2 {\epsilon_n^2 m_{Z,0}^2 \ov M_n^2 - m_{Z,0}^2}  }.
	\label{eq:masssq-perturbed}
}
In this paper, we do not take into account the mass shift at the second order of $\epsilon_n$, in particular, $m_{Z,0} \simeq m_{Z}$.
}
Assuming that all of the dimensionless parameters $\br{ \epsilon_n }_{n=1,2,3,\cdots}$ are sufficiently small,
the unitary transformation  {which} diagonalises the mass terms of Eq.~\eqref{eq:S_eff-Vfree-1} are approximately given by (up to ${\cal O}\fn{\epsilon_n^2}$),
\al{
\wt{Z'}^{\paren{n}}_\mu
	&=
		{Z'}^{\paren{n}}_\mu - t_W {\epsilon_n m_{Z,0}^2 \ov M_n^2 - m_{Z,0}^2} Z_\mu + {\cal O}\fn{\epsilon_n^2}, \\
\wt{Z}_\mu
	&=
		Z_\mu + t_W \sum_n {\epsilon_n m_{Z,0}^2 \ov M_n^2 - m_{Z,0}^2} {Z'}^{\paren{n}}_\mu + {\cal O}\fn{\epsilon_n^2},
}
where the diagonalised form  {in terms of the mass eigenstates} is given as
\al{
{\cal L}_\tx{eff}^{\paren{V\tx{-free}}}
	&=
		-{1\ov 4} \sum_n {Z'}_{\mu\nu}^{\paren{n}} {Z'}^{\paren{n}\mu\nu}
		-{1\ov 4} {Z}_{\mu\nu} {Z}^{\mu\nu}
		-{1\ov 4} A_{\mu\nu} A^{\mu\nu}
		+{1\ov 2} m_{Z,0}^2 Z_\mu Z^\mu\nn
     &\quad   
		+{1\ov 2} \sum_n M_n^2 {Z'}_\mu^{\paren{n}} {Z'}^{\paren{n}\mu}
		+{\cal O}\fn{\epsilon_n^2}.
	\label{eq:L_eff_diagonalised}
}
{In the above expression} $Z_\mu$ and ${Z'}_\mu^{\paren{n}}$ are the $Z$ boson and the $n$-th $U(1)_{L_\mu-L_\tau}$ gauge boson in the mass eigenbasis, respectively.
We summarise the relations between the gauge eigenstates and the mass eigenstates below,
\al{
\wh{B}_\mu
	&=
		s_W Z_\mu + c_W A_\mu + \sum_n \sqbr{ {\epsilon_n\ov c_W} + s_W t_W {\epsilon_n m_{Z,0}^2\ov M_n^2-m_{Z,0}^2} }
			{Z'}^{\paren{n}}_\mu + {\cal O}\fn{\epsilon_n^2},
		\label{eq:KK-mixing-start} \\
W^3_\mu
	&=
		c_W Z_\mu - s_W A_{\mu} + c_W t_W \sum_n {\epsilon_n m_{Z,0}^2 \ov M_n^2 - m_{Z,0}^2} {Z'}^{\paren{n}}_\mu
		+ {\cal O}\fn{\epsilon_n^2}, \\
\wh{Z'}^{\paren{n}}_\mu
	&=
		\wt{Z'}^{\paren{n}}_\mu
		=
		{Z'}^{\paren{n}}_\mu - t_W {\epsilon_n m_{Z,0}^2 \ov M_n^2 - m_{Z,0}^2} Z_\mu + {\cal O}\fn{\epsilon_n^2}.
		\label{eq:KK-mixing-end}
}
In {terms of} the mass eigenstates, the effective Lagrangian {describing} the  interaction {terms takes the following form},
\al{
&{{\cal L}_\tx{eff,int}} \nn
	&=
		\sum_{a=e,\mu,\tau} \Bigg\{
		\ol{l}^a_R i \gamma^\mu \pal_\mu l^a_R
		+ g_1 s_W \ol{l}^a_R \gamma^\mu l^a_R { Z_{\mu}}
		+ e \, \ol{l}^a_R \gamma^\mu l^a_R  { A_{\mu}}\nn
	&\quad
		+ \ol{l}^a_R \gamma^\mu g_1 \sum_n
			\sqbr{ {\epsilon_n \ov c_W} + s_W t_W { \epsilon_n m_{Z,0}^2 \ov M_n^2 - m_{Z,0}^2 } } {Z'}^{\paren{n}}_\mu l^a_R
		+ \ol{l}^a_R \gamma^\mu g' {Q_{a}} \sum_n {f_n}
			\sqbr{ {Z'}^{\paren{n}}_\mu - t_W { \epsilon_n m_{Z,0}^2 \ov M_n^2 - m_{Z,0}^2 } Z_\mu }  l^a_R \nn
	&\quad
		+ \ol{\nu}^a_L i\gamma^\mu \pal_\mu \nu^a_L
		+ \ol{l}^a_L i\gamma^\mu \pal_\mu l^a_L
		+ {g_2\ov \sqrt{2}} \paren{ \ol{\nu}^a_L \gamma^\mu l^a_L W^+_\mu + \ol{l}^a_L \gamma^\mu \nu^a_L W^-_\mu } \nn
	&\quad
		+ \ol{\nu}^a_L {\gamma^\mu}
		\bigg\{
			{1\ov 2} \sqbr{ \paren{g_2c_W + g_1s_W} Z_\mu 
			+ \sum_n \paren{ \paren{ g_2c_W + g_1s_W } t_W {\epsilon_n m_{Z,0}^2 \ov M_n^2 - m_{Z,0}^2} 
			+ g_1 {\epsilon_n \ov c_W} } {Z'}^{\paren{n}}_\mu } \nn
	&\phantom{\quad + \ol{\nu}^a_L \bigg\{}
		+ g' {Q_{a}} \sum_n f_n \paren{ {Z'}^{\paren{n}}_\mu - t_W { \epsilon_n m_{Z,0}^2 \ov M_n^2 - m_{Z,0}^2 } Z_\mu }
		\bigg\} \nu^a_L \nn
	&\quad
		+ \ol{l}^a_L {\gamma^\mu}
		\bigg\{
			e A_\mu
			+ {1\ov 2} \sqbr{ \paren{-g_2c_W + g_1s_W} Z_\mu 
			+ \sum_n \paren{ \paren{ -g_2c_W + g_1s_W } t_W {\epsilon_n m_{Z,0}^2 \ov M_n^2 - m_{Z,0}^2} 
			+ g_1 {\epsilon_n \ov c_W} } {Z'}^{\paren{n}}_\mu } \nn
	&\phantom{\quad + \ol{\nu}^a_L \bigg\{}
		+ g' {Q_{a}} \sum_n f_n \paren{ {Z'}^{\paren{n}}_\mu - t_W { \epsilon_n m_{Z,0}^2 \ov M_n^2 - m_{Z,0}^2 } Z_\mu }
		\bigg\} l^a_L 
	\Bigg\} + {\cal O}\fn{\epsilon_n^2},
	\label{eq:S_eff-Vfree-2}
}
where we introduced the shorthand notation {for the $U(1)_{L_\mu - L_\tau}$ charges}:
\al{
{Q_{a}}
	&=
		\begin{cases}
		0 & \tx{for} \ a = e, \\
		+1 & \tx{for} \ a = \mu, \\
		-1 & \tx{for} \ a = \tau.
		\end{cases}
}
$l^a_{R,L}$ and $\nu^a_L$ ($a = e,\mu,\tau$) represent the charged leptons and neutrinos with definite chiralities in 4D.

\section{Cross section formula for light $Z'$ Bremsstrahlung
\label{sec:cross-section}}

\subsection{4D $U(1)_{L_\mu - L_\tau}$ Case
\label{sec:4D-case}}

\begin{figure}[t]
    \centering
    \includegraphics[width=0.8\linewidth]{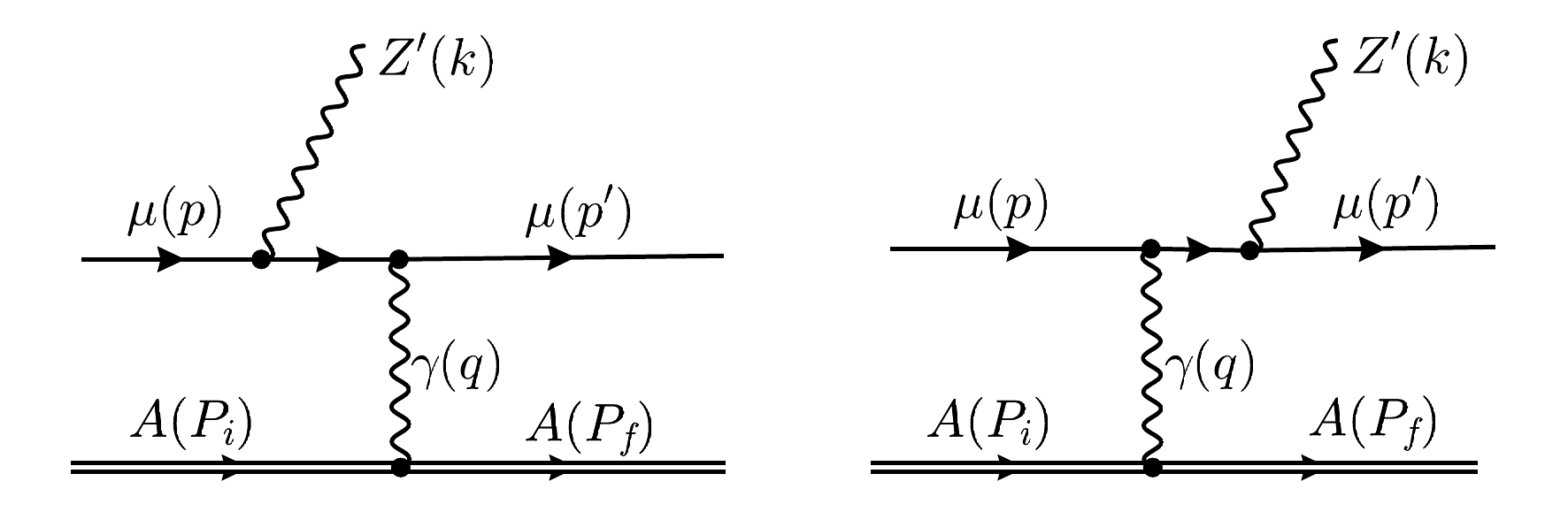}
    \caption{The lowest-order \(2 \to 3\) production process is given by: \(\mu(p) + A(P_i) \rightarrow \mu(p') + A(P_f) + Z'(k)\), where \(p\) and \(P_i\) denote the initial four-momenta of the incoming muon and target nucleus, respectively, and \(p'\) and \(P_f\) are the final-state four-momenta of the outgoing muon and recoiling nucleus. The outgoing \(Z'\) boson carries momentum \(k\), while \(q = P_i - P_f\) represents the momentum of the intermediate virtual photon exchanged in the process. Here, \(A\), \(\gamma\), \(\mu\), and \(Z'\) denote the target nucleus, photon, muon, and the new gauge boson, respectively.}
    \label{fig:2 to 3 cross section}
\end{figure}

In beam-dump experiments for light new particle searches, the significant process is the initial and final-state radiation (Bremsstrahlung) off an accelerated charged lepton, which bombards an atom of a fixed target.
For experiments {involving} an incident muon-beam, the relevant process and the kinematics are given by
\al{
\mu(p) + A(P_i) \rightarrow \mu(p') + A(P_f) + Z'(k),
	\label{eq:full-process-kinematics}
}
where the corresponding Feynman diagrams are {shown} in Fig.~\ref{fig:2 to 3 cross section}.
$A$ denotes an atom of the target, and $p$, $P_i = \paren{M_i, \pmb{0}}$, $p'$, $P_f$ and $k$ are the four momenta in the lab frame, where $M_i$ is the mass of the target atom {which is stationary in the lab frame}.

\begin{figure}[t]
    \centering
    \includegraphics[width=0.9\linewidth]{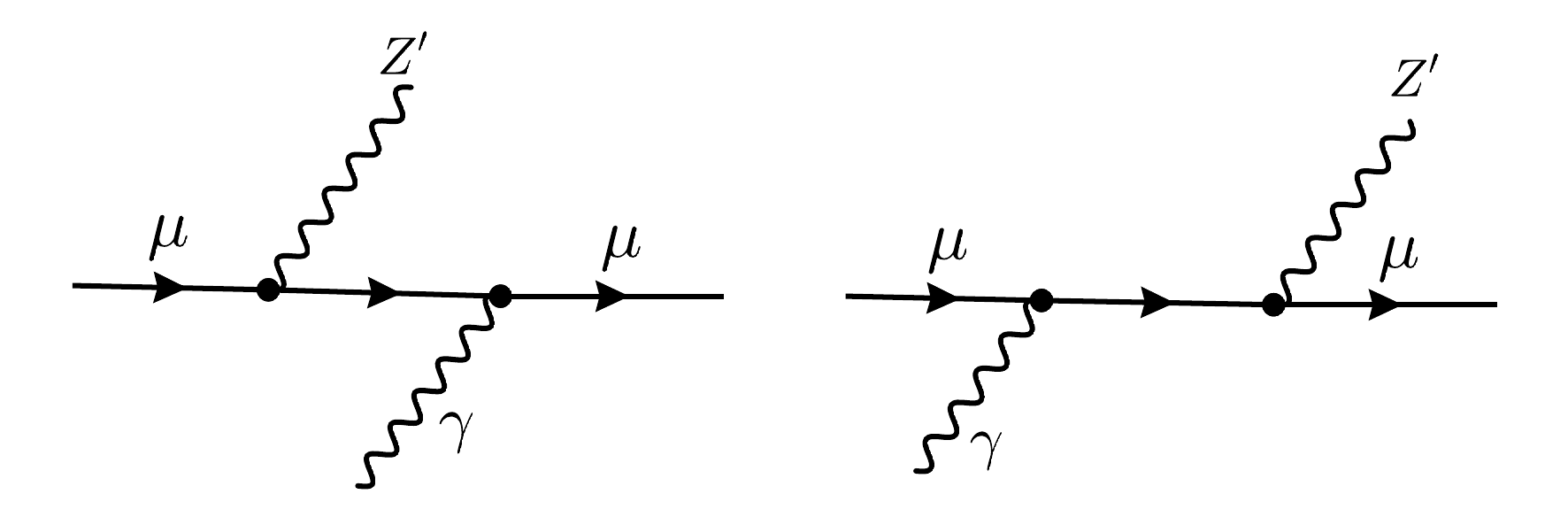}
    \caption{Feynman diagram illustrating a \(2 \to 2\) scattering process mediated by the 4-D gauge boson \(Z'\) associated with the \(U(1)_{L_\mu - L_\tau}\) symmetry.}
    \label{2 to 2 fig}
\end{figure}

The cross section of the $Z'$-boson Bremsstrahlung was formulated in~\cite{Bjorken:2009mm} for the dark photon, based on the discussion on the axion-Bremsstrahlung off an electron beam~\cite{Tsai:1986tx}.
If the initial charged lepton (muon) is sufficiently accelerated, the Weizs\"{a}cker-Williams~(WW) approximation is applicable~\cite{vonWeizsacker:1934nji,Williams:1935dka,Kim:1973he,Tsai:1973py,Tsai:1986tx,Bjorken:2009mm};
see also~\cite{Liu:2016mqv,Liu:2017htz,Kirpichnikov:2021jev,Voronchikhin:2024vfu} for detailed verifications/application limits of this approximation for Bremsstrahlung.\footnote{
It is also known as the method of virtual quanta, a semiclassical technique developed to simplify phase space integration in processes involving high-energy charged particles. It effectively models the interaction of a relativistic charged particle with an external electromagnetic field as a flux of equivalent real photons. Initially designed for bremsstrahlung calculations and later applied in fixed-target and beam-dump studies.
{Furthermore, an investigation into the nonzero momentum transfer in the Bremsstrahlung from the proton is found in~\cite{Gorbunov:2023jnx} (see also~\cite{Blumlein:2013cua,Foroughi-Abari:2021zbm,Foroughi-Abari:2024xlj}).}
}
The WW approximation transforms the full $2 \rightarrow 3$ process (see Eq.~\eqref{eq:full-process-kinematics} and Fig.~\ref{fig:2 to 3 cross section}) into a simpler \(2 \rightarrow 2\) scattering process (Fig.~\ref{2 to 2 fig}) by factoring out the photon flux via the target atom.\footnote{
Note that all of the Feynman diagrams in this manuscript are drawn with {\tt FeynGame\,3.0}~\cite{Harlander:2020cyh,Harlander:2024qbn, Bundgen:2025utt}.
}

{More precisely}, this approximation is valid when the beam energy is much larger than both the muon mass $m_\mu$ and the mass of the new vector boson \(m_{Z'}\). Under these conditions, the differential cross section for the full $2 \rightarrow 3$ process ($d\sigma_{2 \rightarrow 3}^{Z^{\prime}}$) is factorised and can be written as
\begin{equation}
    \frac{1}{E_0^2 x} \frac{d\sigma_{2 \rightarrow 3}^{Z^{\prime}}}{dx\, d\cos{\theta_{Z'}}}
    = \frac{8 \alpha^2 {g'}^2 {{Q_\mu}^2}  {\chi_{Z^{\prime}}}  \beta_{Z'}}{4 \pi}
    \left[
        \frac{1 - x + \frac{x^2}{2}}{u^2}
        + \frac{(1 - x)^2 m_{Z'}^2}{u^4}
        \left(m_{Z'}^2 - \frac{u x}{1 - x}\right)
    \right],
    \label{diffequation}
\end{equation}
where $E_0$ and \(\theta_{Z'}\) are the energy of the incoming single muon beam and the emission angle of the outgoing \(Z'\) boson (taken as relative to the beam axis) in the lab frame.
\(\alpha\) is the electromagnetic fine-structure constant, ${g'}$ is the 4D $U(1)_{L_\mu - L_\tau}$ gauge coupling, and {\(Q_\mu\) is the muon charge of the \(U(1)_{L_\mu - L_\tau}\) gauge interaction ($Q_\mu = +1$).}
The quantity ${\chi_{Z^{\prime}}}$ denotes the effective virtual photon flux, whose details will be provided in the next paragraph. 
The variable
\al{
x := {E_{Z'} \ov E_0}
}
is the fraction of energy carried by the $Z'$, {where  $E_{Z'}$} and $E_0$ denote the energies of $Z'$ and the initial muon, respectively.
$u$ is a kinematic function that depends on both \(x\) and \(\theta_{Z'}\).
Several {relevant} kinematic quantities are defined as {follows,}
\begin{align}
    \beta_{Z'}
    	&:= 
		\sqrt{1 - \frac{m_{Z'}^2}{E_0^2}}, \\
    q
    	&:=
		P_i - P_f, \\
    t 
    	&:= 
		-q^2,\\
    u = u\fn{x, \theta_{Z'}}
    	&:=
		{E_0^2} \theta_{Z'}^2 x + m_{Z'}^2 \left( \frac{1 - x}{x} \right) + m_\mu^2 x,
\end{align}
where $\beta_{Z'}$ is the boost factor of the outgoing \(Z'\) in the lab frame, $q$ represents the four momentum of the virtual photon, respectively.
{Note that, in the numerator of Eq.~\eqref{diffequation}, {except in} $u\fn{x, \theta_{Z'}}$, we {ignored} the terms proportional to $m_\mu^2$.
Also, the form of Eq.~\eqref{diffequation} agrees with that of Eq.~(A12) of~\cite{Bjorken:2009mm}, up to the difference in the types of the $U(1)$ interactions.
}

The photon flux $ {\chi_{Z^{\prime}}}$ appearing in Eq.~\eqref{diffequation} is determined by the electric form factor \(G_2(t)\) as~{\cite{Kim:1973he,Bjorken:2009mm}}
\begin{equation}
    {\chi_{Z^{\prime}}} := \int_{t_{\text{min}}}^{t_{\text{max}}} dt \, \frac{t - t_{\text{min}}}{t^2} \, G_2(t),
    \label{photonFulx}
\end{equation}
with
\al{
t_{\text{min}} &= \left(\frac{m_{Z'}^2}{2 E_0} \right)^2,&
t_{\text{max}} &= m_{Z'}^2.&
	\label{eq:tmin-tmax}
}
The magnetic form factor \(G_1(t)\) is neglected since its contribution is small in the regime of interest~\cite{Bjorken:2009mm}.
The full electric form factor $G_2(t)$ includes both elastic and inelastic components.
The elastic contribution is given by
\al{
    G_{2,\text{el}}(t) = \left(\frac{a^2 t}{1 + a^2 t}\right)^2 \left( \frac{1}{1 + \frac{t}{d}} \right)^2 Z^2,
}
with $a = {111 Z^{-1/3}}/{{m_e}}$ and $d = 0.164~\text{GeV}^2 A^{-2/3}$.
$m_e$ represents the electron mass, and  $Z$ and $A$ denote the atomic number and the atomic mass {number} of the target material, respectively.\footnote{The choice of target material varies across the considered experiments. For the \(\text{NA}64_{\mu}\) experiment, the target is lead with atomic number \(Z = 82\) and mass number \(A = 207.2\). In the MuSIC setup, gold is used as the target material, characterised by \(Z = 79\) and \(A = 196.97\). The \(\text{M}^3\) experiment employs tungsten, with \(Z = 74\) and \(A = 183.84\). Finally, for the proposed future muon beam dump experiment, the target is water, with an effective average atomic number \(Z = 10\) and mass number \(A = 18\)~\cite{ParticleDataGroup:2024cfk}.
}
The inelastic part is expressed as
\al{
    G_{2,\text{in}}(t) = \left(\frac{a'^2 t}{1 + a'^2 t}\right)^2
    \left(\frac{1 + \frac{t}{4 m_p^2}(\mu_p^2 - 1)}{(1 + \frac{t}{0.71~\text{GeV}^2})^4} \right)^2 Z,
}
with \( a' = {773 Z^{-2/3}}/{{m_e}} \), where \( m_p \) is the proton mass and \( \mu_p = 2.79 \) is the  magnetic moment {of the proton (in nuclear magneton)}.
The total electric form factor is given by,
\begin{equation}
     G_2(t) = G_{2,\text{el}}(t) + G_{2,\text{in}}(t).
\end{equation}

{Owing to $t_{\text{max}} = m_{Z'}^2$ in Eq.~\eqref{eq:tmin-tmax}, for heavier $Z'$, the associated momentum transfer to the nucleus $t$ becomes larger, causing the recoil to probe progressively shorter-distance nuclear structures.
For such energetic domains, additional contributions can arise via the photon exchange in the deep-inelastic scattering~(DIS) regime.
A simple but reasonable criterion can take
\al{
t_\tx{min} \gtrsim 1\,\tx{GeV}^2 \quad \tx{(as an estimation of the DIS-active domain)~\cite{Batell:2024cdl} (see also~\cite{Bjorken:2009mm})}.
	\label{eq:DIS-active-domain}
}
Also, we comment on the properties of the photon flux $\chi_{Z'}$ in Eq.~\eqref{photonFulx}:
(i) For a larger $E_0$ (with focusing on a specific $m_{Z'}$), $t_\tx{min}$ is smaller, and eventually $\chi_{Z'}$ becomes larger;
(ii) For a significantly large $m_{Z'}$ (with focusing on a specific $E_0$), $t_\tx{min}$ becomes larger and $\chi_{Z'}$ becomes very small.}

{Integrating $\cos\theta_{Z'}$ from $\theta_{Z' \tx{min}} = 0$ to $\theta_{Z'\tx{max}}$,}
the differential cross section with respect to the energy fraction \(x\) takes the simplified form,\footnote{
{{The integration over the angle is performed by substituting the variable $u$ with $\cos\theta_{Z'}$}.
For details, refer to Appendix A in~\cite{Bjorken:2009mm}.}
}
\al{
    \frac{d\sigma^{{Z'}}_{2 \rightarrow 3}}{dx} = 
    \frac{8 \alpha^2 {g'}^2 {Q_\mu^2} {\chi_{Z'}}}{4\pi} \, \beta_{Z'} \left(m_{Z'}^2 \frac{1 - x}{x} + m_\mu^2 x\right)^{-1}
    \left(1 - x + \frac{x^2}{3}\right),
    \label{diffcrosssecinx}
}
{{where $\theta_{Z'\tx{max}}$} was taken sufficiently large, say, as $\theta_{Z'\tx{max}} > \ \sim {\cal O}\fn{10^{-3}}$.
So, the result is saturated and is insensitive to the value of $\theta_{Z'\tx{max}}$.
This is consistent with Eq.~(5) in~\cite{Cesarotti:2022ttv}; also Eq.~(A14) in~\cite{Bjorken:2009mm} ({after accounting for the} difference in the types of the $U(1)$ interactions).}
The minimum and maximum of $x$ are kinematically determined as~\cite{Kirpichnikov:2021jev}
\al{
x_\tx{min} &= {m_{Z'} \ov E_0},&
x_\tx{max} &= 1 - {{m_{\mu}} \ov E_0}.&
	\label{eq:x-kinetic-range}
}

{{In absence of appreciable} kinetic mixing effect, the decay width of a static $Z'$ boson into a charged lepton-antilepton pair (\(l \bar{l}\), with \(l = \mu, \tau\)) is given by~\cite{Cesarotti:2022ttv}:
\begin{equation}
    \Gamma(Z' \rightarrow l \bar{l}) =
    \frac{ {{g'}^2} {Q_l^2}}{12 \pi} \, m_{Z'}
    \left(1 + \frac{2 m_l^2}{m_{Z'}^2} \right)
    \sqrt{1 - \frac{4 m_l^2}{m_{Z'}^2}} ,
    \label{decayrateformulallbar}
\end{equation}
while the corresponding decay width into a neutrino-antineutrino pair of the second and third generations (\(\nu_\mu, \nu_\tau\)) is given by~\cite{Escudero:2019gzq}:
\begin{equation}
    \Gamma(Z' \rightarrow \nu_\alpha \overline{\nu_\alpha}) =
    \frac{ {{g'}^2} {Q_{\nu_\alpha}^2}}{24 \pi} \, m_{Z'} .
    \label{decayrateformulanunubar}
\end{equation}
Here, \(m_l\) denotes the mass of the muon or tau lepton, and the above expressions are written under the assumption that no additional contributions from kinetic mixing are present.
}

We can also recast the differential cross section in Eq.~\eqref{diffequation} in terms of the boost factor \(\gamma\) and pseudorapidity \(\eta\) of the \(Z'\) {in the lab frame}, using the relations~\cite{Bjorken:2009mm}:\footnote{
{Note that the following expression of the pseudorapidity is equivalent to the definition $\eta = -\ln\sqbr{ \tan\fn{\theta \ov 2} }$, {where $\theta$ is} in the principal value range.}
}
\al{
{\theta_{Z'}} &= \sin^{-1}(\operatorname{sech}\eta),&
x &= \frac{\gamma m_{Z'}}{E_0}.&
	\label{eq:kinetic-variable-change}
}
The integration range for pseudorapidity is set as
\al{
{-\ln\left[ \tan\left( \frac{ \theta_{Z'\tx{max}} }{2} \right) \right]}
	\lesssim \eta \lesssim 
	{+\infty},
\qquad
1 \lesssim \gamma \lesssim \left(1 - \frac{m_\mu}{E_0}\right) \frac{E_0}{m_{Z'}}.
	\label{eq:kinetic-domain-gamma-and-eta}
}

{Finally, we will mention another possible amplitude for the signal process $\mu^- A \to \mu^- A Z'$.
If $Z'$ is attached to quarks via the nonzero kinetic mixing, $Z'$ can be emitted from a proton of the target, and the doubly virtual Compton scattering~(VVCS) off the hadronic state is not always suppressed~\cite{Beranek:2013nqa}.
In the case of the dark photon search in a tantalum-targetted electron beam dump experiment, the authors of~\cite{Beranek:2013nqa} showed that the VVCS signal process provides a $\sim \!\! 10\%$ extra contribution if the mass of the dark photon is sufficiently lighter than the initial beam energy.
Of course, when the kinetic mixing is set to zero, this contribution {\it vanishes.}
However, if we consider the case where the kinetic mixing is nonzero, we should bear this in mind.}\footnote{
In Ref.~\cite{Beranek:2013nqa}, another possibility of the signal process for the final state $e^- A \to e^- A \paren{ l^- l^+ }$ ($l$ being a charged lepton) was examined through a Bethe-Heitler trident production with a {spacelike} $Z'$.
The authors of~\cite{Beranek:2013nqa} concluded that this is negligible.
}

\begin{figure}[t]
    \centering
    \includegraphics[width=0.9\linewidth]{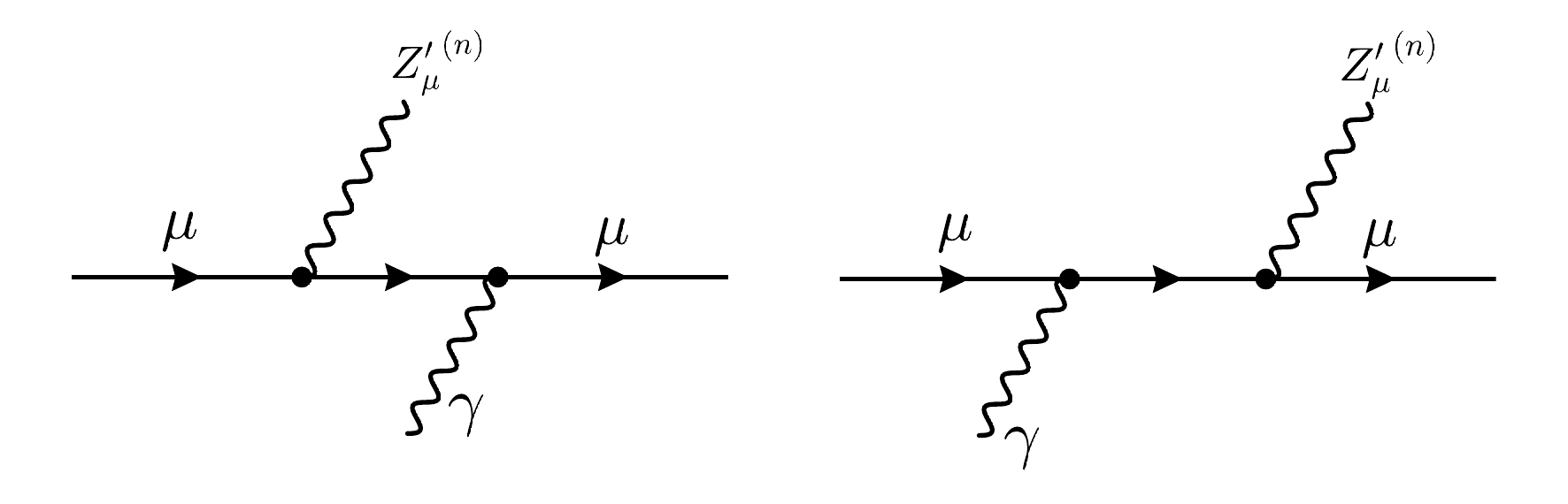}
    \caption{Feynman diagram illustrating a \(2 \to 2\) scattering process mediated by the extra-dimensional gauge boson \(Z^{\prime(n)}_{\mu}\) associated with the \(U(1)_{L_\mu - L_\tau}\) symmetry.}
    \label{fig:2to2ED}
\end{figure}

\subsection{Extra-Dimensional $U(1)_{L_{\mu}-L_{\tau}}$ Case {with kinetic mixing}
\label{sec:ED_cross_sec}}

As introduced in Section~\ref{sec:Model-Infro}, the (minimal) five-dimensional $U(1)_{L_\mu - L_\tau}$ scenario includes multiple $Z'$ bosons as KK vector bosons as $Z'^{(n)}$ $(n = 1,2,3,\cdots)$. As we will see later, their decay widths are proportional to their physical masses, and higher modes rapidly decay compared with lower modes. 
So, the interference effects between different KK modes (as intermediate states of the full chain of
$\mu^- A \to \mu^- A \paren{Z'^{(n)}}^\ast \to \mu^- A f \ol{f}$; $f$ is a $U(1)_{L_\mu - L_\tau}$-charged SM fermion) are negligible. {Further, the narrow 
decay widths of these KK modes as compared to the mass difference of two adjacent states imply that the resonances do not overlap, and that the cross section can be accurately estimated by considering one (on-shell) KK mode at a time, which decays to $ f \ol{f}$.}
Therefore, we need only consider the cross sections of Bremsstrahlung for each KK vector boson separately.

Under the above approximation, the calculation of each cross section is very straightforward.
{{In the following,} we derive a more complete expression that does not neglect the term proportional to the muon mass in the numerator, which we will use in our analysis.}
Using the WW approximation introduced in Section~\ref{sec:4D-case}, the differential cross section of the \(2 \to 3\) process {involving} the \(n\)-th KK mode \(Z'^{(n)}\)  {involving} {in the final state} can be calculated in terms of the energy fraction $x$ and the angle $\theta_{Z^{\prime(n)}}$ {(see Fig.~\ref{fig:2to2ED})} {with the help of {\tt FeynCalc\,10.1.0}~\cite{Mertig:1990an,Shtabovenko:2016sxi,Shtabovenko:2020gxv,Shtabovenko:2023idz}} as
\al{
& \quad \ \frac{1}{E_0^2 x} \frac{d\sigma_{2\rightarrow3}^{Z^{\prime(n)}}}{dx d\cos{\theta_{Z^{\prime(n)}}}} \notag \\
	&=
		\frac{\alpha^2  \chi_{Z^{\prime(n)}} \beta_{Z^{\prime(n)}}}{\pi M_n^2 u^4} \notag \\
	&\quad \times \Bigg\{
		\paren{C_{V}^{(n)}}^2 M_n^2 
		\Big[
		   2 M_n^4 \paren{1-x}^2 - 4 m_\mu^2 ux \paren{1-x} + 4 m_\mu^4x^2 \paren{1-x} \notag \\
	&\qquad \qquad \qquad
		   + u^2 \paren{2 + x\paren{x-2}} + 2M_n^2 \paren{1-x} \Big(-ux + m_\mu^2 \paren{2 + x\paren{x-2}} \Big)  
		\Big] \notag \\
	&\quad \ \ \  +
		\paren{C_{A}^{(n)}}^2
		\Big[
		   2 M_n^6 \paren{1-x}^2 + 2m_\mu^2 u^2 x^2 + M_n^2 \Big( 8 m_\mu^2 ux \paren{1-x} - 8 m_\mu^4 x^2 \paren{1-x}
		   + u^2 \paren{2 + x \paren{x-2}} \Big) \notag \\
	&\qquad \qquad \qquad
		   + 2 M_n^4 \paren{1-x} \Big( -ux + m_\mu^2 \paren{-4 + x \paren{x+4}} \Big)
		\Big]
		\Bigg\},
	\label{eq:diffequation-5d}
}
which results in the following form after the integration over $\cos{\theta_{Z^{\prime(n)}}}$:
\al{
    \frac{d\sigma_{2\rightarrow3}^{Z^{\prime(n)}}}{dx}
    &= \frac{\alpha^2 \chi_{Z^{\prime(n)}} \beta_{Z^{\prime(n)}}  x}
    		{3\pi M_n^2 \left(M_n^2 \left( 1-x \right) + m_\mu^2 x^2 \right)^2} \notag \\
    &\quad 
    \times \Bigg\{
        \paren{C_{V}^{(n)}}^2 M_n^2 
        \Big[
           m_\mu^2 x^2 \paren{x\paren{3x-4}+4} - 2 M_n^2 \paren{x-1} \paren{x\paren{x-3}+3}
        \Big] \notag \\
     &\quad \ \ \  +
        \paren{C_{A}^{(n)}}^2
         \Big[
            6 m_\mu^4 x^4 - 2 M_n^4 \paren{x-1} \paren{x\paren{x-3}+3} + m_\mu^2 M_n^2 x^2 \paren{x-4}\paren{3x-4}
         \Big]
    \Bigg\}.
    \label{diffcrosssecourmodel}
}
The coefficients \( C_{V}^{(n)} \) and \( C_{A}^{(n)} \) represent the vector and axial-vector parts of the interaction, defined as
\al{
    C_{V}^{(n)}
    &:=
        {\frac{1}{2}} \sqbr{ \frac{3}{2}g_1 \frac{\epsilon_n}{c_{{W}}}
        + \frac{1}{2}(3 g_1 s_{{W}} - g_2 c_{{W}}) t_{{W}} \frac{\epsilon_n m_{Z,0}^2}{M_n^2 - m_{Z,0}^2} }
        + f_n g' {Q_{\mu}},
    \label{Cplus} \\
    C_{A}^{(n)}
    &:=
    {\frac{1}{4}} \left[
        g_1 \frac{\epsilon_n}{c_{{W}}}
        + (g_1 s_{{W}} + g_2 c_{{W}}) t_{{W}} \frac{\epsilon_n m_{Z,0}^2}{M_n^2 - m_{Z,0}^2}
    \right].
    \label{CMinus}
}

Here, there are two differences between the formula for the 4D $U(1)_{L_\mu - L_\tau}$ $Z'$ in Eq.~\eqref{diffcrosssecinx} and one of the 5D ones in Eq.~\eqref{diffcrosssecourmodel}.
One is apparent; the gauge-boson mass $m_{Z'}$ should be replaced with the corresponding KK mass $M_n$.\footnote{
Here, we should be careful about the dependence of $m_{Z'}$ in $\beta_{Z'}$, $\chi_{Z'}$, also in $x_\tx{min}$ and $x_\tx{max}$.
}
The other one is more nontrivial.
Due to the kinetic mixing effect, the axial-vector coupling is present, and the form of the differential cross section becomes more involved.
{We point out that, if we take the limits $C^{(n)}_{A} \to 0$ (realised by $\epsilon_n \to 0$) and $m_\mu \to 0$, Eqs.~\eqref{eq:diffequation-5d} and \eqref{diffcrosssecourmodel} are reduced {to} the 4D forms in Eqs.~\eqref{diffequation} and \eqref{diffcrosssecinx} (up to {the} factor $f_n$), respectively.}
Refer to Section~\ref{sec:effective-Lagrangian} for the parameters appearing in Eqs.~\eqref{Cplus} and \eqref{CMinus}.

{The decay width of a static KK-mode \(Z^{\prime(n)}_\mu\) into a charged lepton pair (\(l \bar{l}\), with \(l = e,\mu, \tau\)) is given by  
\begin{equation}
    \Gamma\!\left(Z^{\prime(n)}_\mu \rightarrow l \bar{l}\right)
    =
    \frac{M_n}{24 \pi } \sqrt{1 - \frac{4 m_l^2}{M_n^2}}
    \left[
        \left({C_{Ll}^{(n)}}^2+{C_{Rl}^{(n)}}^2\right) +\frac{m_{{l}}^2}{M_n^2} \left(6  {C_{Ll}^{(n)}} {C_{Rl}^{(n)}}-{C_{Ll}^{(n)}}^2-{C_{Rl}^{(n)}}^2\right)
    \right],
    \label{decayRateZprimellbar}
\end{equation}
while the corresponding decay width into neutrino pairs (\(\nu_\alpha \bar{\nu_\alpha}\), with \(\nu_\alpha = \nu_e,\nu_\mu, \nu_\tau\)) is  
\begin{equation}
    \Gamma\!\left(Z^{\prime(n)}_\mu \rightarrow \nu_\alpha \bar{\nu_\alpha}\right)
    = \frac{M_n}{24 \pi}\, \left(C_{L {\nu_\alpha}}^{(n)}\right)^2 .
    \label{decayRateZprimetonunubar}
\end{equation}
The effective couplings are given by the chiral decomposition (in the 4D {perspective}):
\begin{align}
    C_{R{l}}^{(n)} &=
        g_1 \left(\frac{\epsilon_n}{c_{{W}}} 
        + s_{{W}} t_{{W}} \frac{\epsilon_n m_{Z,0}^2}{M_n^2 - m_{Z,0}^2} \right) 
        + f_n g' Q_{{l}},
    \label{CR} \\
    C_{L{l}}^{(n)} &=
        \frac{1}{2} \left[ 
            (-g_2 c_{{W}} + g_1 s_{{W}})\, t_{{W}} \frac{\epsilon_n m_{Z,0}^2}{M_n^2 - m_{Z,0}^2}
            + g_1 \frac{\epsilon_n}{c_{{W}}} 
        \right] 
        + f_n g' Q_{{l}},
    \label{CL} \\
    C_{L{\nu_\alpha}}^{(n)} &=
        \frac{1}{2} \left[ 
            (g_2 c_{{W}} + g_1 s_{{W}})\, t_{{W}} \frac{\epsilon_n m_{Z,0}^2}{M_n^2 - m_{Z,0}^2}
            + g_1 \frac{\epsilon_n}{c_{{W}}} 
        \right] 
        + f_n g' Q_{{\nu_\alpha}}.
\end{align}
It is straightforward to verify that, in the limit \(\epsilon_n \to 0\) and \(f_n \to 1\), the above expressions reduce to the standard four-dimensional results of Eqs.~\eqref{decayrateformulallbar} and~\eqref{decayrateformulanunubar}.}

\section{Experimental Setups and Signal Estimations
\label{sec:Experiments}}

We focus on the following muon beam dump experiments: NA$64_{\mu}$, M$^3$, MuSIC, and a future muon beam dump,
where the first two and the last two experiments investigate the decay channels $Z' \to \tx{invisible}$ and $Z' \to \mu^- \mu^+$, respectively.
In what follows, one by one, we {describe the methods} for connecting the cross section formula for the emission of $Z'$ or $Z'^{(n)} \, (n=1,2,3,\cdots)$, {as} discussed in Section~\ref{sec:cross-section}, to the number of events in the experiment.

\subsection{NA\texorpdfstring{$64_{\mu}$}{64mu}: Beam-Dump Experiment at CERN
\label{NA64}}

NA$64_{\mu}$ is a fixed-target beam-dump experiment at CERN, employing a 160~GeV muon beam from the Super Proton Synchrotron~(SPS)~\cite{Doble:1994np}. Its primary goal is to search for invisible decays of a light vector boson \( Z' \), produced via a bremsstrahlung process,
\als{
\mu^- A\fn{\tx{Pb}} \rightarrow \mu^- A\fn{\tx{Pb}} Z', \quad Z' \rightarrow \tx{invisible},
}
where the target is made of lead (Pb)~\cite{NA64:2024klw}.
Note that, for the 4D $U(1)_{L_\mu - L_\tau}$ scenario, the invisible final state dominantly consists of $Z' \to \nu_\mu \ol{\nu_\mu}$
and $Z' \to \nu_\tau \ol{\nu_\tau}$.
{In the presence of non-vanishing kinetic mixing}, the channel $Z' \to \nu_e \ol{\nu_e}$ may also be relevant.
For later convenience, we introduce the {notation} $\nu \ol{\nu} := \sum_{\alpha = e,\mu,\tau} \nu_\alpha \ol{\nu_\alpha}$.

The experimental signature is characterised by a single scattered muon carrying less than 80~GeV of energy, with no accompanying activity in the downstream detectors. In the 2022 run, the experiment collected a dataset corresponding to \( (1.98 \pm 0.02) \times 10^{10} \) muons on target ($N_{\text{MOT}}$), {and the future upgrades are expected to achieve \(10^{13}\) muons on target}~\cite{NA64:2024klw}.

In the 4D framework, the expected number of signal events corresponding to invisible decays of the form \( Z' \rightarrow \bar{\nu} \nu \) at the NA$64_\mu$ experiment is given by~\cite{Chen:2018vkr,Chen:2017awl}:
\begin{equation}
N_{\text{signal}}^{{\tx{4D,NA64${}_\mu$}}}
	= 
		\frac{N_{\text{MOT}}\, n_{\text{Pb}}}{\langle dE_{\mu}/dy \rangle} 
		\int_{E_{\mu,\text{min}}}^{E_{\mu,\text{max}}} {dE_{0}}
		\int_{x_{\text{min}}}^{x_{\text{max}}} dx \, 
		\frac{d\sigma_{2\rightarrow3}^{Z^{\prime}}}{dx} \, 
		\text{Br}\fn{Z^{\prime} \rightarrow \bar{\nu} \nu} \, N\left(m_{Z'}\right),
\label{signal_equation_NA64}
\end{equation}
where ${d\sigma_{2\rightarrow3}^{Z^{\prime}}}/{dx}$ and $\text{Br}\fn{Z^{\prime} \rightarrow \bar{\nu} \nu}$ represent the differential cross section and the branching ratio for the event (refer to Section~\ref{sec:cross-section} for the details).
\(N_{\text{MOT}}\) is the total number of muons on target, and \(n_{\text{Pb}} = 3.3 \times 10^{22}\,\text{cm}^{-3}\)~\cite{ParticleDataGroup:2024cfk} denotes the number density of lead nuclei in the target. The quantity \(\langle dE_\mu/dy \rangle\) represents the average energy loss per unit length for muons traversing the lead target. When looking at energies between a few GeV and about 160 GeV, muon energy loss is mostly caused by ionisation processes~\cite{NA64:2024nwj}. It is safe to assume that \(\langle dE_\mu/dy \rangle\)  {is independent of energy}  because it {does not} change much with the muon momentum in this range. For lead, we adopt the value \(\langle dE_\mu/dy \rangle = 12.7 \times 10^{-3}\,\text{GeV/cm}\)~\cite{ParticleDataGroup:2024cfk}.
The integration is performed over the penetrating muon energy \(E_\mu\), ranging from
\al{
E_{\mu,\text{min}} &= E_{\mu,\text{beam}} - L_T \langle dE_\mu/dy \rangle,&
E_{\mu,\text{max}} &= E_{\mu,\text{beam}},&
	\label{eq:E_mu-range-NA64mu}
}
where \(E_{\mu,\text{beam}} \approx 160\,\text{GeV}\) is the initial beam energy and \(L_T = 20\,\text{cm}\) is the total target length~\cite{Chen:2018vkr}.
The integration range for the energy fraction \(x\) is taken as in Eq.~\eqref{eq:x-kinetic-range},
ensuring physical consistency in the production process~\cite{Kirpichnikov:2021jev}.

$N(m_{Z'})$ is a normalisation factor treated as a function of the \(Z'\) mass. This factor accounts for background normalisation and has been calibrated using the 4D results reported by the NA64\(_\mu\) experimental analysis (given in Fig.~3 of~\cite{NA64:2024klw}) for the current bound corresponding to \(N_{\text{MOT}} \approx 2 \times 10^{10}\). For the projected sensitivity with \(N_{\text{MOT}} = 10^{13}\) ({future bound}), we assume a background-free scenario, and thus take \(N(m_{Z'}) = 1\).
A 90\% confidence level exclusion bound is imposed by requiring \(N_{\text{signal}} > 2.3\) in the \((m_{Z'}, g')\) parameter space~\cite{Lista:2016tva,NA64:2024nwj}.

In the case of the 5D scenario, the \(Z'\) boson is replaced by a tower of KK excitations, each denoted \(Z^{\prime(n)}\).
{We introduce the following notation for convenience,}
\al{
{N_{\text{signal}}^{\text{5D}, a}}
	&=
		{
		\sum_{n=1}^{{n_\tx{KKmax}}} N_{\text{signal}}^{(n), a}
		\quad
		\paren{a = \tx{Experiments}}.
		}
	\label{eq:def_5D-totalsignal}
}
The corresponding signal yield is obtained by modifying Eq.~\eqref{signal_equation_NA64} for each KK mode as
\al{
{N_{\text{signal}}^{(n),\tx{NA64${}_\mu$}}}
	&= 
		\frac{N_{\text{MOT}}\, n_{\text{Pb}}}{\langle dE_{\mu}/dy \rangle} 
		\int_{{E_{\mu,\text{min}}}}^{{E_{\mu,\text{max}}}} {dE_{0}}
		\int_{{x^{Z^{\prime(n)}}_{\text{min}}}}^{{x^{Z^{\prime(n)}}_{\text{max}}}} dx \, 
		\frac{d\sigma_{2\rightarrow3}^{Z^{\prime(n)}}}{dx} \, 
		\text{Br}\fn{Z^{\prime(n)} \rightarrow \bar{\nu} \nu} 
		N\fn{{M_n}},
		\label{eq:N_nKK_NA64mu}
}
where, as discussed in the first part of this section, the total number of signal events is provided as the sum of each $Z'^{(n)}$'s contribution, up to the $n_\tx{KKmax}$-mode.
Here, the variable for integration $x$ should be understood as $x = E_{Z'^{(n)}}/E_0$, where its domain {is} appropriately recognised as ${x^{Z^{\prime(n)}}_{\text{min}}} = M_n/E_0$ and ${x^{Z^{\prime(n)}}_{\text{max}}} = 1- {m_{\mu}}/E_0$ (see Eq.~\eqref{eq:x-kinetic-range}).

$N\fn{{M_n}}$ is the normalisation factor, defined as a function of the KK-mode mass \({M_n}\).
For the current bound, this factor is taken to be the same as in the 4D scenario in order to account for background contributions.
For the projected future bound under the assumption of a background-free analysis, we set $N\fn{{M_n}} = 1$.

\subsection{$\text{M}^3$: Muon Missing Momentum Experiment at Fermilab}

The M$^3$ experiment is a proposed fixed-target initiative at the Fermilab Accelerator Complex, designed to explore new muon-philic particles that decay invisibly~\cite{Kahn:2018cqs,Chen:2017awl}. The experimental concept builds upon the design principles of the LDMX experiment~\cite{LDMX:2018cma, Mans:2017vej}, utilising a high-intensity \(15\,\mathrm{GeV}\) muon beam to probe final states characterised by substantial missing momentum, which may signal the emission of undetected particles such as dark photons or new gauge bosons.
The signal process is
\als{
\mu^- A\fn{\tx{W}} \rightarrow \mu^- A\fn{\tx{W}} Z', \quad Z' \rightarrow \tx{invisible},
}
where the target is made of tungsten (W)~\cite{Kahn:2018cqs}.

The detector architecture features a precision tracking system positioned both upstream and downstream of a thick tungsten target of approximately $50 \, X_0$, enabling accurate measurement of the incoming and outgoing muon momenta. A downstream electromagnetic calorimeter (ECAL) and hadronic calorimeter (HCAL) system serves to veto visible final-state activity and suppress SM backgrounds.
The experimental program is expected to proceed in two phases~\cite{Kahn:2018cqs}:
\begin{itemize}
    \item \textbf{Phase 1:} \(N_{\text{MOT}} = 10^{10}\) muons on target and is primarily sensitive to light mediators with masses \(m_{Z'} \lesssim 500\,\mathrm{MeV}\).
    \item \textbf{Phase 2:} \(N_{\text{MOT}} = 10^{13}\) muons with enhanced veto capabilities, targeting background rejection at the \(10^{-13}\) level.
\end{itemize}

\noindent In the 4D framework, the expected number of signal events for invisible decays of the type \(Z' \rightarrow \bar{\nu} \nu\) is given by~\cite{Kahn:2018cqs,Chen:2017awl}:
\al{
N_{\text{signal}}^{{\tx{4D,M${}^3$}}}
	= 
		\frac{N_{\text{MOT}}\, n_{\text{W}}}{\langle dE_{\mu}/dy \rangle} 
		\int_{E_{\mu,\text{min}}}^{E_{\mu,\text{max}}}
		{dE_{0}}
		\int_{x_{\text{min}}}^{x_{\text{max}}} dx\, \frac{d\sigma_{2\rightarrow3}^{Z'}}{dx} \, 
		\text{Br}(Z' \rightarrow \bar{\nu} \nu) \, \alpha(x, m_{Z'}),
\label{signal_equation}
}
which takes a similar form as in Eq.~\eqref{signal_equation_NA64} for the NA$64_{\mu}$ experiment.
\(n_{\text{W}} = 6.3199 \times 10^{22}\,\mathrm{cm}^{-3}\) denotes the number density of tungsten nuclei in the target material~\cite{ParticleDataGroup:2024cfk}, while \(\langle dE_{\mu}/dy \rangle = 22.1 \times 10^{-3}\,\mathrm{GeV/cm}\) represents the average energy loss per unit length for muons traversing tungsten~\cite{Chen:2017awl}. 
The integration over the penetrating muon energy \(E_\mu\) extends as in Eq.~\eqref{eq:E_mu-range-NA64mu}, whereas the incident muon beam energy is taken to be \(E_{\mu,\text{beam}} \approx 15\,\mathrm{GeV}\)~\cite{Kahn:2018cqs}.
The target thickness is set to \(L_T = 17.52\,\mathrm{cm}\), corresponding to a material configuration with \(100\%\) tungsten purity and equivalent to a radiation length of \(50\,X_0\)~\cite{Chen:2018vkr,ParticleDataGroup:2024cfk}. 
The integration limits for the energy fraction $x$ are determined as in Eq.~\eqref{eq:x-kinetic-range}, which ensure the physical validity of the production process~\cite{Kirpichnikov:2021jev}.

The factor \(\alpha(x, m_{Z'})\) represents the detector acceptance as a function of \(x\) and \(m_{Z'}\), extracted by interpolating the data provided in Fig.~5 of~\cite{Kahn:2018cqs}.
In this analysis, we approximate the decay probability density per unit length as unity, justified by the thick-target regime, the invisibility of the final-state particles, and the fact that all geometric and detector efficiency effects are already incorporated into the acceptance factor \(\alpha(x, m_{Z'})\), as described in~\cite{Kahn:2018cqs}.
A 90\% confidence level exclusion bound is imposed by requiring the predicted number of signal events to satisfy \(N_{\text{signal}} > 2.3\)~\cite{Rohlf:1994wy}.

In ED extensions, the \(Z'\) boson is generalised to a tower of KK excitations, denoted by \(Z^{\prime(n)}\). The corresponding signal yield generalizes Eq.~\eqref{signal_equation} as:
\al{
{N_{\text{signal}}^{(n),\tx{M${}^3$}}}
	=
		\frac{N_{\text{MOT}}\, n_{\text{W}}}{\langle dE_{\mu}/dy \rangle}
		\int_{E_{\mu,\text{min}}}^{E_{\mu,\text{max}}}
		{dE_{0}}
		\int_{{x^{Z^{\prime(n)}}_{\text{min}}}}^{{x^{Z^{\prime(n)}}_{\text{max}}}} dx \, 
		\frac{d\sigma_{2\rightarrow3}^{Z^{\prime(n)}}}{dx} \, 
		\text{Br}(Z^{\prime(n)} \rightarrow \bar{\nu} \nu)\, \alpha(x, {M_n}),
		\label{eq:N_nKK_M3}
}
{where the total signal is defined as in Eq.~\eqref{eq:def_5D-totalsignal}.}
The factor \(\alpha(x, {M_n})\) denotes the detector acceptance as a function of the energy fraction \(x\) and the $n$-th KK boson mass ${M_n}$, and is extracted using the same interpolation procedure employed in the 4D case.

\subsection{MuSIC: Muon Synchrotron Ion Collider}

The Muon Synchrotron Ion Collider~(MuSIC) is a proposed high-energy muon-proton collider designed to explore physics beyond the SM, particularly in the muon sector~\cite{Acosta:2022ejc,Acosta:2021qpx}.
It builds upon existing infrastructure at Brookhaven National Laboratory~(BNL) and has been envisioned as a successor to the Electron-Ion Collider (EIC) following its projected completion in the 2040s.
In addition to probing new regimes of deep inelastic scattering in nuclear physics, MuSIC offers a promising opportunity to investigate rare muon processes, precision measurements, and the origin of the muon anomalous magnetic moment \( (g-2)_\mu \).

We focus on a similar scenario that the baseline configuration of MuSIC assumes a centre-of-mass energy of \( {E_0} = 8.9\,\text{TeV} \) and an integrated luminosity of \( L_I = 2000\,\text{fb}^{-1} \).\footnote{
{We have verified that, for this configuration of $E_0$ and $L_I$, the result obtained by taking only the $n_\text{KK} = 1$ case of our five-dimensional $U(1)_{L_\mu - L_\tau}$ theory (that will be shown in Fig.~\ref{fig:yzero_epsilon-zero}, for $\eps_4 = 0$) is in close agreement with the result for the four-dimensional effective $Z'$ description coupled with the muons calculated in Fig.~2 of~\cite{Davoudiasl:2024fiz}.}
}
One compelling physics target is the search for light muon-philic vector bosons \( Z' \), which may couple to \( L_\mu - L_\tau \), and decay visibly to a muon pair.
The \( Z' \) boson can be produced in muon-nucleus collisions via a bremsstrahlung-like process:
\als{
\mu^-\,A\fn{\text{Au}} \rightarrow \mu^-\,A\fn{\text{Au}}\, Z', \quad Z' \rightarrow \mu^- \mu^+,
}
where a high-energy muon interacts with a stationary gold (Au) nucleus as a target~\cite{Davoudiasl:2024fiz}.

In the 4D case, the number of observable events from displaced vertex signatures is given by~\cite{Davoudiasl:2024fiz},
\begin{equation}
N_{\text{signal}}^{{\tx{4D,MuSIC}}}
	= 
		L_I \int d\gamma\, d\eta\, \frac{d\sigma_{2\rightarrow3}^{Z^{\prime}}}{d\gamma\, d\eta} \,
		\text{Br}(Z' \rightarrow \mu^+ \mu^-) \, \epsilon_{\text{det}}\, 
		\mathcal{P}\fn{\ell_{\text{min}}, \ell_{\text{max}}},
\end{equation}
where $\gamma$ and $\eta$ are the Lorentz boost factor and the pseudorapidity of $Z'$ in the lab frame, respectively.
{The relations and the kinematically allowed domains have been given by  Eq.~\eqref{eq:kinetic-variable-change} and Eq.~\eqref{eq:kinetic-domain-gamma-and-eta}, respectively}.
Note that the differential production cross section in \((\gamma, \eta)\) coordinates is expressed as
\begin{equation}
\frac{d\sigma}{d\gamma\, d\eta} = \frac{m_{Z'}}{E_0} \operatorname{sech} \eta \sin \theta\, \frac{d\sigma}{dx\, d\cos \theta},
\end{equation}
with \( x = \gamma m_{Z'}/E_0 \) and \( \theta = \sin^{-1}(\operatorname{sech} \eta) \).
{Refer to Eq.~\eqref{diffequation} for ${d\sigma_{2\rightarrow3}^{Z^{\prime}}}/\paren{dx\, d\cos\theta}$.}

The detection efficiency is assumed to be \( \epsilon_{\text{det}} = 0.86 \) for the dimuon channel~\cite{Bandyopadhyay:2022klg}.
Since the measurement of $Z' \to \mu^- \mu^+$ is made by a displaced detector far from the target, we should take into account the reduction of the signals due to the decay of $Z'$ during the propagation.
The decay probability density is modelled as~\cite{Davoudiasl:2024fiz}:
\al{
P(\ell) = \frac{e^{-\ell/\ell_{Z'}}}{\ell_{Z'}}, \quad \text{with} \quad \ell_{Z'} := \frac{\gamma}{\Gamma_{Z'}},
\label{probabilityMuSIC}
}
where $\Gamma_{Z'}$ is the total decay width of a static $Z'$, and the boost factor $\gamma$ describes the lifetime dilation for observers in the lab frame. 
The probability that the \(Z'\) boson decays within the fiducial volume of interest, denoted by
$\mathcal{P}\fn{\ell_{\text{min}}, \ell_{\text{max}}}$, is obtained by integrating Eq.~\eqref{probabilityMuSIC} as
\al{
\mathcal{P}\fn{\ell_{\text{min}}, \ell_{\text{max}}}
	&:=
		\int_{\ell_\tx{min}}^{\ell_\tx{max}} d\ell P(\ell),
}
with the minimum distance \(\ell_{\text{min}} = 1\,\text{mm}\) and the maximum distance \(\ell_{\text{max}} = 30\,\text{m}\), corresponding to the location of the downstream detector capable of reconstructing the decay products~\cite{Davoudiasl:2024fiz}.
We impose a threshold of \( N_{\text{signal}} > 5 \) signal events to define the exclusion limit for the MuSIC experiment~\cite{Cesarotti:2022ttv}.

In the 5D case, the vector boson \( Z' \) originates from a tower of KK excitations \( Z^{\prime(n)} \), each contributing to the signal. The total number of displaced dimuon events is obtained, {as in Eq.~\eqref{eq:def_5D-totalsignal},} by summing over {each KK mode:}
\al{
{N_{\text{signal}}^{(n),\tx{MuSIC}}}
	= 
		L_I 
		\int d\gamma\, d\eta\, \frac{d\sigma_{2\rightarrow3}^{Z^{\prime (n)}}}{d\gamma\, d\eta} \, 
		\text{Br}(Z^{\prime(n)} \rightarrow \mu^+ \mu^-) \,
		\epsilon_{\text{det}}\, \mathcal{P}^{(n)}\fn{\ell_{\text{min}}, \ell_{\text{max}}},
		\label{eq:N_nKK_MuSIC}
}
where the kinetic variables $\gamma$ and $\eta$ should be understood as those for the vector boson with mass $M_n$.
{Refer to Eq.~\eqref{eq:diffequation-5d} for the form of the differential cross section without ignoring the muon mass.}
We should also be careful about the corresponding decay probability $\mathcal{P}^{(n)}\fn{\ell_{\text{min}}, \ell_{\text{max}}}$.
The superscript \((n)\) indicates that both the production cross section and decay probability are mode-dependent. In particular, the decay length for each KK mode is
\al{
\ell_{Z^{\prime(n)}} := \frac{\gamma}{\Gamma_{Z^{\prime(n)}}},
	\label{eq:ell_KKZprime}
}
which impacts the probability \(\mathcal{P}^{(n)}(\ell_{\text{min}}, \ell_{\text{max}})\) for decays occurring within the detector volume. Due to increasing masses and typically shorter lifetimes of higher KK modes, only a subset of modes contributes appreciably within the experimental geometry.

\subsection{Future High-Energy Muon Beam Dump Experiment}

A high-energy muon beam dump setup provides a cost-effective and complementary strategy for probing weakly coupled new physics, particularly in regimes that may be inaccessible at high-luminosity colliders~\cite{Bjorken:2009mm,Cesarotti:2022ttv}. In this configuration, a 1.5~TeV muon beam is directed onto a dense, thick target such as water~\cite{Cesarotti:2022ttv}. This enhances the probability of producing new light mediators like the \(Z'\) boson via a bremsstrahlung-like process. These mediators can travel downstream before decaying into visible particles, which are subsequently detected by a spectrometer system placed after the shielding.
The signal process is
\als{
\mu^-\,A\fn{\tx{H}_2 \tx{O}} \rightarrow \mu^-\,A\fn{\tx{H}_2 \tx{O}}\, Z', \quad Z' \rightarrow \mu^- \mu^+,
}
where the target is made of water ($\tx{H}_2 \tx{O}$)~\cite{Cesarotti:2022ttv}.

For the 4D scenario, the number of signal events along the beamline is given by~\cite{Cesarotti:2022ttv}
\al{
N_{\text{signal}}^{{\tx{4D,FMBD}}}
	= 
		N_{\mu}
		\int_{x_{\text{min}}}^{x_{\text{max}}} dx \, 
		\frac{\rho \, \ell_{Z'}}{m_T} \, \frac{d\sigma_{2\rightarrow3}^{Z^{\prime}}}{dx} \,
		\text{Br}(Z^{\prime} \rightarrow \mu^{+}\mu^{-}) \times
		\left(e^{L_{\text{tar}}/\ell_{Z'}} - 1\right) 
		e^{-(L_{\text{tar}} + L_{\text{sh}})/\ell_{Z'}} 
		\left(1 - e^{-L_{\text{dec}}/\ell_{Z'}}\right),
}
where \( N_{\mu} = 10^{22} \) is the total number of muons incident on the target~\cite{Cesarotti:2022ttv},
the quantity \( m_T = A / N_0 \) represents the mass of a target atom, with \( A \) being the atomic mass number and \( N_0 \) Avogadro’s number, while {$\rho \, ( = 0.997\,\tx{g}/\tx{cm}^3)$} is the target density of $\tx{H}_2 \tx{O}$~\cite{ParticleDataGroup:2024cfk}.
See Eq.~\eqref{probabilityMuSIC} for the definition of $\ell_{Z'}$.

Like the MuSIC experiment, the signal part $Z' \to \mu^- \mu^+$ is measured by a displaced detector.
{We adopt} the specific form of the decay probability for this experimental setup {as} discussed in~\cite{Cesarotti:2022ttv}.
The lengths \(L_{\text{tar}}\), \(L_{\text{sh}}\), and \(L_{\text{dec}}\) correspond to the target, shielding, and decay regions, respectively. For numerical analysis, we adopt benchmark values \(L_{\text{tar}} = 10\,\text{m}\), \(L_{\text{sh}} = 10\,\text{m}\), and \(L_{\text{dec}} = 100\,\text{m}\)~\cite{Cesarotti:2022ttv}. 
We impose a threshold of \( N_{\text{signal}} > 5 \) signal events to define the exclusion limit for this experiment~\cite{Cesarotti:2022ttv}.
In this regime, {radiative} energy losses are assumed to be negligible in this setup, allowing the energy distribution function to be treated as sharply peaked around the initial muon energy.

In the case of the 5D extension, the signal receives contributions from a tower of KK modes \(Z^{\prime(n)}\), {as in Eq.~\eqref{eq:def_5D-totalsignal}.}
{The contribution from the $n$th mode is given as}
\al{
{N_{\text{signal}}^{(n),\tx{FMBD}}}
	&=  
		N_{\mu}
		\int_{{x^{Z^{\prime(n)}}_{\text{min}}}}^{{x^{Z^{\prime(n)}}_{\text{max}}}} dx  \, 
		\frac{\rho \, \ell_{Z'^{(n)}}}{m_T} \, 
 		\frac{d\sigma_{2\rightarrow3}^{Z^{\prime(n)}}}{dx} \,
		\text{Br}(Z^{\prime(n)} \rightarrow \mu^{+}\mu^{-})  \nonumber \\
	&\quad \times 
		\left(e^{L_{\text{tar}}/\ell_{Z'^{(n)}}} - 1\right) 
		e^{-(L_{\text{tar}} + L_{\text{sh}})/\ell_{Z'^{(n)}}} 
		\left(1 - e^{-L_{\text{dec}}/\ell_{Z'^{(n)}}}\right),
		\label{eq:N_nKK_FMBD}
}
where the variable for integration $x$ should be understood as $x = E_{Z'^{(n)}}/E_0$ (also for ${{x^{Z^{\prime(n)}}_{\text{min}}}}$ and ${{x^{Z^{\prime(n)}}_{\text{max}}}}$; see Eq.~\eqref{eq:x-kinetic-range}).
Refer to Eq.~\eqref{eq:ell_KKZprime} for the definition of $\ell_{Z'^{(n)}}$.

\section{Results
\label{sec:Results}}

\begin{figure}[t]
\centering
\includegraphics[width=0.38\textwidth]{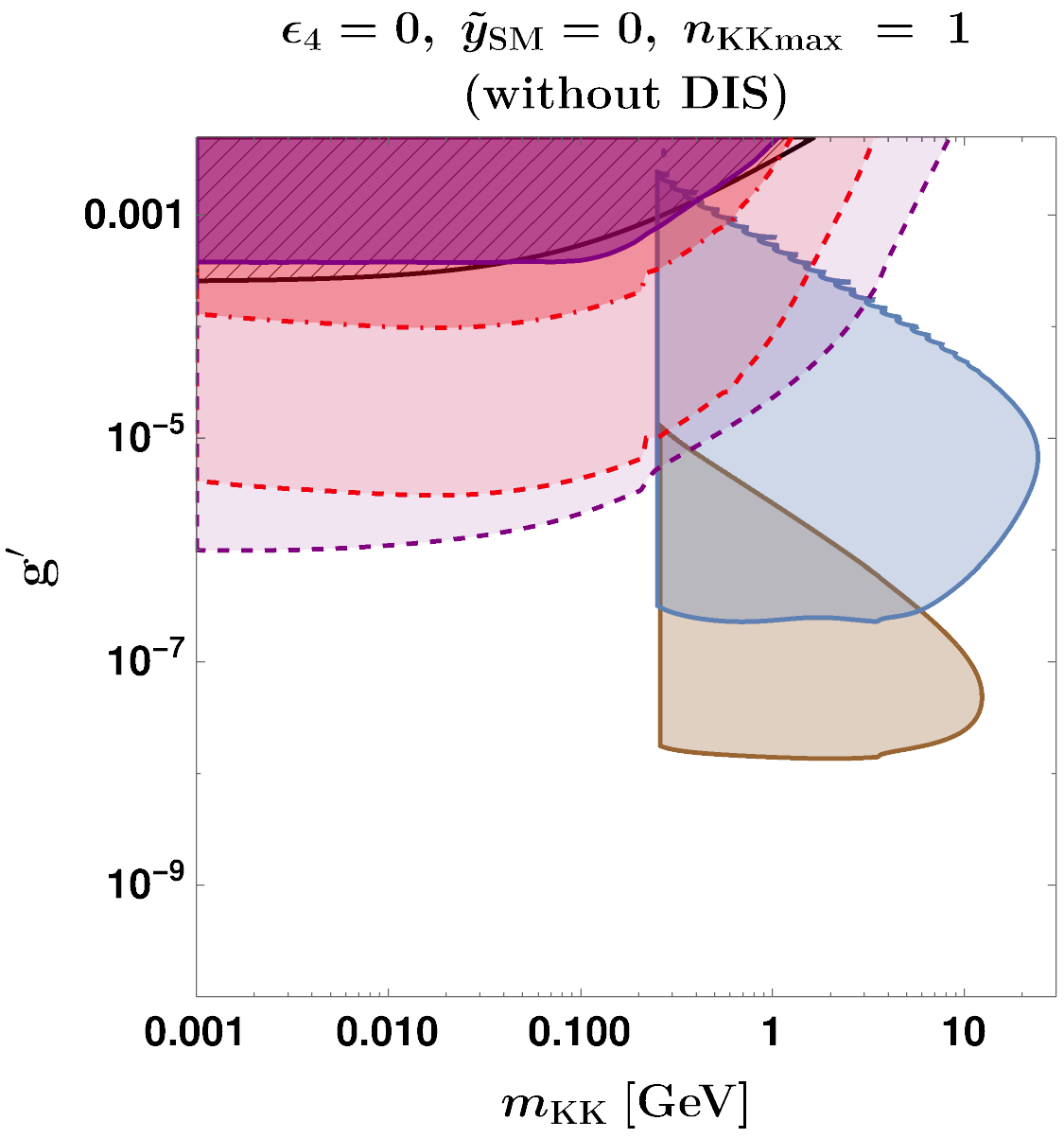} \
\includegraphics[width=0.38\textwidth]{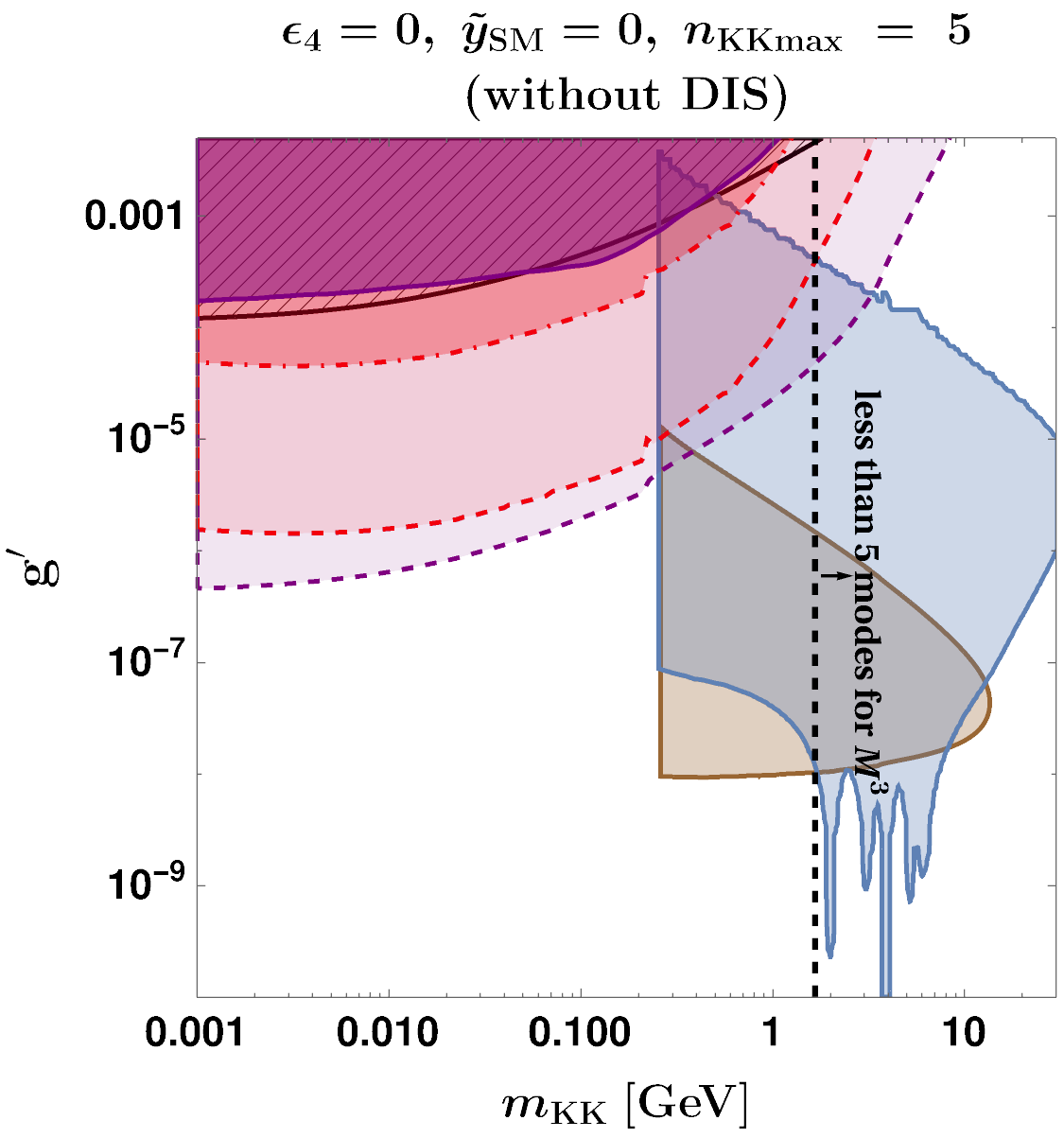}
\raisebox{40pt}{\includegraphics[width=0.20\textwidth]{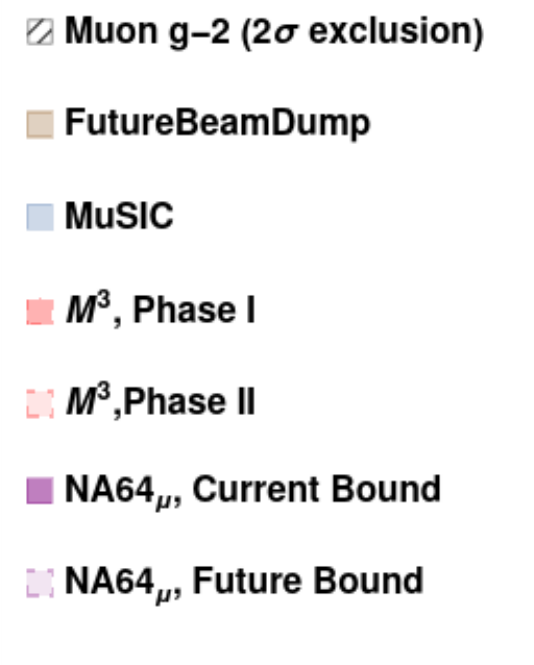}}
\caption{
Summary of {current/future} exclusion limits in the \( (m_{\text{KK}},\, g') \) parameter space for the 5D \(U(1)_{L_\mu - L_\tau}\) model under the benchmark choices \(\epsilon_4 = 0\) and \(\wt{y}_\tx{SM} = 0\).
They illustrate the sensitivity of various experiments for two KK mode truncations: \(n_{\text{KKmax}} = 1\) (Left Panel) and \(n_{\text{KKmax}} = 5\) (Right Panel).
For the meaning of the vertical black dotted line in the right panel, refer to Footnote~\ref{foot:KK-treatments}.
{For the definitions of each exclusion limit via the beam dump experiments, refer to Section~\ref{sec:Experiments}.}
}
\label{fig:yzero_epsilon-zero}
\end{figure}

\begin{figure}[t]
  \centering
  \begin{minipage}[c]{0.495\textwidth}
    \centering
    \begin{minipage}[c]{0.75\textwidth}
      \includegraphics[width=\textwidth]{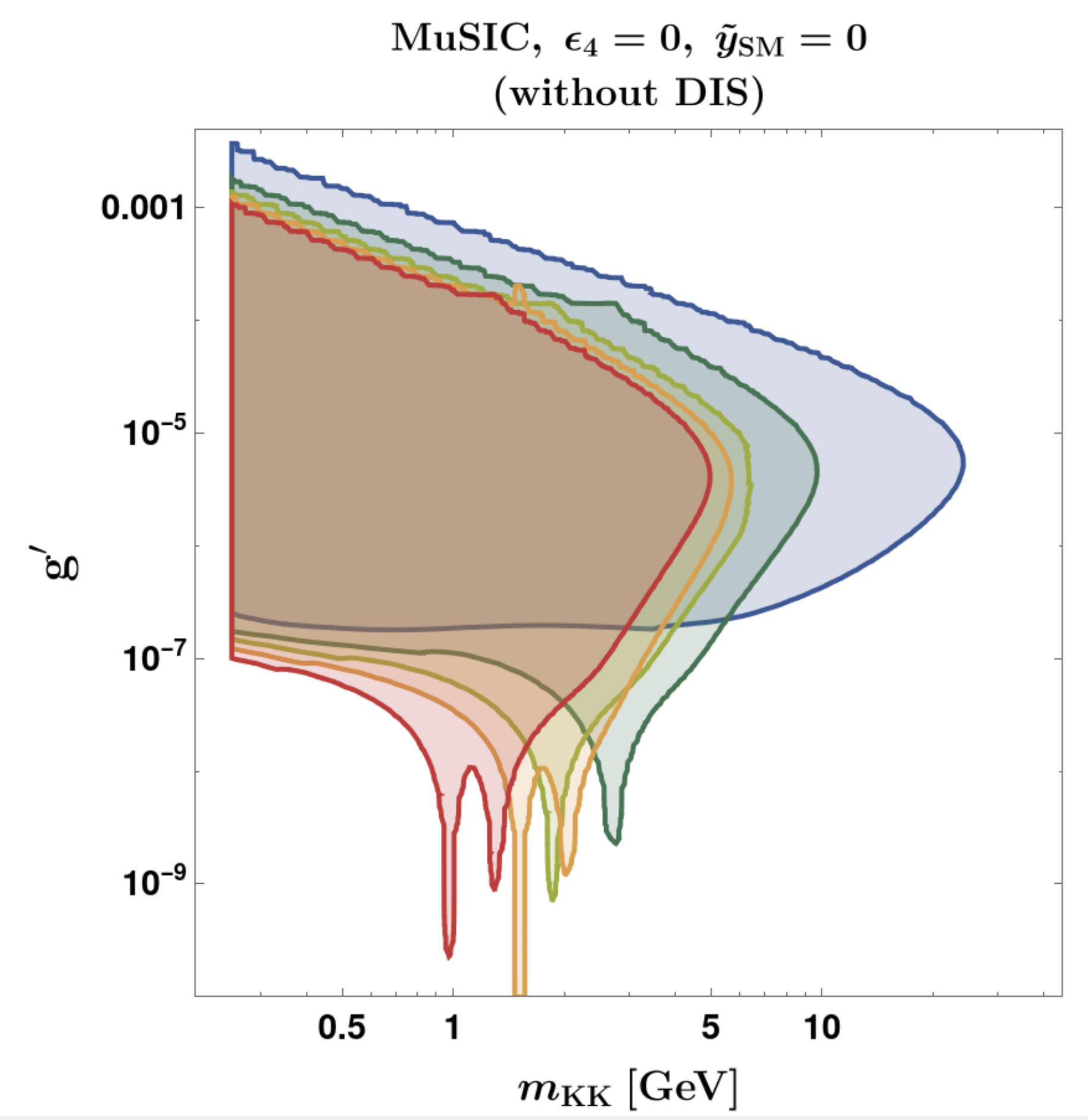}
    \end{minipage}%
    \hspace{-0.4em}
    \begin{minipage}[c]{0.20\textwidth}
      \includegraphics[width=\textwidth]{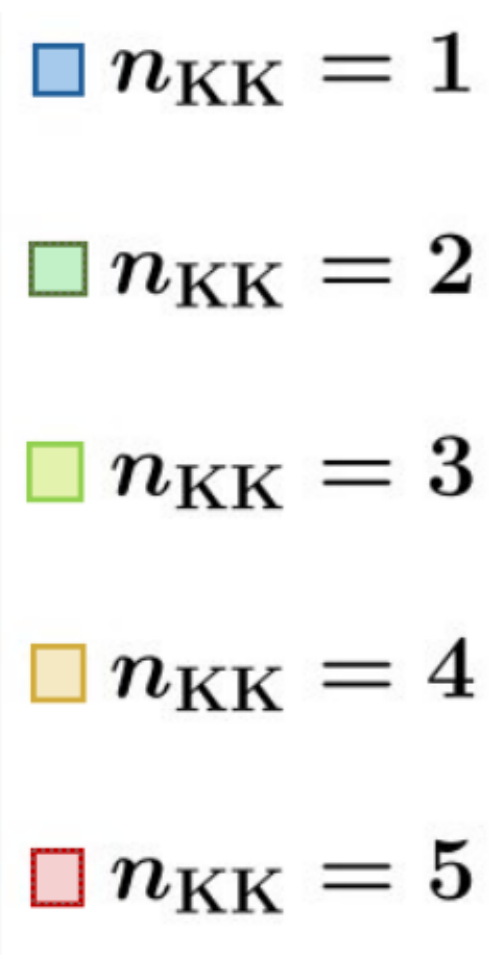}
    \end{minipage}
  \end{minipage}
  \begin{minipage}[c]{0.495\textwidth}
    \centering
    \begin{minipage}[c]{0.75\textwidth}
      \includegraphics[width=\textwidth]{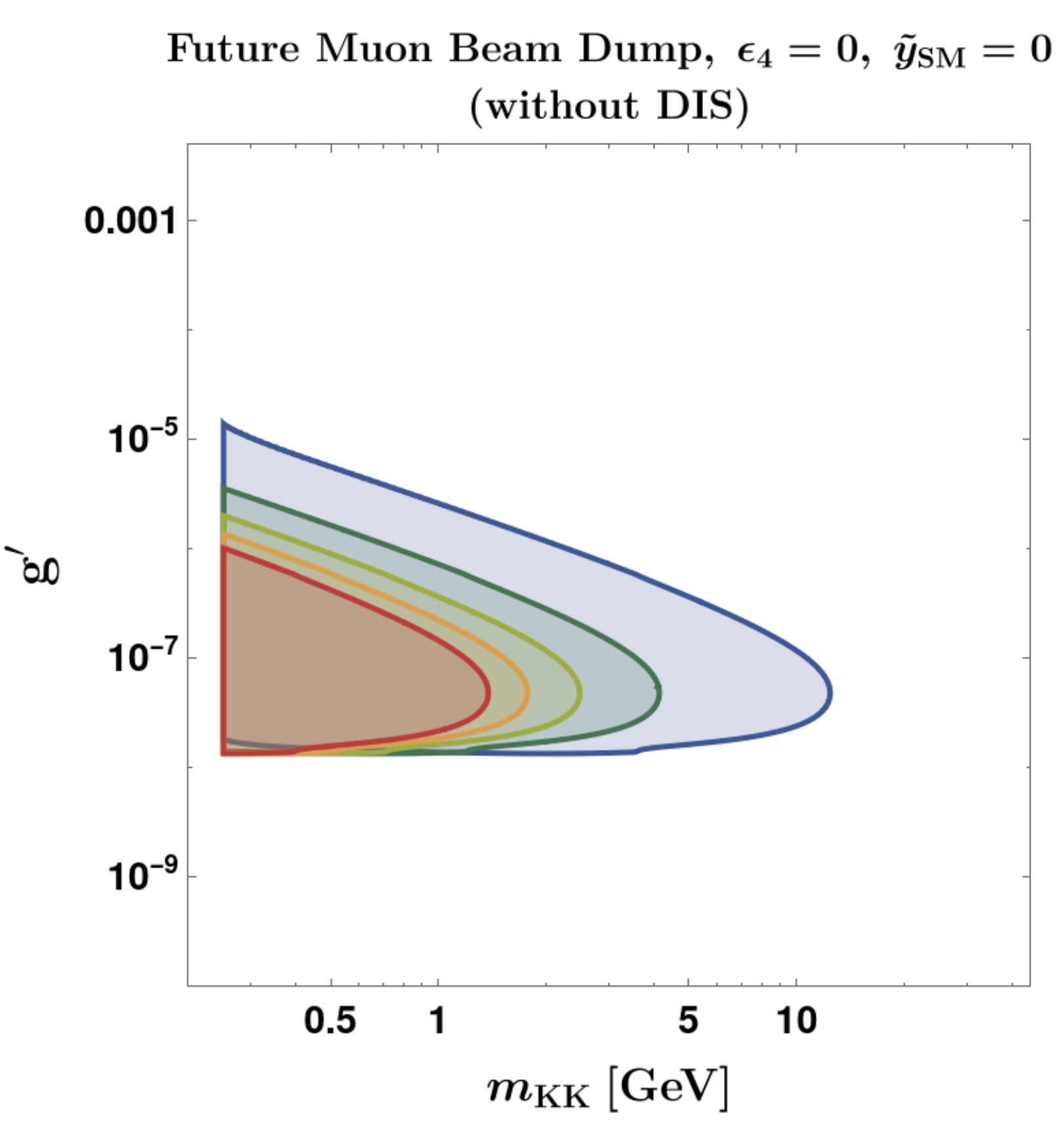}
    \end{minipage}%
    \hspace{-0.4em}
    \begin{minipage}[c]{0.20\textwidth}
      \includegraphics[width=\textwidth]{./figures/caption_nKK.pdf}
    \end{minipage}
  \end{minipage}
\caption{Comparison of {MuSIC (Left) and future beam dump (Right) sensitivities} for individual $n_{\text{KK}}$ modes for $\wt{y}_\tx{SM} = 0$ and $\epsilon_4=0$.}
\label{fig:combinedPlotWithLegends}
\end{figure}

\begin{figure}[t]
\centering
\includegraphics[width=0.3\textwidth]{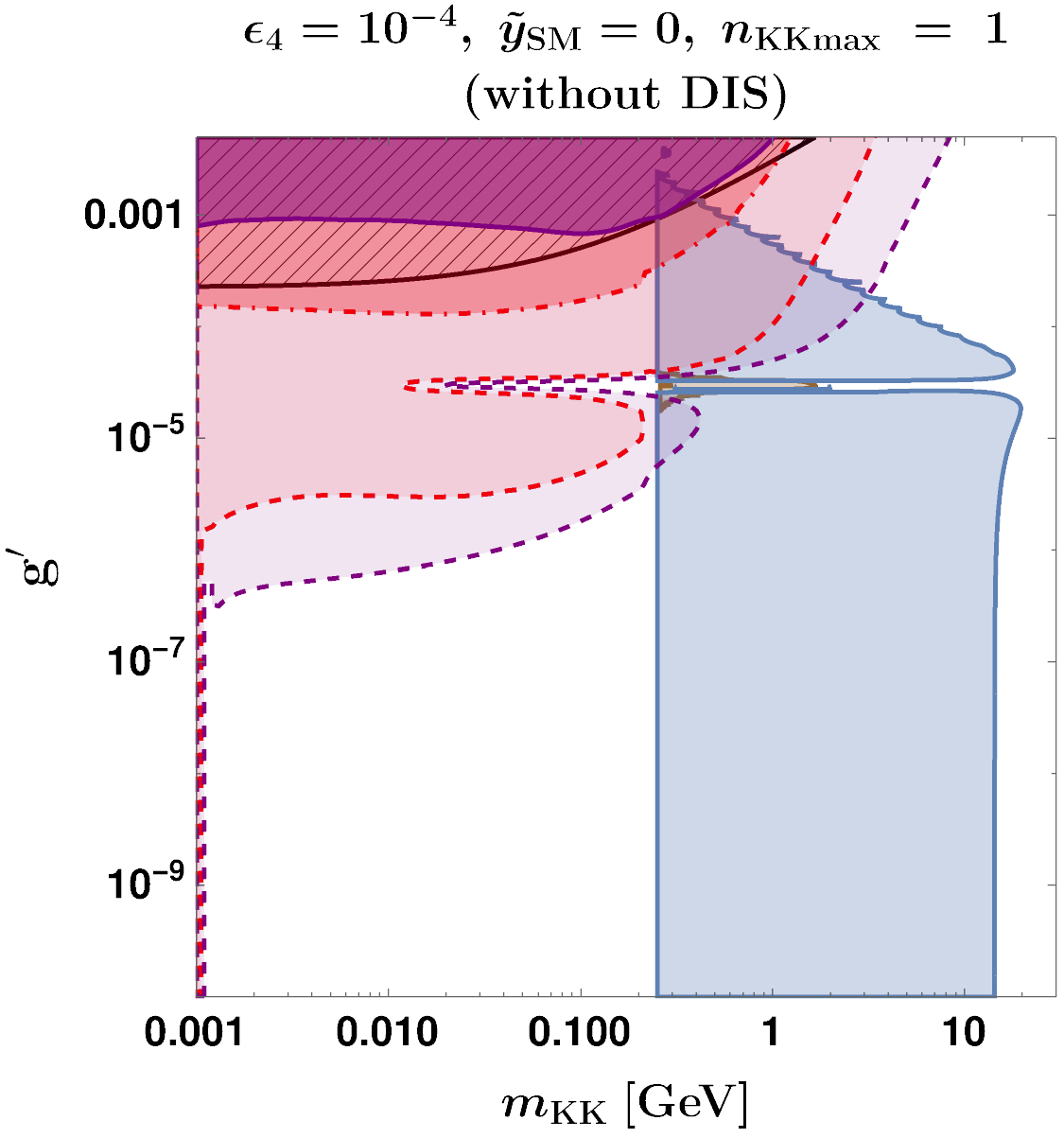} \
\includegraphics[width=0.3\textwidth]{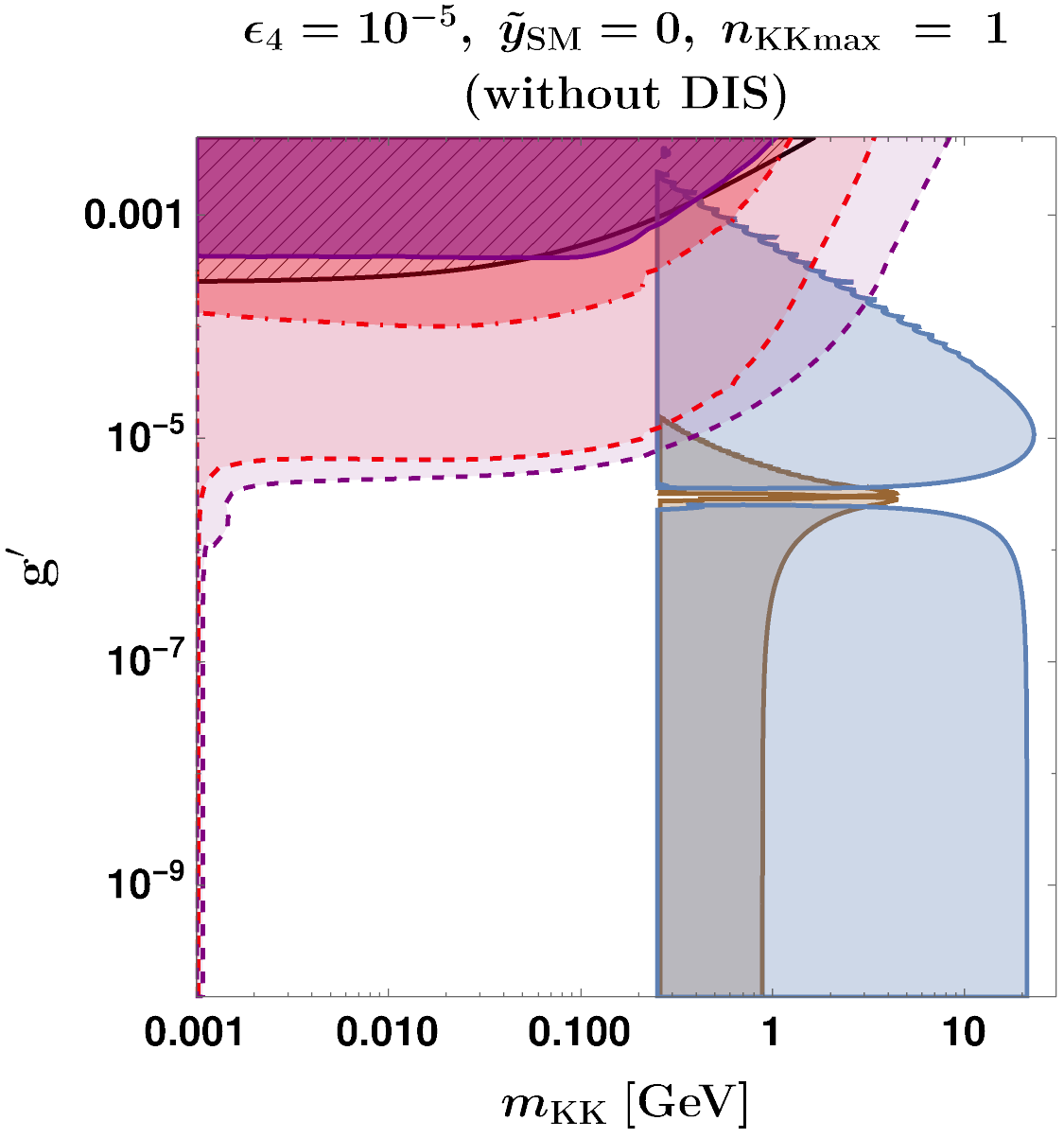} \
\includegraphics[width=0.3\textwidth]{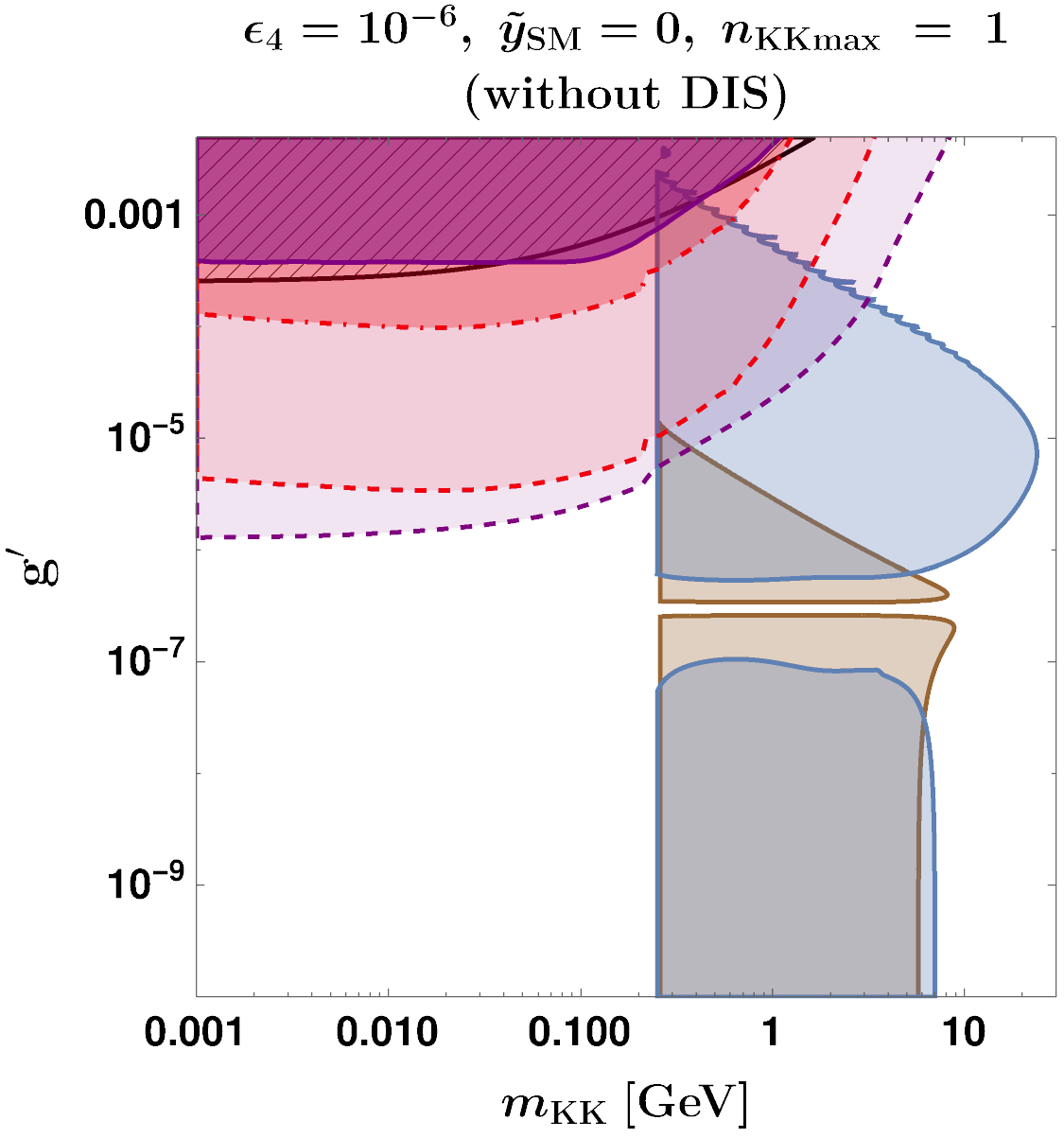} \\[5pt]
\includegraphics[width=0.3\textwidth]{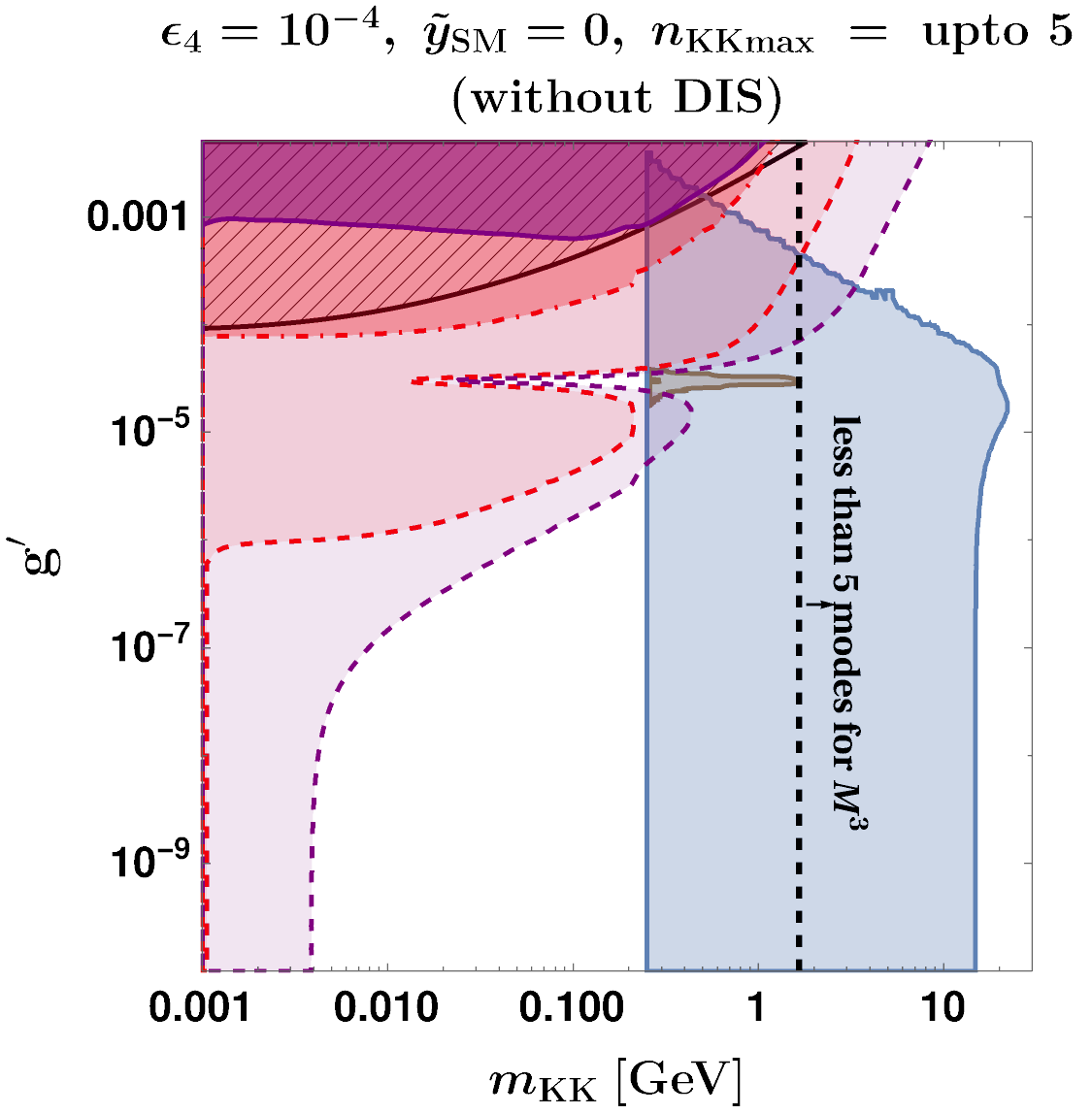} \
\includegraphics[width=0.3\textwidth]{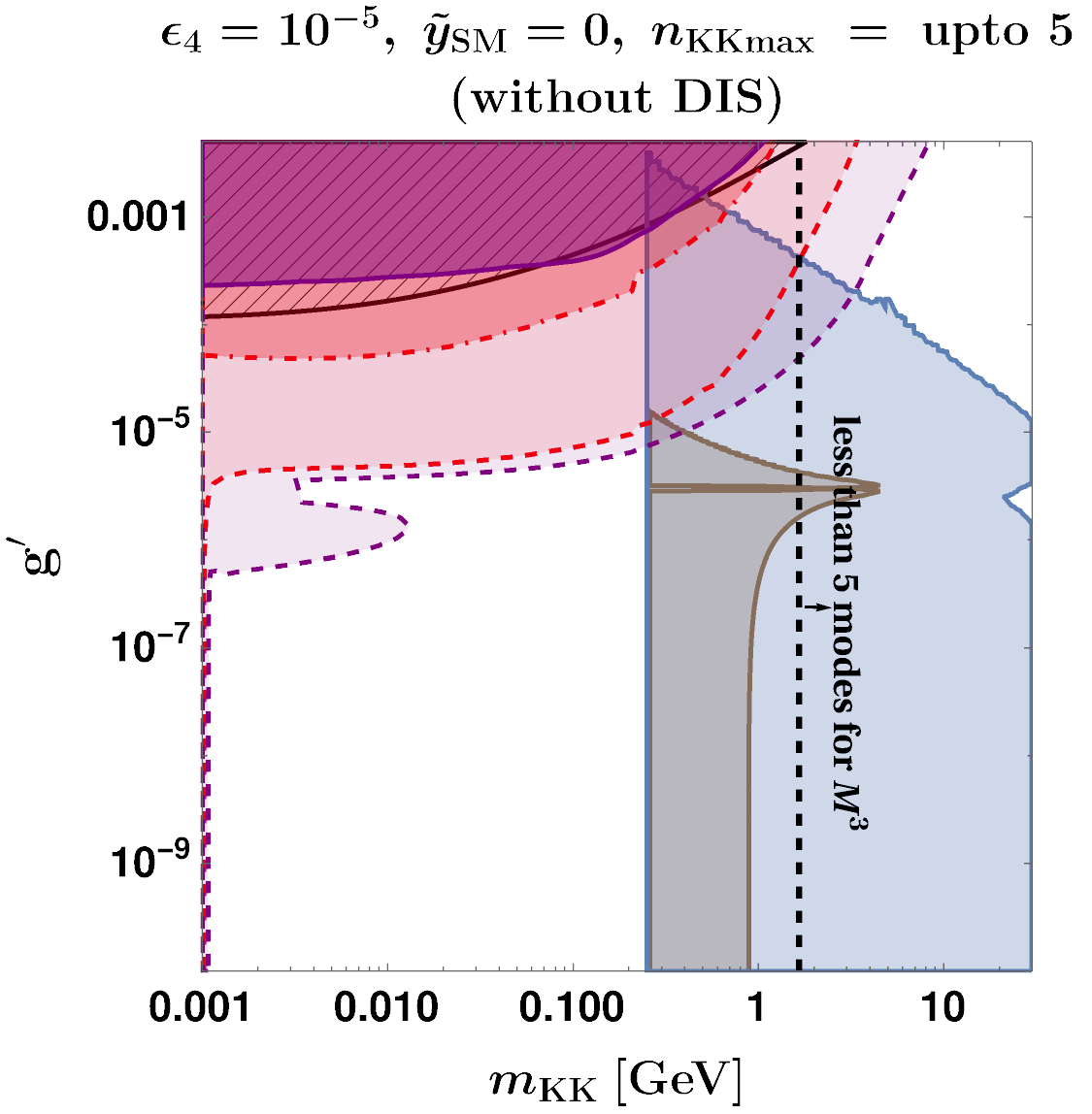} \
\includegraphics[width=0.3\textwidth]{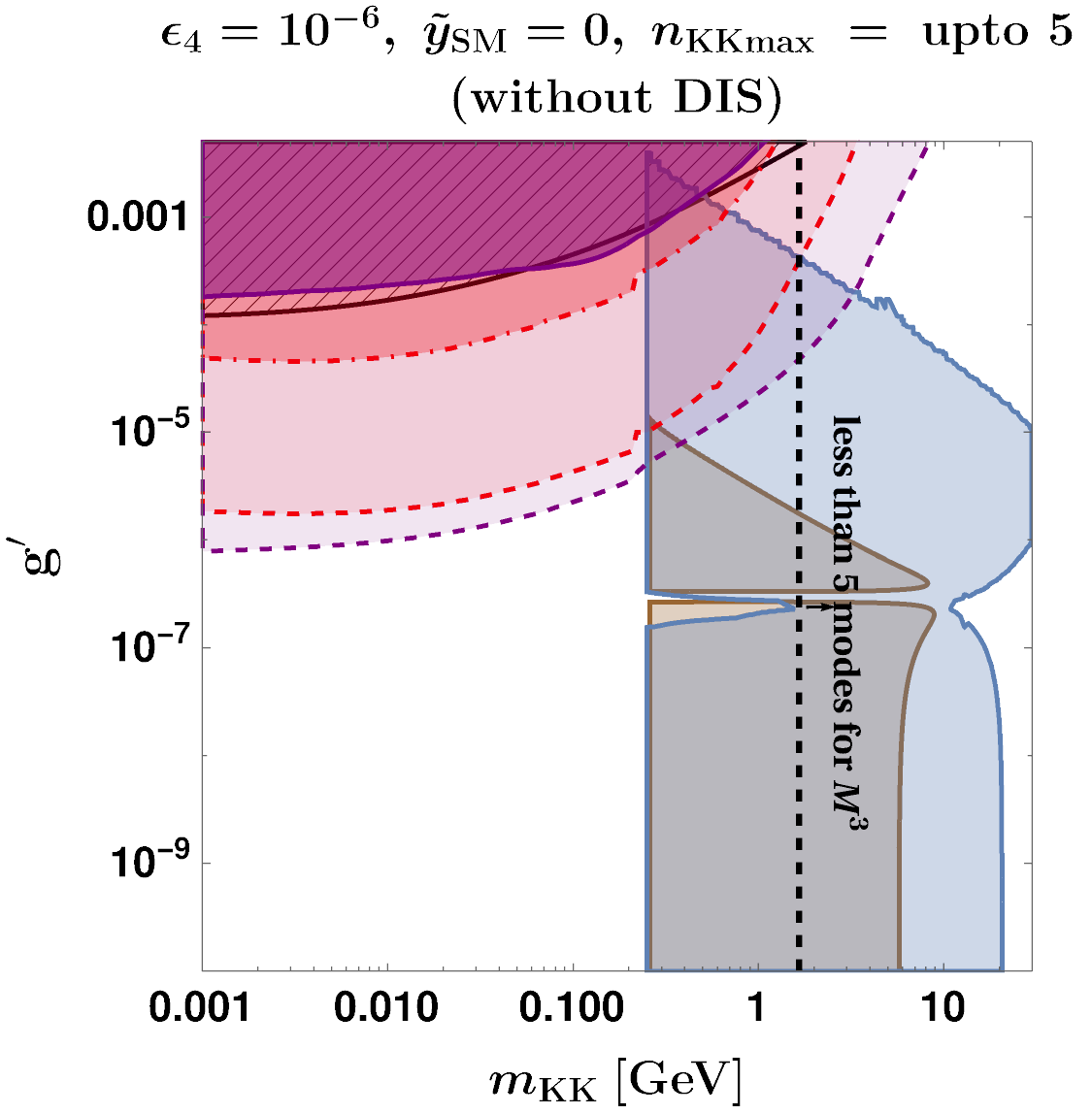} 
\caption{
Summary of {current/future} exclusion limits in the \((m_{\text{KK}},\, g')\) parameter space for the ED \(U(1)_{L_\mu - L_\tau}\) model under the benchmark choices $\epsilon_4 = 10^{-4}$, $10^{-5}$ and $10^{-6}$ (from Left to Right columns) under \(\wt{y}_\tx{SM} = 0\).
The first and second row correspond to \(n_{\text{KKmax}} = 1\) and \(n_{\text{KKmax}} = {\text{upto }5} \), respectively.
Here, we follow the same colour conventions as those for Fig.~\ref{fig:yzero_epsilon-zero}.
{The meaning of `upto 5' is as follows.}
We imposed the ad hoc cutoff $M_{i} \lesssim 45\,\tx{GeV}\,\paren{i=1,2,3,4,5}$ to eliminate heavy modes that are phenomenologically unfavourable and exceed the current mixing approximation's validity range.
Refer to Footnote~\ref{foot:KK-treatments} for details.
}
\label{fig:yzero_epsilon-Nonzero}
\end{figure}

We report our {results}, illustrating the exclusion sensitivity of NA$64_{\mu}$, M$^3$, MuSIC, and a forthcoming muon beam dump experiment to the $U(1)_{L_\mu - L_\tau}$ gauge boson in 5D.
The analysis is performed for two benchmark scenarios: \(n_{\text{KKmax}} = 1\), signifying the  lowest KK excitation, and \(n_{\text{KKmax}} = 5\), including the {first five} KK modes.
We remind {readers} that the following four effective parameters affect results, $g'$, $m_\tx{KK}$, $\epsilon_4$, and $y_\tx{SM}$
(in addition to $n_{\text{KKmax}}$).
For simplicity, we introduce the notation $\wt{y}_\tx{SM} := y_\tx{SM}/R$.
{Here, we explicitly state that the DIS effect mentioned in Section~\ref{sec:4D-case} is ignored in all of the following plots.
For the legitimate reason for this, refer to the final paragraph of this section.}

{In Figure~\ref{fig:yzero_epsilon-zero}}, The benchmarks are presented assuming \(\epsilon_4 = 0\) and \(\wt{y}_\tx{SM} = 0\).\footnote{
Complementary plots for \(\epsilon_4 = 0\) but \(\wt{y}_\tx{SM} \not= 0\) is available as Fig.~\ref{fig:yNonzero_epsilon-zero} in Appendix~\ref{sec:Additional-Plots}, where a weak dependence on $\wt{y}_\tx{SM}$ is observed.
}
Note that the details of our signal estimations in the four experiments have been propounded in Section~\ref{sec:Experiments}, based on the signal cross-section formulas in Section~\ref{sec:cross-section}.\footnote{
Regarding several minor handling points, we will comment on them below.
(i) Since the beam energy in the M${}^3$ experiment is 15 GeV, it is kinematically impossible to directly produce up to the fifth KK mode when the $m_\tx{KK}$ is greater than $\sim \! 2 \, \tx{GeV}$, corresponding to the area on the right separated by the vertical dotted line in the right panel of Fig.~\ref{fig:yzero_epsilon-zero}.
In this region, we consider the contribution of all kinematically possible KK vector bosons with $n_\tx{KK}$ less than five;
(ii) As introduced in Section~\ref{sec:Model-Infro}, in calculating the mixing of the fields, we performed an expansion using the epsilon parameter and retained only the lowest-order term.
However, this is not necessarily justified when the mass of a given unperturbed KK particle approaches the mass of the $Z$ boson sufficiently closely.
{If $\epsilon_4 \not= 0$, this failure is materialised in calculations arising from the factor \(\left(\frac{\epsilon_n m_{Z,0}^2}{M_n^2 - m_{Z,0}^2}\right)\), which manifests in mixing (see Eqs.~\eqref{eq:KK-mixing-start} to \eqref{eq:KK-mixing-end}).
To avoid the catastrophe, in the calculations with $\epsilon_4 \not= 0$ of the MuSIC and the future beam dump, we will impose the following ad hoc cutoffs for the KK modes, {$M_{i} \lesssim 45\,\tx{GeV}\,\paren{i=1,2,3,4,5}$}.
Via the mode-by-mode reaches for $\epsilon_4 = 0$ (up to the 5th mode) in Fig.~\ref{fig:combinedPlotWithLegends}, namely
{$m_\tx{KK} \lesssim 10\,\tx{GeV}$ in MuSIC and the future muon beam dump}, the introduction of the above ad hoc cutoffs may change the results very little.}
{To improve this situation, it is necessary to take into account higher-order perturbations relating to mixing; however, as this improvement is only meaningful when $m_\text{KK}$ is very large, we shall not engage in any further quantitative discussion of this point in this paper.}
%
%
\label{foot:KK-treatments}
}
{
Here, we provide some additional comments from a kinematic perspective regarding the results of experiments observing muon pairs in decays.
When observing the $Z^{\prime(n)} \to \mu^-\mu^-$ signal experimentally, it is common practice to place the detector at a point considerably distant from the collision point in order to suppress background events.
In this case, a signal can only be detected if each $Z^{\prime(n)}$ particle decays inside a displaced detector, where the averaged lifetime of $Z^{\prime(n)}$ should be within a certain range.
This is the cause of the ‘island structure’ observed in the experimental results, as seen, e.g., in Fig.~\ref{fig:yzero_epsilon-zero}.
To clarify further, in $Z^{\prime(n)} \to \mu^-\mu^+$, the reason there is no restriction in the region where $m_\text{KK} < 2m_{\mu} \sim 0.2\,\text{GeV}$ is that the decay process of interest ceases to occur on-shell in this region.
}

An appreciable improvement in sensitivity is noted with the addition of KK modes. For example, increasing \(n_{\text{KKmax}}\) from one to {five} significantly enhances the exclusion reach for both NA$64_{\mu}$ and M$^3$.
Note that we have established that truncating the KK tower at \(n_{\text{KKmax}} = 5\) provides a sufficiently precise and convergent approximation {for our purpose, to use the accuracy of the log-log plot to ascertain the current global status and future projections of the parameter space.
You can find some experimental circumstances in NA$64_{\mu}$, M$^3$ in Figs.~\ref{fig:convergencePlot} and \ref{fig:Diffyt} in the Appendix~\ref{sec:Additional-Plots}.
Furthermore, please refer to Appendix~\ref{sec:for-nmax-choice} regarding the accuracy of the adopted $n_\text{KKmax}=5$.}

We will briefly summarise the information obtained from the latest results on muon $(g-2)$.
As a sensational announcement, the long-standing anomaly between the SM prediction and the experimental results was resolved as $a_\mu^\tx{exp} - a_\mu^\tx{SM} = 38(63) \times 10^{-11}$, which is consistent within $\sim \! 0.6\,\sigma$~\cite{Aliberti:2025beg}.
Currently, this experimental value can be used to exclude specific areas of new physics coupled with muons.
As shown in Fig.~\ref{fig:yzero_epsilon-zero}, it is notable that the constraint on our scenario is comparable to the current experimental constraint from NA64${}_\mu$.
Note that we used the formula for evaluating KK gauge boson contributions to the muon $(g-2)$ available in Section~2.4 of~\cite{Chakraborty:2024xxc}.

{
Here, we comment further on the constraints from neutrino-trident production (via the {Chicago-Columbia-Fermilab-Rochester}~(CCFR) measurement~\cite{CCFR:1991lpl} of the neutrino trident cross-section) on our scenario.
As shown in Fig.~2 of~\cite{Altmannshofer:2014pba}, (prior to its experimental exclusion last year), in a four-dimensional model containing a single corresponding $Z'$, a portion of the parameter space capable of explaining the muon $(g-2)$ anomaly evaded the CCFR experimental constraints; even within the excluded region, it was not particularly far removed from the CCFR constraint boundary.
Furthermore, as the measurement uncertainties have been improved in the update of last year’s muon $(g-2)$ results, we reach the notion that the constraints derived from the CCFR experimental data via neutrino trident production are at least not significantly better than those derived from the latest muon $(g-2)$ data.
We therefore decided not to discuss the current constraints on neutrino trident production in this paper, as they require substantial calculations and do not affect our main conclusions based on current experimental constraints.\footnote{
{In the future, once experimental results have been updated---for example, from the DUNE experiment---it is expected that further constraints will be obtained from neutrino trident production.
See~\cite{Altmannshofer:2019zhy,Ballett:2019xoj,Shimomura:2020tmg} for details of prospects for the 4D scenario.}
}
}

The most phenomenologically interesting prospect arises from the future observation of massive gauge boson decays to muon pairs at a displaced detector in upcoming experiments, MuSIC and a forthcoming muon beam dump experiment.
These effective search regions are determined by the lifetimes of massive gauge bosons, where feebly interacting light massive particles near $\sim \! 1\,\tx{GeV}$ are the primary targets for exploration.

{With improved sensitivity, compared to the single-mode case, adding more KK modes to the MuSIC experiment creates a fascinating region in the parameter space.}
The enhancement can be attributed to the configuration of the mode functions \(f_n\), which gain significance with the inclusion of additional KK modes (see Ref.~\cite{Chakraborty:2024xxc} for further information), {also, for a given mass, a kinematic resonance arises due to ultra-high incident energy as shown in Figure~\ref{fig:nKKmaxratios} in Appendix~\ref{sec:for-nmax-choice}.}\footnote{
{
To understand this, it is best to focus on Eq.~\eqref{diffcrosssecourmodel}, which is the form obtained after integrating over $\theta_{Z^{\prime(n)}}$ in the differential cross section expressed in terms of the kinematic parameters $\theta_{Z^{\prime(n)}}$ and $x$ (although the analysis in the MuSIC experiment is performed using different kinematic parameters, $\eta$ and $\gamma$; see the second last paragraph of Section~\ref{sec:4D-case}).
Here, the factor $\left(M_n^2 \left( 1-x \right) + m_\mu^2 x^2 \right)$ appearing in the denominator becomes zero at a point where $x$ is very slightly greater than unity, provided that $M_n \gg m_\mu$.
The maximum possible value of $x$ in terms of kinematics is given by Eq.~\eqref{eq:x-kinetic-range}, and this approaches unity as the beam’s incident energy increases.
Note that the focused energy in the MuSIC experiment ($E_0 = 8.9\,\tx{TeV}$) is significantly higher than in other experiments.
Since the value of $x_\text{max}$ is always less than unity, no divergence occurs; only kinematic enhancement takes place.
Furthermore, when considering the final result---the number of signals---it is necessary to account for the $x$-integral and the effects of decay before reaching the displaced detector (as formulated in Eq.~\eqref{eq:N_nKK_MuSIC} for the MuSIC experiment).
}
}
Conversely, in the forthcoming muon beam dump experiment, the incorporation of elevated KK modes does not result in any significant enhancement in sensitivity.
This difference can be explained by focusing on the future charted domains by individual KK modes.
In Fig.~\ref{fig:combinedPlotWithLegends}, we figure out the `$N_{\text{signal}} > 5$' regions for the first five KK modes separately.

In the experimental setup of the future muon beam dump experiment, as shown in the {right} panel of Fig.~\ref{fig:combinedPlotWithLegends}, the domain of the $(i+1)$-mode is located just inside of that of the $i$-mode ($i=1,2,3,4$).
Therefore, considering the general property that the cross section for generating the high KK mode is suppressed compared to that for the low KK mode, it is most reasonable that the restriction from the first mode alone provides a good approximation, as shown in Fig.~\ref{fig:yzero_epsilon-zero}.

In contrast, in the MuSIC experiment setup, the initial muon beam energy is extremely high ($= 8.9\,\tx{TeV}$), so the production cross sections for high KK modes do not decrease so rapidly.
Furthermore, this high initial energy may enable clear identification of differences in the lifetimes of KK modes.
These observations can be clearly discerned from the {left} panel of Fig.~\ref{fig:combinedPlotWithLegends}.
Therefore, in the MuSIC experiment, a significant increase in the searchable region based on the existence of multiple KK modes is clearly observable.

Moreover, the fact that the muon pair signal allows the parent particle's mass to be reconstructed holds significant importance in our scenario.
In the overlapping regions of the searchable areas for each mode in Fig.~\ref{fig:combinedPlotWithLegends}, the existence of multiple massive gauge bosons with different masses can be confirmed from the reconstruction of muon pair signals after accumulating the targeted amount of data.
Therefore, muon pair signal surveys using displaced detectors provide a powerful tool for directly distinguishing whether the $U(1)_{L_\mu - L_\tau}$ interaction originates in five dimensions or in four dimensions.

Finally, we discuss future exploration limits when the typical size of kinetic mixing is nonzero.
Figure~\ref{fig:yzero_epsilon-Nonzero} provides the future expected reaches for $\wt{y}_\tx{SM} = 0$ under the presence of nonzero kinetic mixing parameter, with $\epsilon_4 = 10^{-4}$ (Left column), $\epsilon_4 = 10^{-5}$ (Centre column), and $\epsilon_4 = 10^{-6}$ (Right column), respectively.  
As can be seen immediately from the figure, there is an expansion of the searchable region, particularly where $g'$ is small.
This can be easily understood because a new contribution is added to the effective coupling.\footnote{
{We provide a comment on why the constraint via the future muon dump experiment becomes tiny for $\epsilon_4 = 10^{-4}$ in Fig.~\ref{fig:yzero_epsilon-Nonzero}.
Under the `sizable' value of $\epsilon_4$, each KK mode couples with SM particles quite strongly and tends to decay immediately.
So, most of the KK vector bosons evanesce in the large decay volume before reaching the displaced detector.
In the case of MuSIC, the size of the decay volume is relatively small, and still, the emitted KK particles are able to reach the detector region.}
}

It is noteworthy that for each specific $\epsilon_4$ coordination, a narrow region of $g'$ that cannot be probed has been reported.
This phenomenon, reported in a previous paper for E$\nu$ES~\cite{Chakraborty:2024xxc}, is thought to be realised by the contributions from $\epsilon_4$ and $g'$ coincidentally cancelling each other out.
What has become clear here is that muon beam dump experiments also possess strong detection capabilities for even minute $g'$ values at tiny but nonzero $\epsilon_4$.

Here, we will discuss the contribution of VVCS mentioned at the end of Section~{\ref{sec:4D-case}}.
As stated in paper~\cite{Beranek:2013nqa}, when all interactions are provided by kinetic mixing, the VVCS contributes an additional $\sim \! 10\%$, if the initial charged lepton energy is significantly greater than the mass of a massive gauge boson.
When $g'$ is significantly larger than $\epsilon_4$, we can conclude that this contribution is negligible.
Conversely, when $\epsilon_4$ is significantly larger than $g'$, this contribution cannot necessarily be ignored; however, when the incident lepton's energy is sufficiently high, this contribution amounts to about $10\%$.
This parameter region, currently under investigation $\epsilon_4 \gg g'$, is part of the focus of the MuSIC and the future muon beam dump experiments.
However, since the initial muon beam energies in these two experiments are extremely high, exceeding $1\,\tx{TeV}$, the application of the aforementioned excess estimate is justified.
Therefore, it can be concluded that the estimation of the searchable region in Fig.~\ref{fig:yzero_epsilon-Nonzero} is sufficiently accurate, under the outlook that a subleading extra contribution of approximately $10\%$ in the signal cross section can be neglected.

Also, we provide a comment on the DIS-active domain.
Based on the criterion in Eq.~\eqref{eq:DIS-active-domain}, in each experiment, the vector bosons in
$m_{Z'} \gtrsim 18\,\tx{GeV}$ (NA64${}_\mu$),
$m_{Z'} \gtrsim 6\,\tx{GeV}$ (M${}^3$),
$m_{Z'} \gtrsim 133\,\tx{GeV}$ (MuSIC),
$m_{Z'} \gtrsim 55\,\tx{GeV}$ (Future Muon Beam Dump)
reaches the DIS-active domain, and additional gains can entre into the picture.
Up to the 5th KK mode, the above conditions can be rephrased in the sense of $m_\tx{KK}$ as
$m_\tx{KK} \gtrsim 2\,\tx{GeV}$ (NA64${}_\mu$),
$m_\tx{KK} \gtrsim 0.7\,\tx{GeV}$ (M${}^3$),
$m_\tx{KK} \gtrsim 15\,\tx{GeV}$ (MuSIC),
$m_\tx{KK} \gtrsim 6\,\tx{GeV}$ (Future Muon Beam Dump), respectively.
In this particular domain, we believe we are underestimating the constraints and predictions in the experiments.
However, as can be seen from Figs.~\ref{fig:yzero_epsilon-zero} and \ref{fig:combinedPlotWithLegends} (cf. Fig.~\ref{fig:yzero_epsilon-Nonzero}), in each experiment, this only applies near the heavier boundary of the corresponding exclusion limits.
Therefore, within the scope discussed in this manuscript, it can be concluded that additional signals from deep-inelastic scattering can appear significantly only at the very edge of the probeable region, affecting only a small fraction of the total area, and thus have little effect on the global shapes of the exclusion limits.

\section{Conclusions
\label{sec:Summary}}

In this paper, we have investigated how to survey the nature of the 5D $U(1)_{L_\mu - L_\tau}$ interactions in present and future muon dump experiments, namely, NA64$_\mu$, M$^3$, MuSIC, and a future muon dump experiment.
These experiments fall into two categories: the first two can probe processes where feebly interacting massive particles go into invisible channels, while the latter two can probe processes where they go into muon pairs.
These two types of experiments are complementary in that they allow exploration of different parameter regions of a model.
In our scenario, the presence of multiple massive gauge bosons as KK particles leads to a tendency for more signals to appear compared to the corresponding four-dimensional one.
In particular, the decay process into muon pairs enables mass reconstruction of the parent particle, making it possible to directly demonstrate the existence of multiple KK particles in at least some parameter regions.
This provides clear evidence that the origin of the $U(1)_{L_\mu - L_\tau}$ interaction lies in five dimensions.
Furthermore, the muon $(g-2)$ value, which is now consistent with the SM, can be used to exclude specific parameter regions for new particles interacting with muons.
{We have further discussed the non-trivial effects arising from kinetic mixing (which many other related studies have not addressed).
As succinctly illustrated in Figure~\ref{fig:yzero_epsilon-Nonzero}, kinetic mixing produces interesting effects, such as the extension of restrictions to regions with small $g'$ values and the disappearance of certain restricted areas due to the cancellation of impacts between $g'$ and $\epsilon_4$.}

The motivation to explain the (once-observed) muon $(g-2)$ anomaly has since been lost, but as briefly mentioned in the introduction, the $U(1)_{L_\mu - L_\tau}$ interaction is still considered a meaningful interaction.
For example, it is phenomenologically interesting to investigate the role of the mediator for thermal WIMP dark matter at the MeV and GeV mass scales in the current five-dimensional case.
Furthermore, it is a significant issue to discuss the parameter regions that can explain the experimental results of the Hubble tension in the case of 5D $U(1)_{L_\mu - L_\tau}$, as well as those that can be ruled out due to strong inconsistencies.
Also, future muon collider experiments currently in planning can also impose constraints on KK gauge bosons, which are light and interact very weakly with muons~\cite{Chakraborty:2026ocj}.

\section*{Acknowledgements}
{D.C. thanks CSIR, India, for financial support through the Senior Research Fellowship (Direct) (File No: 09/1128(23623)/2025-EMR-I).}
K.N. thanks Takeshi Araki and Kento Asai for discussions on beam dump experiments.
We thank Shiv Nadar Institution of Eminence for providing us with workstations for numerical calculations.
The authors thank the Yukawa Institute for Theoretical Physics at Kyoto University. Discussions during the YITP workshop YITP-W-24-09 on ``Progress in Particle Physics 2024" were useful to complete this work.


\section*{Appendix}
\appendix

\section{Supplemental and Complementary Plots
\label{sec:Additional-Plots}}

\begin{figure}[H]
\centering
\includegraphics[width=0.38\textwidth]{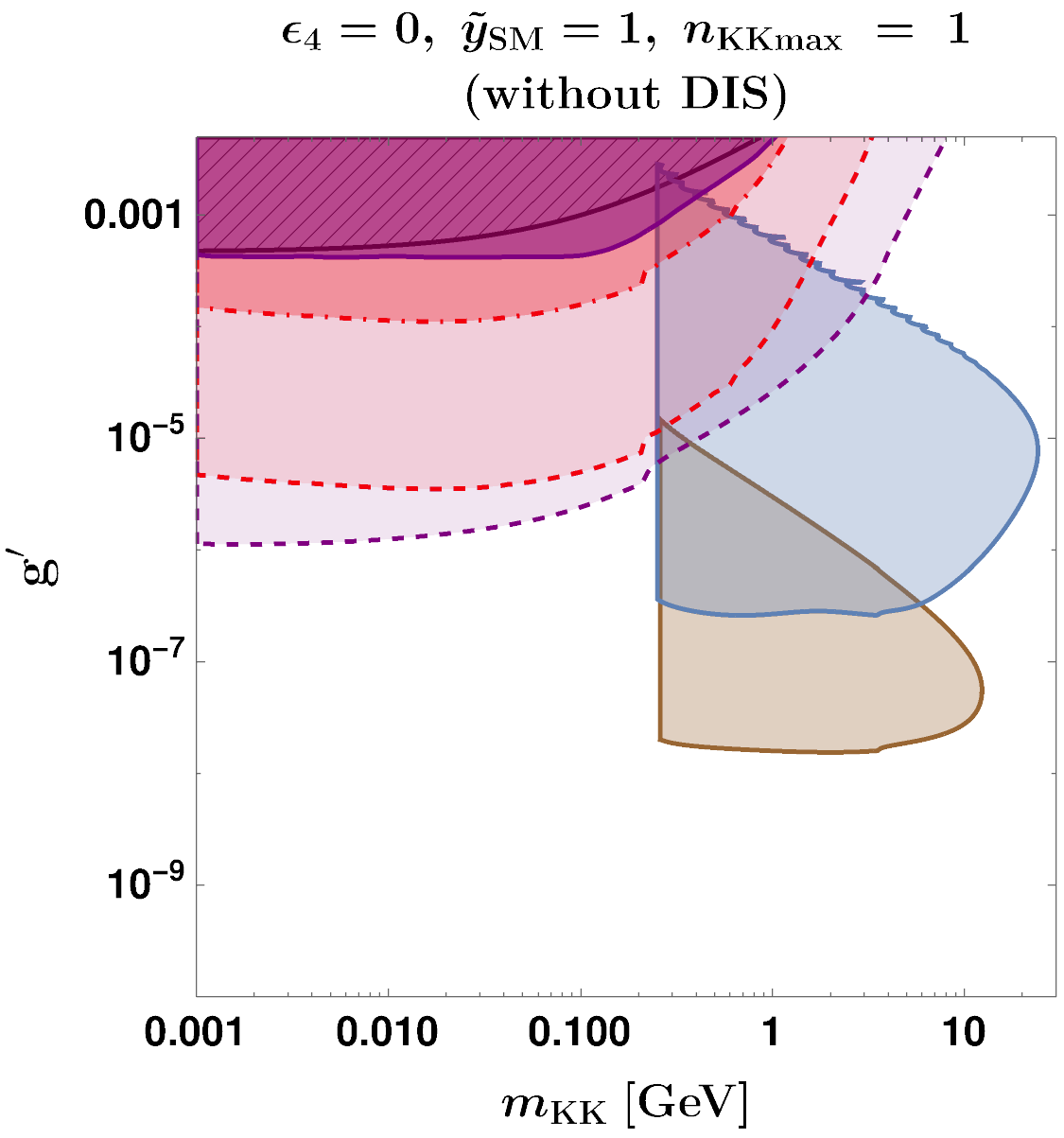} \
\includegraphics[width=0.38\textwidth]{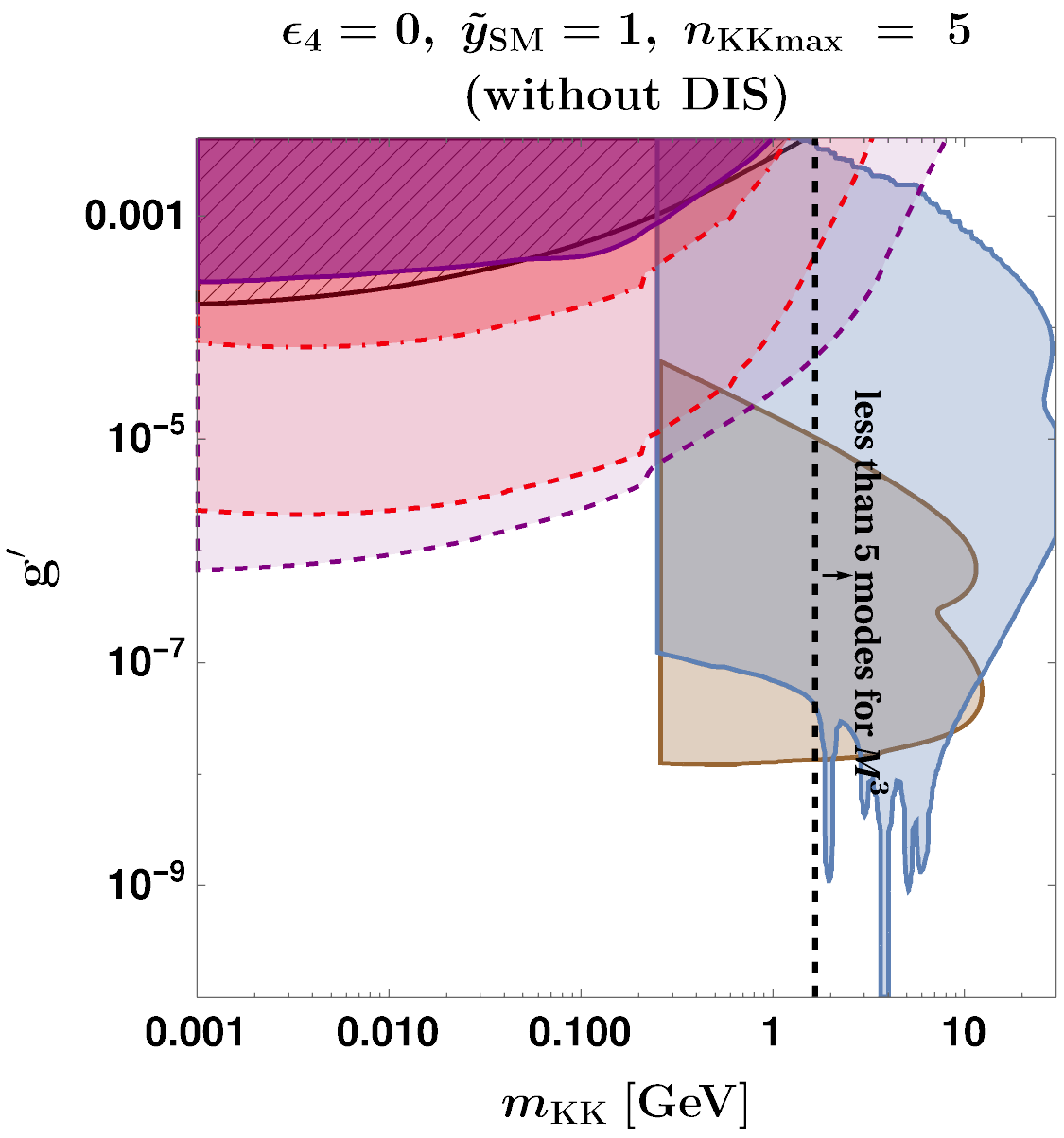}
\raisebox{40pt}{\includegraphics[width=0.20\textwidth]{./figures/caption_fourexps.pdf}} \\[5pt]
\includegraphics[width=0.38\textwidth]{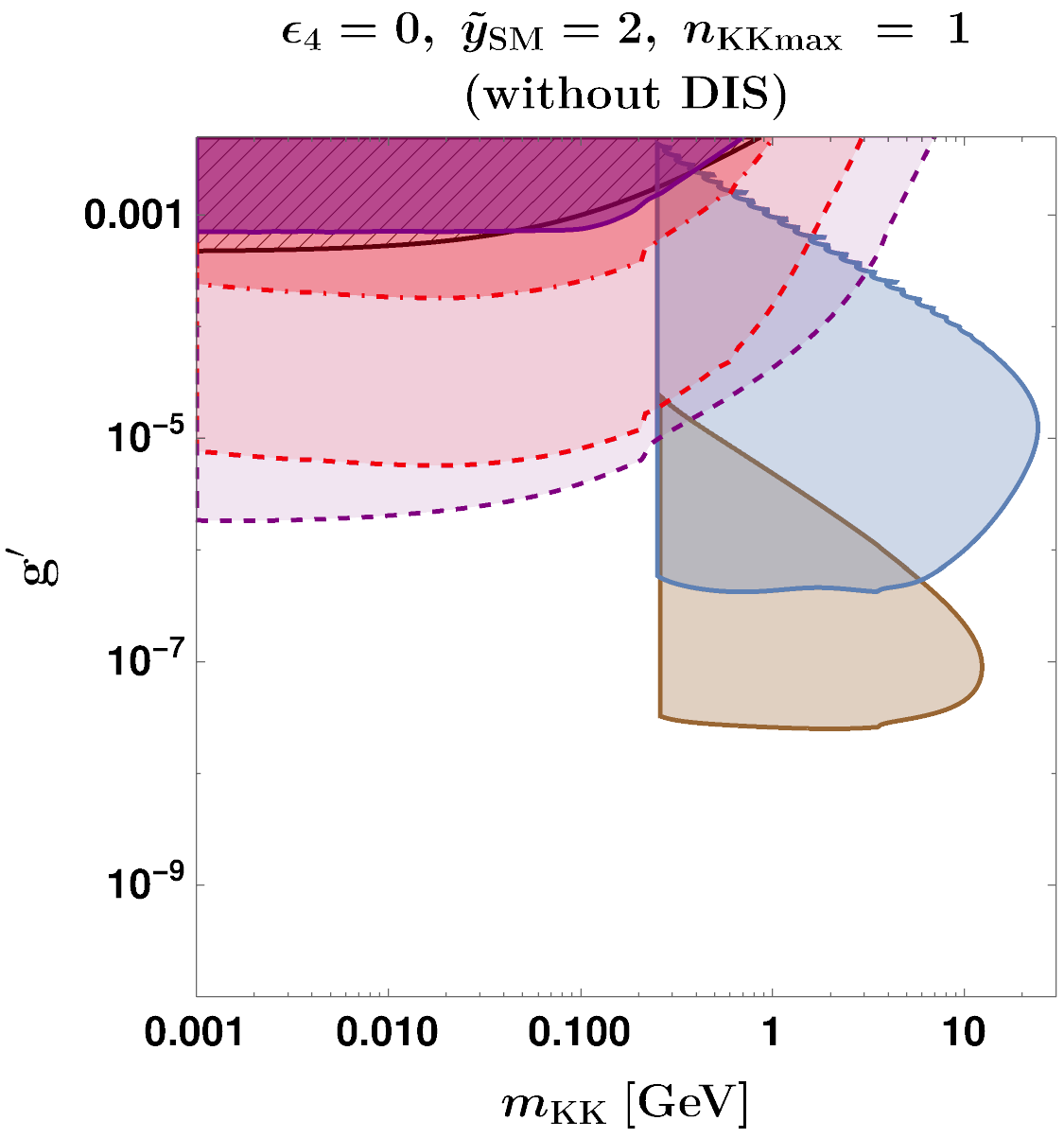} \
\includegraphics[width=0.38\textwidth]{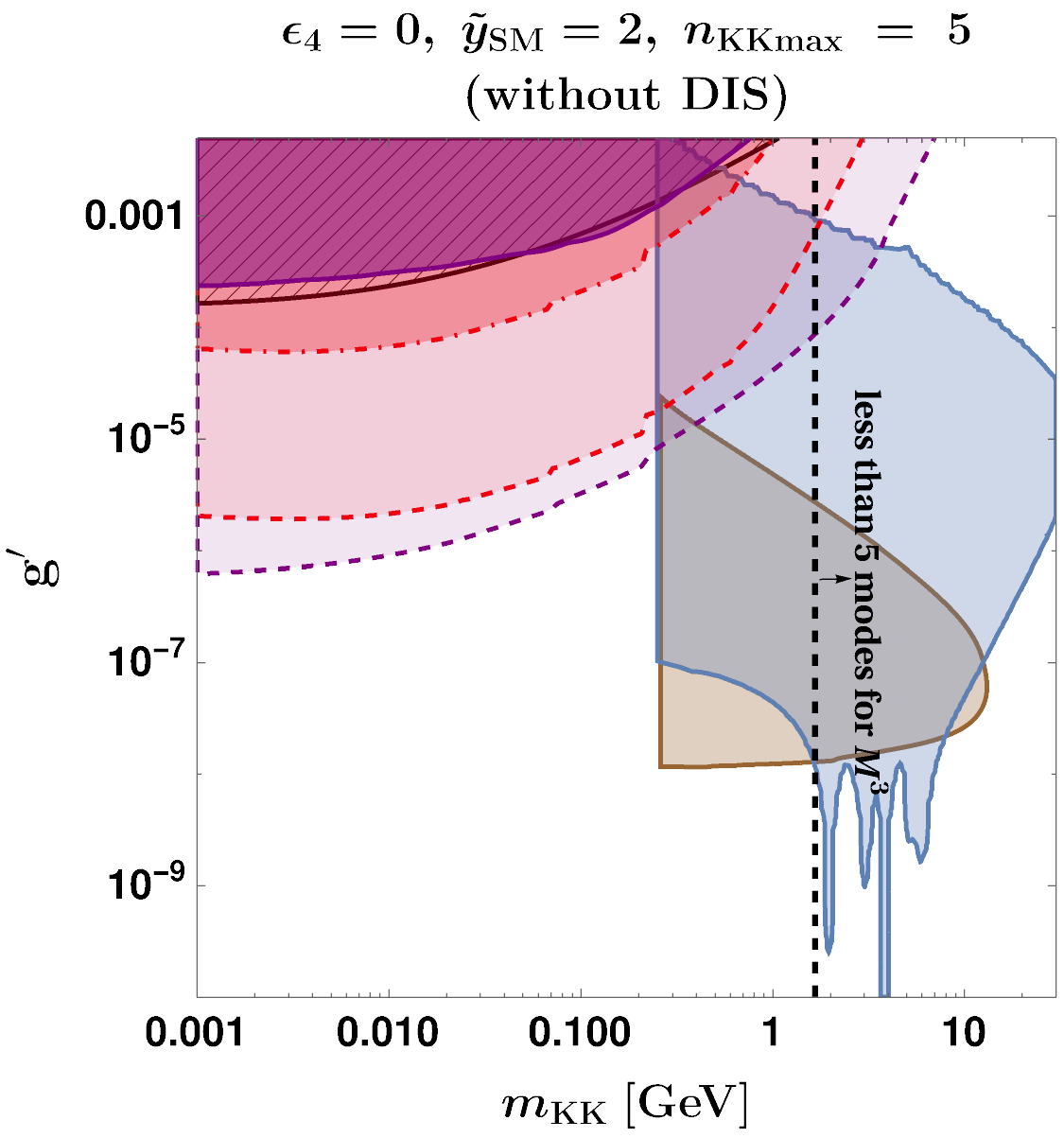}
\hspace{98pt}
\caption{
Summary of {current/future} exclusion limits in the \( (m_{\text{KK}},\, g') \) parameter space for the 5D \(U(1)_{L_\mu - L_\tau}\) model under the benchmark choices \(\epsilon_4 = 0\) and \(\wt{y}_\tx{SM} = 1\) and $2$.
They illustrate the sensitivity of various experiments for two KK mode truncations: \(n_{\text{KKmax}} = 1\) (Left Panels) and \(n_{\text{KKmax}} = 5\) (Right Panels).
Here, we follow the same colour conventions as those for Fig.~\ref{fig:yzero_epsilon-zero}.
}
\label{fig:yNonzero_epsilon-zero}
\end{figure}

\begin{figure}[H]
\centering
\includegraphics[width=0.4\textwidth]{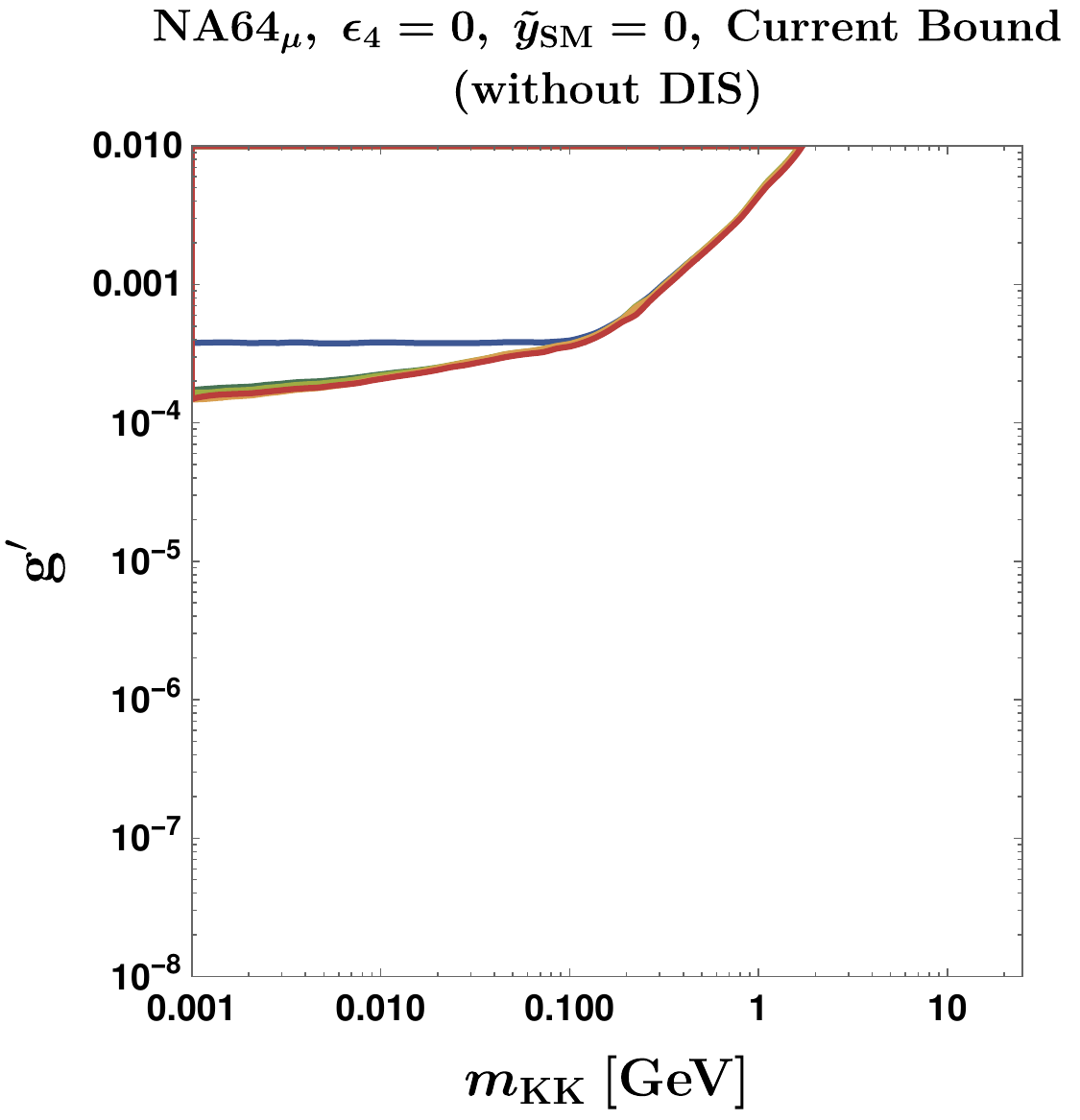} \
\includegraphics[width=0.38\textwidth]{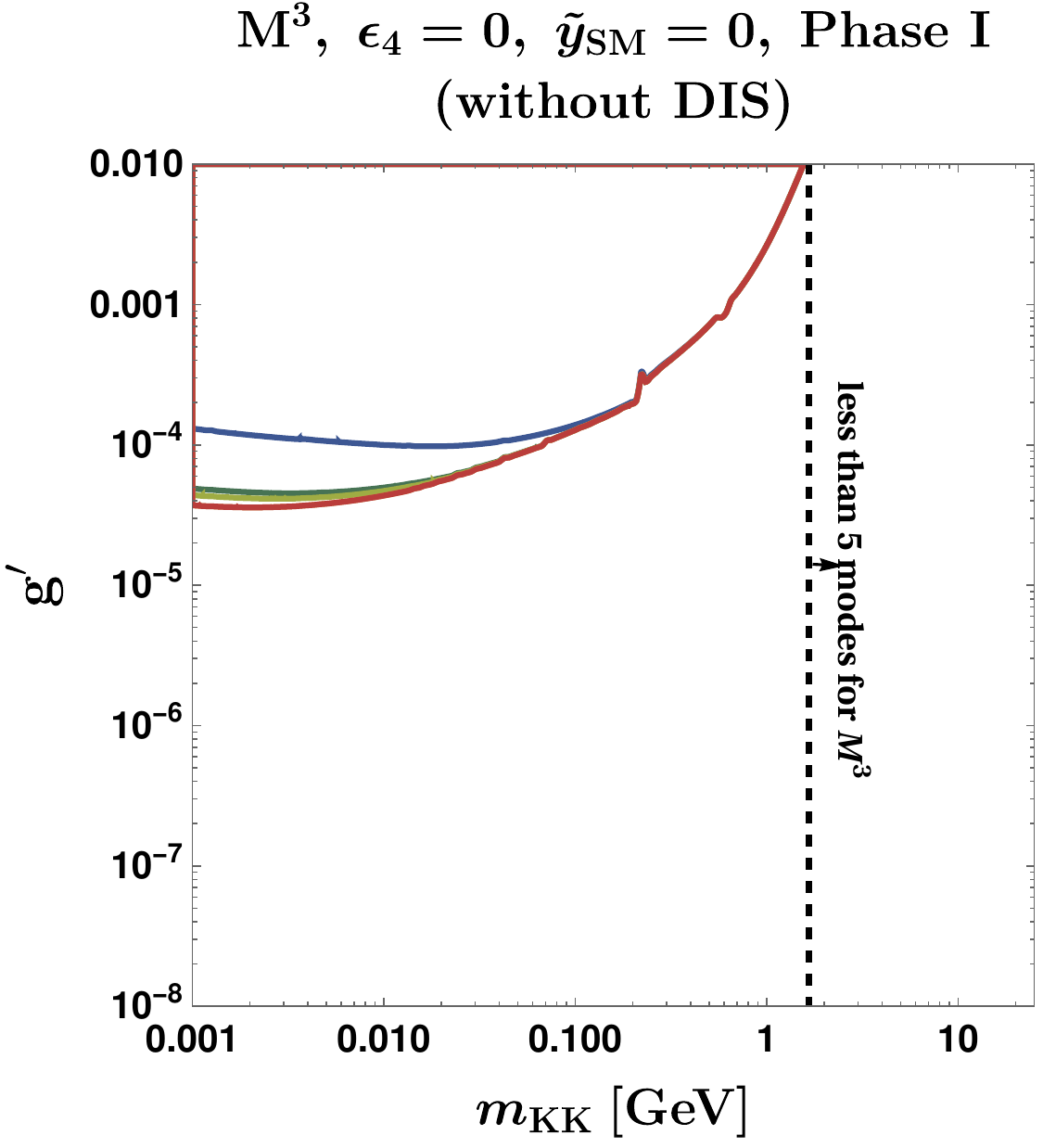}
\raisebox{40pt}{\includegraphics[width=0.15\textwidth]{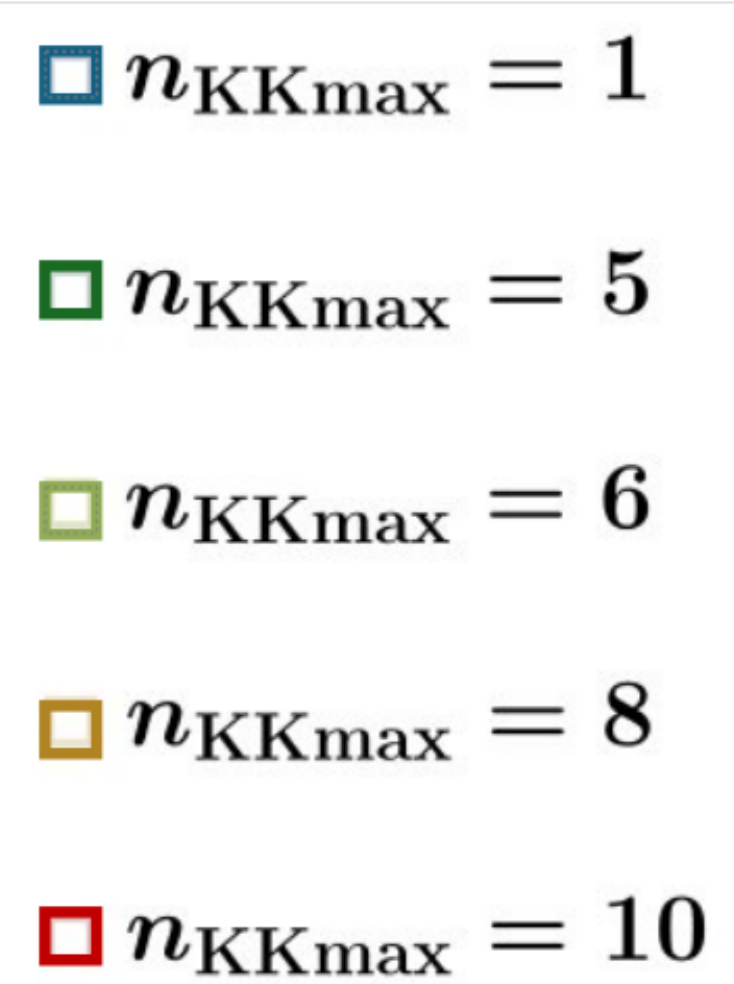}}
\caption{Convergence for $\text{NA}64_{\mu}$ and $\text{M}^3$ experimnts of $n_{\text{KKmax}}$ modes.}
\label{fig:convergencePlot}
\end{figure}

\begin{figure}[H]
  \centering
  \begin{minipage}[c]{0.49\textwidth}
    \centering
    \begin{minipage}[c]{0.7\textwidth}
      \includegraphics[width=\textwidth]{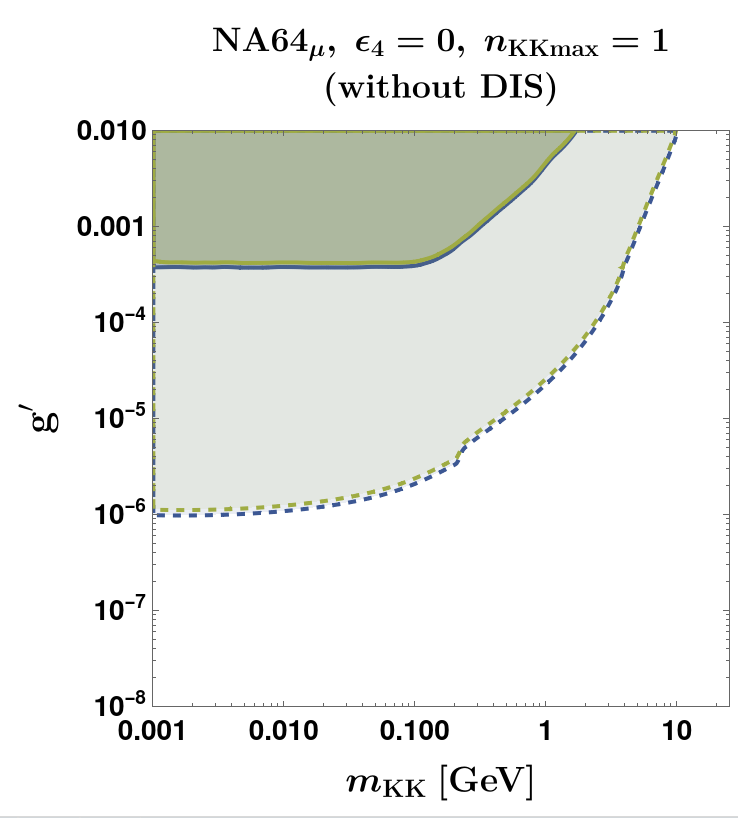}
    \end{minipage}%
    \hspace{-0.4em}
    \begin{minipage}[c]{0.3\textwidth}
      \includegraphics[width=\textwidth]{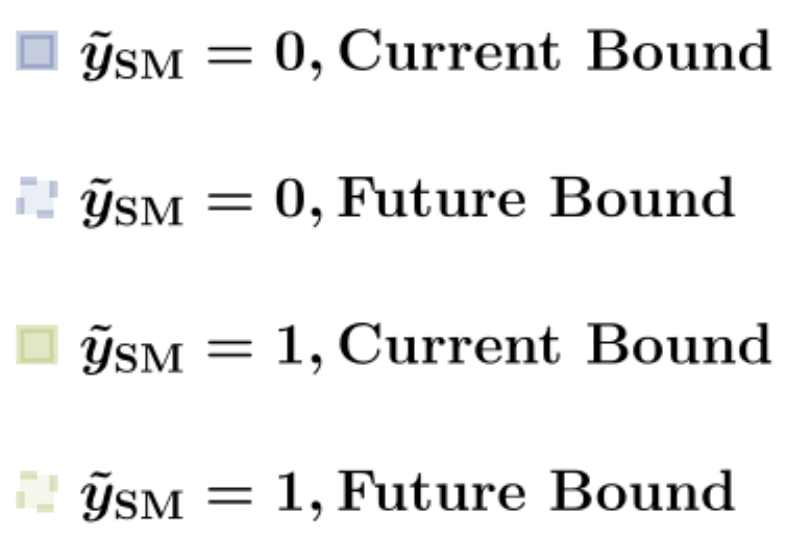}
    \end{minipage}
  \end{minipage}
  \begin{minipage}[c]{0.47\textwidth}
    \centering
    \begin{minipage}[c]{0.7\textwidth}
      \includegraphics[width=\textwidth]{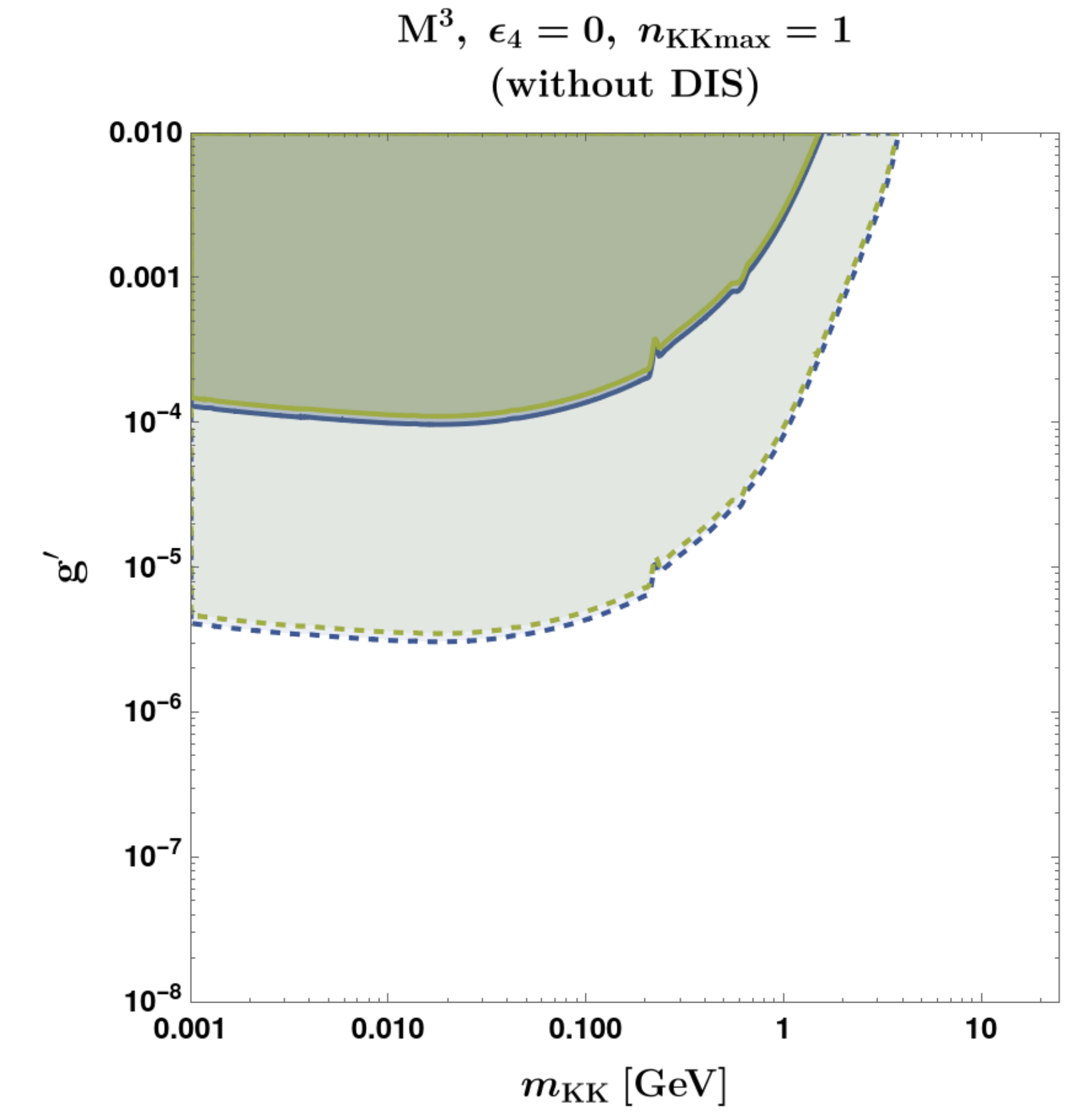}
    \end{minipage}%
    \hspace{-0.4em}
    \begin{minipage}[c]{0.3\textwidth}
      \includegraphics[width=\textwidth]{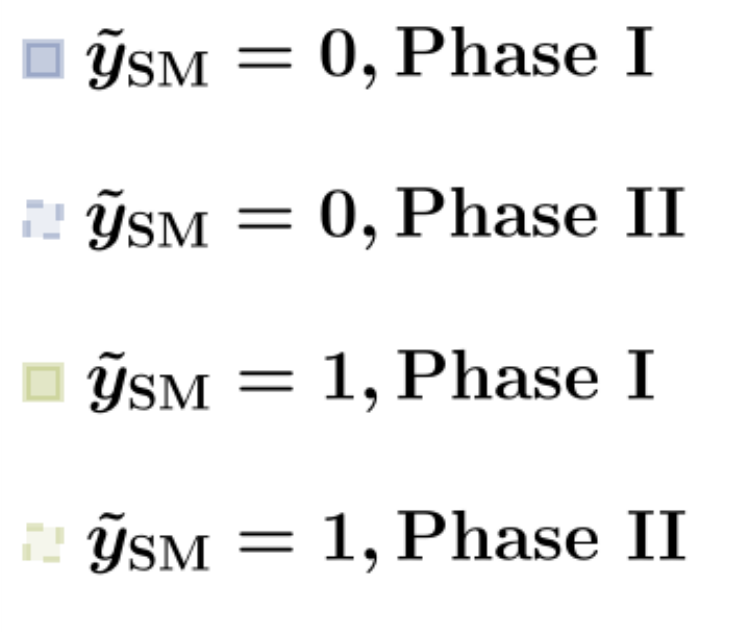}
    \end{minipage}
  \end{minipage}
\caption{Comparison of different $\wt{y}_\tx{SM}$ for $\text{NA}64_{\mu}$ (Left) and {M${}^3$} (Right) for $n_{\text{KKmax}}=1$.}
\label{fig:Diffyt}
\end{figure}

In this Appendix, we provide some supplemental plots that provide relevant additional details and provide a better understanding of our analysis.
Figure~\ref{fig:yNonzero_epsilon-zero} provides a complementary analysis with those in Fig.~\ref{fig:yzero_epsilon-zero}, where the value of $\wt{y}_\tx{SM}$ is nonzero.
Figures~\ref{fig:convergencePlot} and \ref{fig:Diffyt} show the convergence of our numerical analyses of NA64${}_\mu$ and M${}^3$.

\section{{On our choice of $n_\text{KKmax} = 5$}
\label{sec:for-nmax-choice}}

For our 5D scenario, the expected number of signals in each experiment is estimated as the sum of the contributions of the $n$th KK mode, as shown in Eq.~\eqref{eq:def_5D-totalsignal}. 
As $n$ increases, the mass of the corresponding KK particles becomes heavier, the production cross-section decreases, and their contribution to the signal count diminishes.
$n_\tx{KKmax}$ is introduced as a cutoff of the mode summation.

To investigate this issue caused by the variation in $n_\tx{KKmax}$, we plot the distributions of $N_{\text{signal}}^{(n_\text{KK}), a}$ for the four experiments ($a =$ NA64${}_\mu$, M${}^3$, MuSIC, FMBD), based on Eqs.~\eqref{eq:N_nKK_NA64mu}, \eqref{eq:N_nKK_M3}, \eqref{eq:N_nKK_MuSIC} and \eqref{eq:N_nKK_FMBD}, with the focused couplings and the individual target luminosities, respectively.
For NA64${}_\mu$, M${}^3$ and FMBD, if $m_\text{KK}$ is heavier than $\sim 10\,\tx{MeV}$, signals via higher modes immediately decouple, and the difference by ignoring more than $n_\text{KK} = 5$ modes will be around 10\%, which seems to be therefore more than sufficient for log-log plots.
Conversely, if $m_\text{KK}$ is $\sim 10\,\tx{MeV}$ or less, a difference of the order of ${\cal O}(10)\%$ or even higher may arise; however, this difference is not so large that it is barely discernible on log-log plots, and we consider $n_\text{KKmax} = 5$ to be sufficient for our purposes also for this region.

\begin{figure}[t]
\centering
\includegraphics[width=0.45\textwidth]{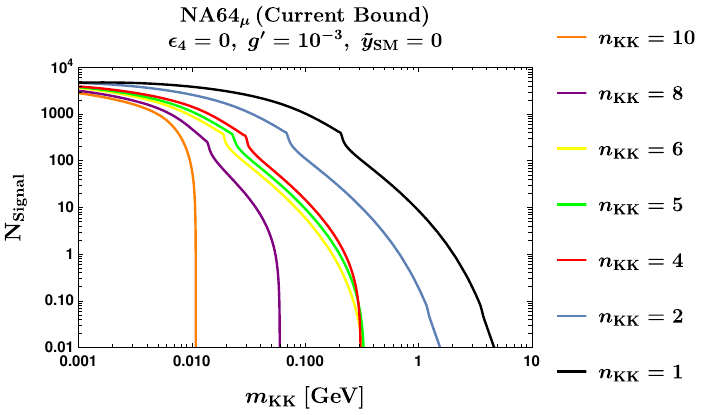} \
\includegraphics[width=0.45\textwidth]{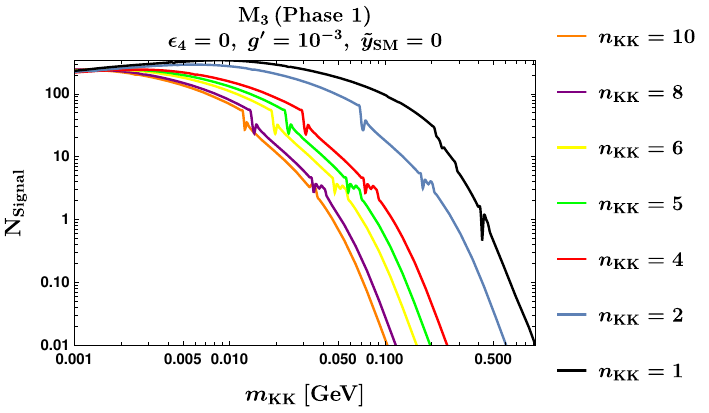} \\[4pt]
\includegraphics[width=0.45\textwidth]{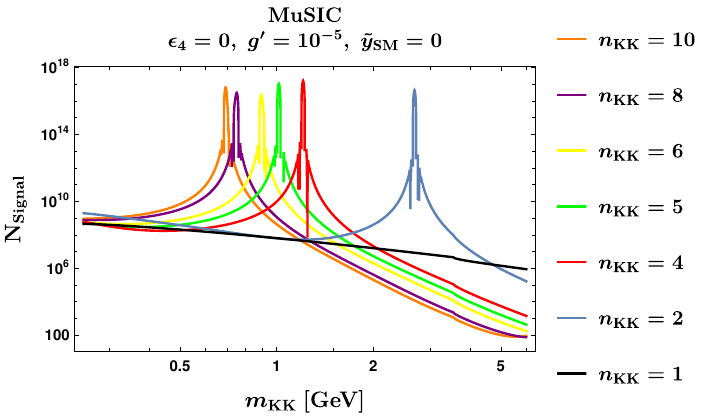} \
\includegraphics[width=0.45\textwidth]{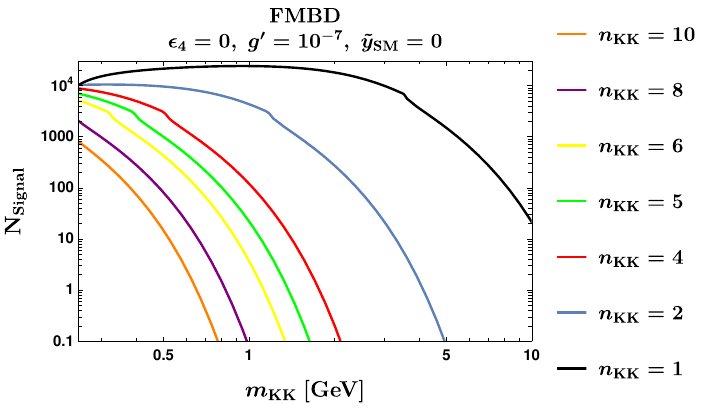}
\caption{Disctibutions of $N_{\text{signal}}^{(n_\text{KK}),a}$ for the four experiments ($a =$ NA64${}_\mu$, M${}^3$, MuSIC, FMBD),
based on Eqs.~\eqref{eq:N_nKK_NA64mu}, \eqref{eq:N_nKK_M3}, \eqref{eq:N_nKK_MuSIC} and \eqref{eq:N_nKK_FMBD}, with the focused couplings and the individual target luminosities, respectively.}
\label{fig:nKKmaxratios}
\end{figure}

In the MuSIC experiment, the situation is broadly similar, but there is one significant difference.
As shown in Figure~\ref{fig:nKKmaxratios}, for a given mass, a resonance signal arises due to ultra-high incident energy.
By ignoring modes above $n_\text{KK} = 5$, these resonance signals are also ignored; however, as can be seen from Figure~\ref{fig:yzero_epsilon-zero}, the new contributions generated by these resonance signals tend to act to exclude regions where $g'$ is very small.
Conversely, unless one requires extremely high precision in the exclusion of regions where $g'$ is particularly small, ignoring these resonance signals with high $n_\text{KK}$ values does not affect the global structure of the exclusion region where $g'$ is above a certain level.

From the above analysis, we can conclude that setting $n_\text{KKmax} = 5$ is necessary and sufficient for determining the regions to be excluded in each experiment (excluding the resonance region where $n_\text{KK} > 5$) from log-log plots.

\bibliographystyle{utphys}
{\small \bibliography{muon_beam_dump_project,ref_Kenji_beam-dump-general}}

\providecommand{\href}[2]{#2}\begingroup\raggedright\begin{thebibliography}{100}

\bibitem{Antel:2023hkf}
C.~Antel {\em et~al.}, ``{Feebly-interacting particles: FIPs 2022 Workshop
  Report},'' \href{http://dx.doi.org/10.1140/epjc/s10052-023-12168-5}{{\em Eur.
  Phys. J. C} {\bfseries 83} no.~12, (2023) 1122},
  \href{http://arxiv.org/abs/2305.01715}{{\ttfamily arXiv:2305.01715
  [hep-ph]}}.

\bibitem{CHARM:1985nku}
{\bfseries CHARM} Collaboration, F.~Bergsma {\em et~al.}, ``{A Search for
  Decays of Heavy Neutrinos in the Mass Range 0.5-{GeV} to 2.8-{GeV}},''
  \href{http://dx.doi.org/10.1016/0370-2693(86)91601-1}{{\em Phys. Lett. B}
  {\bfseries 166} (1986) 473--478}.

\bibitem{Konaka:1986cb}
A.~Konaka {\em et~al.}, ``{Search for Neutral Particles in Electron Beam Dump
  Experiment},'' \href{http://dx.doi.org/10.1103/PhysRevLett.57.659}{{\em Phys.
  Rev. Lett.} {\bfseries 57} (1986) 659}.

\bibitem{Riordan:1987aw}
E.~M. Riordan {\em et~al.}, ``{A Search for Short Lived Axions in an Electron
  Beam Dump Experiment},''
  \href{http://dx.doi.org/10.1103/PhysRevLett.59.755}{{\em Phys. Rev. Lett.}
  {\bfseries 59} (1987) 755}.

\bibitem{Bjorken:1988as}
J.~D. Bjorken, S.~Ecklund, W.~R. Nelson, A.~Abashian, C.~Church, B.~Lu, L.~W.
  Mo, T.~A. Nunamaker, and P.~Rassmann, ``{Search for Neutral Metastable
  Penetrating Particles Produced in the SLAC Beam Dump},''
  \href{http://dx.doi.org/10.1103/PhysRevD.38.3375}{{\em Phys. Rev. D}
  {\bfseries 38} (1988) 3375}.

\bibitem{Davier:1989wz}
M.~Davier and H.~Nguyen~Ngoc, ``{An Unambiguous Search for a Light Higgs
  Boson},'' \href{http://dx.doi.org/10.1016/0370-2693(89)90174-3}{{\em Phys.
  Lett. B} {\bfseries 229} (1989) 150--155}.

\bibitem{Bross:1989mp}
A.~Bross, M.~Crisler, S.~H. Pordes, J.~Volk, S.~Errede, and J.~Wrbanek, ``{A
  Search for Shortlived Particles Produced in an Electron Beam Dump},''
  \href{http://dx.doi.org/10.1103/PhysRevLett.67.2942}{{\em Phys. Rev. Lett.}
  {\bfseries 67} (1991) 2942--2945}.

\bibitem{LSND:1997vqj}
{\bfseries LSND} Collaboration, C.~Athanassopoulos {\em et~al.}, ``{Evidence
  for muon-neutrino ---{\ensuremath{>}} electron-neutrino oscillations from
  pion decay in flight neutrinos},''
  \href{http://dx.doi.org/10.1103/PhysRevC.58.2489}{{\em Phys. Rev. C}
  {\bfseries 58} (1998) 2489--2511},
  \href{http://arxiv.org/abs/nucl-ex/9706006}{{\ttfamily
  arXiv:nucl-ex/9706006}}.

\bibitem{NOMAD:2001eyx}
{\bfseries NOMAD} Collaboration, P.~Astier {\em et~al.}, ``{Search for heavy
  neutrinos mixing with tau neutrinos},''
  \href{http://dx.doi.org/10.1016/S0370-2693(01)00362-8}{{\em Phys. Lett. B}
  {\bfseries 506} (2001) 27--38},
  \href{http://arxiv.org/abs/hep-ex/0101041}{{\ttfamily arXiv:hep-ex/0101041}}.

\bibitem{NA64:2018lsq}
{\bfseries NA64} Collaboration, D.~Banerjee {\em et~al.}, ``{Search for a
  Hypothetical 16.7 MeV Gauge Boson and Dark Photons in the NA64 Experiment at
  CERN},'' \href{http://dx.doi.org/10.1103/PhysRevLett.120.231802}{{\em Phys.
  Rev. Lett.} {\bfseries 120} no.~23, (2018) 231802},
  \href{http://arxiv.org/abs/1803.07748}{{\ttfamily arXiv:1803.07748
  [hep-ex]}}.

\bibitem{SHiP:2015vad}
{\bfseries SHiP} Collaboration, M.~Anelli {\em et~al.}, ``{A facility to Search
  for Hidden Particles (SHiP) at the CERN SPS},''
  \href{http://arxiv.org/abs/1504.04956}{{\ttfamily arXiv:1504.04956
  [physics.ins-det]}}.

\bibitem{Alekhin:2015byh}
S.~Alekhin {\em et~al.}, ``{A facility to Search for Hidden Particles at the
  CERN SPS: the SHiP physics case},''
  \href{http://dx.doi.org/10.1088/0034-4885/79/12/124201}{{\em Rept. Prog.
  Phys.} {\bfseries 79} no.~12, (2016) 124201},
  \href{http://arxiv.org/abs/1504.04855}{{\ttfamily arXiv:1504.04855
  [hep-ph]}}.

\bibitem{FASER:2018bac}
{\bfseries FASER} Collaboration, A.~Ariga {\em et~al.}, ``{Technical Proposal
  for FASER: ForwArd Search ExpeRiment at the LHC},''
  \href{http://arxiv.org/abs/1812.09139}{{\ttfamily arXiv:1812.09139
  [physics.ins-det]}}.

\bibitem{FASER:2018eoc}
{\bfseries FASER} Collaboration, A.~Ariga {\em et~al.},
  ``{FASER{\textquoteright}s physics reach for long-lived particles},''
  \href{http://dx.doi.org/10.1103/PhysRevD.99.095011}{{\em Phys. Rev. D}
  {\bfseries 99} no.~9, (2019) 095011},
  \href{http://arxiv.org/abs/1811.12522}{{\ttfamily arXiv:1811.12522
  [hep-ph]}}.

\bibitem{Feng:2022inv}
J.~L. Feng {\em et~al.}, ``{The Forward Physics Facility at the High-Luminosity
  LHC},'' \href{http://dx.doi.org/10.1088/1361-6471/ac865e}{{\em J. Phys. G}
  {\bfseries 50} no.~3, (2023) 030501},
  \href{http://arxiv.org/abs/2203.05090}{{\ttfamily arXiv:2203.05090
  [hep-ex]}}.

\bibitem{ATLAS:2014fzk}
{\bfseries ATLAS} Collaboration, G.~Aad {\em et~al.}, ``{Search for long-lived
  neutral particles decaying into lepton jets in proton-proton collisions at $
  \sqrt{s}=8 $ TeV with the ATLAS detector},''
  \href{http://dx.doi.org/10.1007/JHEP11(2014)088}{{\em JHEP} {\bfseries 11}
  (2014) 088}, \href{http://arxiv.org/abs/1409.0746}{{\ttfamily arXiv:1409.0746
  [hep-ex]}}.

\bibitem{ATLAS:2015itk}
{\bfseries ATLAS} Collaboration, G.~Aad {\em et~al.}, ``{A search for prompt
  lepton-jets in $pp$ collisions at $\sqrt{s}=$ 8 TeV with the ATLAS
  detector},'' \href{http://dx.doi.org/10.1007/JHEP02(2016)062}{{\em JHEP}
  {\bfseries 02} (2016) 062}, \href{http://arxiv.org/abs/1511.05542}{{\ttfamily
  arXiv:1511.05542 [hep-ex]}}.

\bibitem{ATLAS:2022bll}
{\bfseries ATLAS} Collaboration, ``{Search for light long-lived neutral
  particles that decay to collimated pairs of leptons or light hadrons in $pp$
  collisions at $\sqrt{s}=13$ TeV with the ATLAS detector},''.

\bibitem{ATLAS:2022izj}
{\bfseries ATLAS} Collaboration, G.~Aad {\em et~al.}, ``{Search for light
  long-lived neutral particles that decay to collimated pairs of leptons or
  light hadrons in pp collisions at $ \sqrt{s} $ = 13 TeV with the ATLAS
  detector},'' \href{http://dx.doi.org/10.1007/JHEP06(2023)153}{{\em JHEP}
  {\bfseries 06} (2023) 153}, \href{http://arxiv.org/abs/2206.12181}{{\ttfamily
  arXiv:2206.12181 [hep-ex]}}.

\bibitem{ATLAS:2023cjw}
{\bfseries ATLAS} Collaboration, G.~Aad {\em et~al.}, ``{Search for light
  long-lived neutral particles from Higgs boson decays via vector-boson-fusion
  production from pp collisions at $\sqrt{s}=13$ TeV with the ATLAS
  detector},'' \href{http://dx.doi.org/10.1140/epjc/s10052-024-12902-7}{{\em
  Eur. Phys. J. C} {\bfseries 84} no.~7, (2024) 719},
  \href{http://arxiv.org/abs/2311.18298}{{\ttfamily arXiv:2311.18298
  [hep-ex]}}.

\bibitem{ATLAS:2024zxk}
{\bfseries ATLAS} Collaboration, G.~Aad {\em et~al.}, ``{Search for light
  neutral particles decaying promptly into collimated pairs of electrons or
  muons in pp collisions at $\sqrt{s}$ = 13 $\text {T}\text
  {e}\hspace{-1.00006pt}\text {V}$ with the ATLAS detector},''
  \href{http://dx.doi.org/10.1140/epjc/s10052-025-13916-5}{{\em Eur. Phys. J.
  C} {\bfseries 85} no.~3, (2025) 335},
  \href{http://arxiv.org/abs/2407.09168}{{\ttfamily arXiv:2407.09168
  [hep-ex]}}.

\bibitem{LHCb:2019vmc}
{\bfseries LHCb} Collaboration, R.~Aaij {\em et~al.}, ``{Search for
  $A'\to\mu^+\mu^-$ Decays},''
  \href{http://dx.doi.org/10.1103/PhysRevLett.124.041801}{{\em Phys. Rev.
  Lett.} {\bfseries 124} no.~4, (2020) 041801},
  \href{http://arxiv.org/abs/1910.06926}{{\ttfamily arXiv:1910.06926
  [hep-ex]}}.

\bibitem{LHCb:2020ysn}
{\bfseries LHCb} Collaboration, R.~Aaij {\em et~al.}, ``{Searches for low-mass
  dimuon resonances},'' \href{http://dx.doi.org/10.1007/JHEP10(2020)156}{{\em
  JHEP} {\bfseries 10} (2020) 156},
  \href{http://arxiv.org/abs/2007.03923}{{\ttfamily arXiv:2007.03923
  [hep-ex]}}.

\bibitem{Kanemura:2015cxa}
S.~Kanemura, T.~Moroi, and T.~Tanabe, ``{Beam dump experiment at future
  electron{\textendash}positron colliders},''
  \href{http://dx.doi.org/10.1016/j.physletb.2015.10.002}{{\em Phys. Lett. B}
  {\bfseries 751} (2015) 25--28},
  \href{http://arxiv.org/abs/1507.02809}{{\ttfamily arXiv:1507.02809
  [hep-ph]}}.

\bibitem{Marsicano:2018krp}
L.~Marsicano, M.~Battaglieri, M.~Bondi', C.~D.~R. Carvajal, A.~Celentano,
  M.~De~Napoli, R.~De~Vita, E.~Nardi, M.~Raggi, and P.~Valente, ``{Dark photon
  production through positron annihilation in beam-dump experiments},''
  \href{http://dx.doi.org/10.1103/PhysRevD.98.015031}{{\em Phys. Rev. D}
  {\bfseries 98} no.~1, (2018) 015031},
  \href{http://arxiv.org/abs/1802.03794}{{\ttfamily arXiv:1802.03794
  [hep-ex]}}.

\bibitem{Nardi:2018cxi}
E.~Nardi, C.~D.~R. Carvajal, A.~Ghoshal, D.~Meloni, and M.~Raggi, ``{Resonant
  production of dark photons in positron beam dump experiments},''
  \href{http://dx.doi.org/10.1103/PhysRevD.97.095004}{{\em Phys. Rev. D}
  {\bfseries 97} no.~9, (2018) 095004},
  \href{http://arxiv.org/abs/1802.04756}{{\ttfamily arXiv:1802.04756
  [hep-ph]}}.

\bibitem{Berryman:2019dme}
J.~M. Berryman, A.~de~Gouvea, P.~J. Fox, B.~J. Kayser, K.~J. Kelly, and J.~L.
  Raaf, ``{Searches for Decays of New Particles in the DUNE Multi-Purpose Near
  Detector},'' \href{http://dx.doi.org/10.1007/JHEP02(2020)174}{{\em JHEP}
  {\bfseries 02} (2020) 174}, \href{http://arxiv.org/abs/1912.07622}{{\ttfamily
  arXiv:1912.07622 [hep-ph]}}.

\bibitem{Berlin:2020uwy}
A.~Berlin, P.~deNiverville, A.~Ritz, P.~Schuster, and N.~Toro, ``{Sub-GeV dark
  matter production at fixed-target experiments},''
  \href{http://dx.doi.org/10.1103/PhysRevD.102.095011}{{\em Phys. Rev. D}
  {\bfseries 102} no.~9, (2020) 095011},
  \href{http://arxiv.org/abs/2003.03379}{{\ttfamily arXiv:2003.03379
  [hep-ph]}}.

\bibitem{Celentano:2020vtu}
A.~Celentano, L.~Darm{\'e}, L.~Marsicano, and E.~Nardi, ``{New production
  channels for light dark matter in hadronic showers},''
  \href{http://dx.doi.org/10.1103/PhysRevD.102.075026}{{\em Phys. Rev. D}
  {\bfseries 102} no.~7, (2020) 075026},
  \href{http://arxiv.org/abs/2006.09419}{{\ttfamily arXiv:2006.09419
  [hep-ph]}}.

\bibitem{Asai:2021xtg}
K.~Asai, T.~Moroi, and A.~Niki, ``{Leptophilic Gauge Bosons at ILC Beam Dump
  Experiment},'' \href{http://dx.doi.org/10.1016/j.physletb.2021.136374}{{\em
  Phys. Lett. B} {\bfseries 818} (2021) 136374},
  \href{http://arxiv.org/abs/2104.00888}{{\ttfamily arXiv:2104.00888
  [hep-ph]}}.

\bibitem{Asai:2021ehn}
K.~Asai, S.~Iwamoto, Y.~Sakaki, and D.~Ueda, ``{New physics searches at the ILC
  positron and electron beam dumps},''
  \href{http://dx.doi.org/10.1007/JHEP09(2021)183}{{\em JHEP} {\bfseries 09}
  (2021) 183}, \href{http://arxiv.org/abs/2105.13768}{{\ttfamily
  arXiv:2105.13768 [hep-ph]}}.

\bibitem{Moroi:2022qwz}
T.~Moroi and A.~Niki, ``{Leptophilic gauge bosons at lepton beam dump
  experiments},'' \href{http://dx.doi.org/10.1007/JHEP05(2023)016}{{\em JHEP}
  {\bfseries 05} (2023) 016}, \href{http://arxiv.org/abs/2205.11766}{{\ttfamily
  arXiv:2205.11766 [hep-ph]}}.

\bibitem{Asai:2022zxw}
K.~Asai, A.~Das, J.~Li, T.~Nomura, and O.~Seto, ``{Chiral Z' in FASER, FASER2,
  DUNE, and ILC beam dump experiments},''
  \href{http://dx.doi.org/10.1103/PhysRevD.106.095033}{{\em Phys. Rev. D}
  {\bfseries 106} no.~9, (2022) 095033},
  \href{http://arxiv.org/abs/2206.12676}{{\ttfamily arXiv:2206.12676
  [hep-ph]}}.

\bibitem{Battaglieri:2022dcy}
M.~Battaglieri {\em et~al.}, ``{Dark matter search with the BDX-MINI
  experiment},'' \href{http://dx.doi.org/10.1103/PhysRevD.106.072011}{{\em
  Phys. Rev. D} {\bfseries 106} no.~7, (2022) 072011},
  \href{http://arxiv.org/abs/2208.01387}{{\ttfamily arXiv:2208.01387
  [hep-ex]}}.

\bibitem{Asai:2023dzs}
K.~Asai, S.~Iwamoto, M.~Perelstein, Y.~Sakaki, and D.~Ueda, ``{Sub-GeV dark
  matter search at ILC beam dumps},''
  \href{http://dx.doi.org/10.1007/JHEP02(2024)129}{{\em JHEP} {\bfseries 02}
  (2024) 129}, \href{http://arxiv.org/abs/2301.03816}{{\ttfamily
  arXiv:2301.03816 [hep-ph]}}.

\bibitem{Blinov:2024pza}
N.~Blinov, P.~J. Fox, K.~J. Kelly, P.~A.~N. Machado, and R.~Plestid, ``{Dark
  fluxes from electromagnetic cascades},''
  \href{http://dx.doi.org/10.1007/JHEP07(2024)022}{{\em JHEP} {\bfseries 07}
  (2024) 022}, \href{http://arxiv.org/abs/2401.06843}{{\ttfamily
  arXiv:2401.06843 [hep-ph]}}.

\bibitem{Curtin:2023bcf}
D.~Curtin, Y.~Kahn, and R.~Nguyen, ``{Dark photons from charged pion
  bremsstrahlung at proton beam experiments},''
  \href{http://dx.doi.org/10.1103/PhysRevD.108.095039}{{\em Phys. Rev. D}
  {\bfseries 108} no.~9, (2023) 095039},
  \href{http://arxiv.org/abs/2305.19309}{{\ttfamily arXiv:2305.19309
  [hep-ph]}}.

\bibitem{Schulthess:2025tct}
I.~Schulthess and F.~Meloni, ``{New Physics Search with the Optical Dump
  Concept at Future Colliders},''
  \href{http://arxiv.org/abs/2503.20996}{{\ttfamily arXiv:2503.20996
  [hep-ph]}}.

\bibitem{Chen:2017awl}
C.-Y. Chen, M.~Pospelov, and Y.-M. Zhong, ``{Muon Beam Experiments to Probe the
  Dark Sector},'' \href{http://dx.doi.org/10.1103/PhysRevD.95.115005}{{\em
  Phys. Rev. D} {\bfseries 95} no.~11, (2017) 115005},
  \href{http://arxiv.org/abs/1701.07437}{{\ttfamily arXiv:1701.07437
  [hep-ph]}}.

\bibitem{Kahn:2018cqs}
Y.~Kahn, G.~Krnjaic, N.~Tran, and A.~Whitbeck, ``{M$^{3}$: a new muon missing
  momentum experiment to probe $(g-2)_{\mu}$ and dark matter at Fermilab},''
  \href{http://dx.doi.org/10.1007/JHEP09(2018)153}{{\em JHEP} {\bfseries 09}
  (2018) 153}, \href{http://arxiv.org/abs/1804.03144}{{\ttfamily
  arXiv:1804.03144 [hep-ph]}}.

\bibitem{Gninenko:2018ter}
S.~N. Gninenko, D.~V. Kirpichnikov, and N.~V. Krasnikov, ``{Probing
  millicharged particles with NA64 experiment at CERN},''
  \href{http://dx.doi.org/10.1103/PhysRevD.100.035003}{{\em Phys. Rev. D}
  {\bfseries 100} no.~3, (2019) 035003},
  \href{http://arxiv.org/abs/1810.06856}{{\ttfamily arXiv:1810.06856
  [hep-ph]}}.

\bibitem{Gninenko:2020hbd}
S.~N. Gninenko, N.~V. Krasnikov, and V.~A. Matveev, ``{Search for dark sector
  physics with NA64},'' \href{http://dx.doi.org/10.1134/S1063779620050044}{{\em
  Phys. Part. Nucl.} {\bfseries 51} no.~5, (2020) 829--858},
  \href{http://arxiv.org/abs/2003.07257}{{\ttfamily arXiv:2003.07257
  [hep-ph]}}.

\bibitem{Acosta:2021qpx}
D.~Acosta and W.~Li, ``{A muon{\textendash}ion collider at BNL: The future QCD
  frontier and path to a new energy frontier of
  {\ensuremath{\mu}}+{\ensuremath{\mu}}{\ensuremath{-}} colliders},''
  \href{http://dx.doi.org/10.1016/j.nima.2022.166334}{{\em Nucl. Instrum. Meth.
  A} {\bfseries 1027} (2022) 166334},
  \href{http://arxiv.org/abs/2107.02073}{{\ttfamily arXiv:2107.02073
  [physics.acc-ph]}}.

\bibitem{Kirpichnikov:2021jev}
D.~V. Kirpichnikov, H.~Sieber, L.~M. Bueno, P.~Crivelli, and M.~M. Kirsanov,
  ``{Probing hidden sectors with a muon beam: Total and differential cross
  sections for vector boson production in muon bremsstrahlung},''
  \href{http://dx.doi.org/10.1103/PhysRevD.104.076012}{{\em Phys. Rev. D}
  {\bfseries 104} no.~7, (2021) 076012},
  \href{http://arxiv.org/abs/2107.13297}{{\ttfamily arXiv:2107.13297
  [hep-ph]}}.

\bibitem{Cesarotti:2022ttv}
C.~Cesarotti, S.~Homiller, R.~K. Mishra, and M.~Reece, ``{Probing New Gauge
  Forces with a High-Energy Muon Beam Dump},''
  \href{http://dx.doi.org/10.1103/PhysRevLett.130.071803}{{\em Phys. Rev.
  Lett.} {\bfseries 130} no.~7, (2023) 071803},
  \href{http://arxiv.org/abs/2202.12302}{{\ttfamily arXiv:2202.12302
  [hep-ph]}}.

\bibitem{Acosta:2022ejc}
D.~Acosta, E.~Barberis, N.~Hurley, W.~Li, O.~Miguel~Colin, Y.~Wang, D.~Wood,
  and X.~Zuo, ``{The potential of a TeV-scale muon-ion collider},''
  \href{http://dx.doi.org/10.1088/1748-0221/18/09/P09025}{{\em JINST}
  {\bfseries 18} no.~09, (2023) P09025},
  \href{http://arxiv.org/abs/2203.06258}{{\ttfamily arXiv:2203.06258
  [hep-ex]}}.

\bibitem{Zhevlakov:2022vio}
A.~S. Zhevlakov, D.~V. Kirpichnikov, and V.~E. Lyubovitskij, ``{Implication of
  the dark axion portal for the EDM of fermions and dark matter probing with
  NA64e, NA64{\ensuremath{\mu}}, LDMX, M3, and BaBar},''
  \href{http://dx.doi.org/10.1103/PhysRevD.106.035018}{{\em Phys. Rev. D}
  {\bfseries 106} no.~3, (2022) 035018},
  \href{http://arxiv.org/abs/2204.09978}{{\ttfamily arXiv:2204.09978
  [hep-ph]}}.

\bibitem{Voronchikhin:2022rwc}
I.~V. Voronchikhin and D.~V. Kirpichnikov, ``{Probing hidden spin-2 mediator of
  dark matter with NA64e, LDMX, NA64{\ensuremath{\mu}}, and M3},''
  \href{http://dx.doi.org/10.1103/PhysRevD.106.115041}{{\em Phys. Rev. D}
  {\bfseries 106} no.~11, (2022) 115041},
  \href{http://arxiv.org/abs/2210.00751}{{\ttfamily arXiv:2210.00751
  [hep-ph]}}.

\bibitem{Forbes:2022bvo}
D.~Forbes, C.~Herwig, Y.~Kahn, G.~Krnjaic, C.~Mantilla~Suarez, N.~Tran, and
  A.~Whitbeck, ``{New searches for muonphilic particles at proton beam dump
  spectrometers},'' \href{http://dx.doi.org/10.1103/PhysRevD.107.116026}{{\em
  Phys. Rev. D} {\bfseries 107} no.~11, (2023) 116026},
  \href{http://arxiv.org/abs/2212.00033}{{\ttfamily arXiv:2212.00033
  [hep-ph]}}.

\bibitem{Radics:2023tkn}
B.~Radics, L.~Molina-Bueno, L.~Fields., H.~Sieber, and P.~Crivelli,
  ``{Sensitivity potential to a light flavor-changing scalar boson with DUNE
  and NA64$\mu $},''
  \href{http://dx.doi.org/10.1140/epjc/s10052-023-11891-3}{{\em Eur. Phys. J.
  C} {\bfseries 83} no.~9, (2023) 775},
  \href{http://arxiv.org/abs/2306.07405}{{\ttfamily arXiv:2306.07405
  [hep-ex]}}.

\bibitem{Zhevlakov:2023jzt}
A.~S. Zhevlakov, D.~V. Kirpichnikov, and V.~E. Lyubovitskij, ``{Lepton flavor
  violating dark photon},''
  \href{http://dx.doi.org/10.1103/PhysRevD.109.015015}{{\em Phys. Rev. D}
  {\bfseries 109} no.~1, (2024) 015015},
  \href{http://arxiv.org/abs/2307.10771}{{\ttfamily arXiv:2307.10771
  [hep-ph]}}.

\bibitem{Cesarotti:2023sje}
C.~Cesarotti and R.~Gambhir, ``{The new physics case for beam-dump experiments
  with accelerated muon beams},''
  \href{http://dx.doi.org/10.1007/JHEP05(2024)283}{{\em JHEP} {\bfseries 05}
  (2024) 283}, \href{http://arxiv.org/abs/2310.16110}{{\ttfamily
  arXiv:2310.16110 [hep-ph]}}.

\bibitem{Batell:2024cdl}
B.~Batell, H.~Davoudiasl, R.~Marcarelli, E.~T. Neil, and S.~Trojanowski,
  ``{Lepton-flavor-violating ALP signals with TeV-scale muon beams},''
  \href{http://dx.doi.org/10.1103/PhysRevD.110.075039}{{\em Phys. Rev. D}
  {\bfseries 110} no.~7, (2024) 075039},
  \href{http://arxiv.org/abs/2407.15942}{{\ttfamily arXiv:2407.15942
  [hep-ph]}}.

\bibitem{Voronchikhin:2024ygo}
I.~V. Voronchikhin and D.~V. Kirpichnikov, ``{The bremsstrahlung-like
  production of the massive spin-2 dark matter mediator},''
  \href{http://arxiv.org/abs/2412.10150}{{\ttfamily arXiv:2412.10150
  [hep-ph]}}.

\bibitem{Li:2025yzb}
H.~Li, Z.~Liu, and N.~Song, ``{Probing axion and muon-philic new physics with
  muon beam dump},'' \href{http://arxiv.org/abs/2501.06294}{{\ttfamily
  arXiv:2501.06294 [hep-ph]}}.

\bibitem{Sieber:2021fue}
H.~Sieber, D.~Banerjee, P.~Crivelli, E.~Depero, S.~N. Gninenko, D.~V.
  Kirpichnikov, M.~M. Kirsanov, V.~Poliakov, and L.~Molina~Bueno, ``{Prospects
  in the search for a new light Z' boson with the NA64{\ensuremath{\mu}}
  experiment at the CERN SPS},''
  \href{http://dx.doi.org/10.1103/PhysRevD.105.052006}{{\em Phys. Rev. D}
  {\bfseries 105} no.~5, (2022) 052006},
  \href{http://arxiv.org/abs/2110.15111}{{\ttfamily arXiv:2110.15111
  [hep-ex]}}.

\bibitem{NA64:2024klw}
{\bfseries NA64} Collaboration, Y.~M. Andreev {\em et~al.}, ``{First Results in
  the Search for Dark Sectors at NA64 with the CERN SPS High Energy Muon
  Beam},'' \href{http://dx.doi.org/10.1103/PhysRevLett.132.211803}{{\em Phys.
  Rev. Lett.} {\bfseries 132} no.~21, (2024) 211803},
  \href{http://arxiv.org/abs/2401.01708}{{\ttfamily arXiv:2401.01708
  [hep-ex]}}.

\bibitem{Foot:1990mn}
R.~Foot, ``{New Physics From Electric Charge Quantization?},''
  \href{http://dx.doi.org/10.1142/S0217732391000543}{{\em Mod. Phys. Lett. A}
  {\bfseries 6} (1991) 527--530}.

\bibitem{He:1990pn}
X.~G. He, G.~C. Joshi, H.~Lew, and R.~R. Volkas, ``{NEW Z-prime
  PHENOMENOLOGY},'' \href{http://dx.doi.org/10.1103/PhysRevD.43.R22}{{\em Phys.
  Rev. D} {\bfseries 43} (1991) 22--24}.

\bibitem{He:1991qd}
X.-G. He, G.~C. Joshi, H.~Lew, and R.~R. Volkas, ``{Simplest Z-prime model},''
  \href{http://dx.doi.org/10.1103/PhysRevD.44.2118}{{\em Phys. Rev. D}
  {\bfseries 44} (1991) 2118--2132}.

\bibitem{Foot:1994vd}
R.~Foot, X.~G. He, H.~Lew, and R.~R. Volkas, ``{Model for a light Z-prime
  boson},'' \href{http://dx.doi.org/10.1103/PhysRevD.50.4571}{{\em Phys. Rev.
  D} {\bfseries 50} (1994) 4571--4580},
  \href{http://arxiv.org/abs/hep-ph/9401250}{{\ttfamily arXiv:hep-ph/9401250}}.

\bibitem{Baek:2008nz}
S.~Baek and P.~Ko, ``{Phenomenology of U(1)(L(mu)-L(tau)) charged dark matter
  at PAMELA and colliders},''
  \href{http://dx.doi.org/10.1088/1475-7516/2009/10/011}{{\em JCAP} {\bfseries
  10} (2009) 011}, \href{http://arxiv.org/abs/0811.1646}{{\ttfamily
  arXiv:0811.1646 [hep-ph]}}.

\bibitem{Baek:2015fea}
S.~Baek, ``{Dark matter and muon $(g-2)$ in local
  $U(1)_{L_\mu-L_\tau}$-extended Ma Model},''
  \href{http://dx.doi.org/10.1016/j.physletb.2016.02.062}{{\em Phys. Lett. B}
  {\bfseries 756} (2016) 1--5},
  \href{http://arxiv.org/abs/1510.02168}{{\ttfamily arXiv:1510.02168
  [hep-ph]}}.

\bibitem{Patra:2016shz}
S.~Patra, S.~Rao, N.~Sahoo, and N.~Sahu, ``{Gauged $U(1)_{L_\mu - L_\tau}$
  model in light of muon $g-2$ anomaly, neutrino mass and dark matter
  phenomenology},''
  \href{http://dx.doi.org/10.1016/j.nuclphysb.2017.02.010}{{\em Nucl. Phys. B}
  {\bfseries 917} (2017) 317--336},
  \href{http://arxiv.org/abs/1607.04046}{{\ttfamily arXiv:1607.04046
  [hep-ph]}}.

\bibitem{Biswas:2016yan}
A.~Biswas, S.~Choubey, and S.~Khan, ``{Neutrino Mass, Dark Matter and Anomalous
  Magnetic Moment of Muon in a $U(1)_{L_{\mu}-L_{\tau}}$ Model},''
  \href{http://dx.doi.org/10.1007/JHEP09(2016)147}{{\em JHEP} {\bfseries 09}
  (2016) 147}, \href{http://arxiv.org/abs/1608.04194}{{\ttfamily
  arXiv:1608.04194 [hep-ph]}}.

\bibitem{Biswas:2016yjr}
A.~Biswas, S.~Choubey, and S.~Khan, ``{FIMP and Muon ($g-2$) in a
  U$(1)_{L_{\mu}-L_{\tau}}$ Model},''
  \href{http://dx.doi.org/10.1007/JHEP02(2017)123}{{\em JHEP} {\bfseries 02}
  (2017) 123}, \href{http://arxiv.org/abs/1612.03067}{{\ttfamily
  arXiv:1612.03067 [hep-ph]}}.

\bibitem{Asai:2017ryy}
K.~Asai, K.~Hamaguchi, and N.~Nagata, ``{Predictions for the neutrino
  parameters in the minimal gauged U(1)$_{L_\mu-L_\tau}$ model},''
  \href{http://dx.doi.org/10.1140/epjc/s10052-017-5348-x}{{\em Eur. Phys. J. C}
  {\bfseries 77} no.~11, (2017) 763},
  \href{http://arxiv.org/abs/1705.00419}{{\ttfamily arXiv:1705.00419
  [hep-ph]}}.

\bibitem{Arcadi:2018tly}
G.~Arcadi, T.~Hugle, and F.~S. Queiroz, ``{The Dark $L_\mu - L_\tau$ Rises via
  Kinetic Mixing},''
  \href{http://dx.doi.org/10.1016/j.physletb.2018.07.028}{{\em Phys. Lett. B}
  {\bfseries 784} (2018) 151--158},
  \href{http://arxiv.org/abs/1803.05723}{{\ttfamily arXiv:1803.05723
  [hep-ph]}}.

\bibitem{Kamada:2018zxi}
A.~Kamada, K.~Kaneta, K.~Yanagi, and H.-B. Yu, ``{Self-interacting dark matter
  and muon $g-2$ in a gauged U$(1)_{L_{\mu} - L_{\tau}}$ model},''
  \href{http://dx.doi.org/10.1007/JHEP06(2018)117}{{\em JHEP} {\bfseries 06}
  (2018) 117}, \href{http://arxiv.org/abs/1805.00651}{{\ttfamily
  arXiv:1805.00651 [hep-ph]}}.

\bibitem{Foldenauer:2018zrz}
P.~Foldenauer, ``{Light dark matter in a gauged $U(1)_{L_\mu-L_\tau}$ model},''
  \href{http://dx.doi.org/10.1103/PhysRevD.99.035007}{{\em Phys. Rev. D}
  {\bfseries 99} no.~3, (2019) 035007},
  \href{http://arxiv.org/abs/1808.03647}{{\ttfamily arXiv:1808.03647
  [hep-ph]}}.

\bibitem{Asai:2019ciz}
K.~Asai, ``{Predictions for the neutrino parameters in the minimal model
  extended by linear combination of $U(1)_{L_e-L_\mu}$, $U(1)_{L_\mu-L_\tau}$
  and U(1)$_{B-L}$ gauge symmetries},''
  \href{http://dx.doi.org/10.1140/epjc/s10052-020-7622-6}{{\em Eur. Phys. J. C}
  {\bfseries 80} no.~2, (2020) 76},
  \href{http://arxiv.org/abs/1907.04042}{{\ttfamily arXiv:1907.04042
  [hep-ph]}}.

\bibitem{Okada:2019sbb}
N.~Okada and O.~Seto, ``{Inelastic extra $U(1)$ charged scalar dark matter},''
  \href{http://dx.doi.org/10.1103/PhysRevD.101.023522}{{\em Phys. Rev. D}
  {\bfseries 101} no.~2, (2020) 023522},
  \href{http://arxiv.org/abs/1908.09277}{{\ttfamily arXiv:1908.09277
  [hep-ph]}}.

\bibitem{Asai:2020qlp}
K.~Asai, S.~Okawa, and K.~Tsumura, ``{Search for $ \mathrm{U}{(1)}_{L_{\mu
  }-{L}_{\tau }} $ charged dark matter with neutrino telescope},''
  \href{http://dx.doi.org/10.1007/JHEP03(2021)047}{{\em JHEP} {\bfseries 03}
  (2021) 047}, \href{http://arxiv.org/abs/2011.03165}{{\ttfamily
  arXiv:2011.03165 [hep-ph]}}.

\bibitem{Holst:2021lzm}
I.~Holst, D.~Hooper, and G.~Krnjaic, ``{Simplest and Most Predictive Model of
  Muon g-2 and Thermal Dark Matter},''
  \href{http://dx.doi.org/10.1103/PhysRevLett.128.141802}{{\em Phys. Rev.
  Lett.} {\bfseries 128} no.~14, (2022) 141802},
  \href{http://arxiv.org/abs/2107.09067}{{\ttfamily arXiv:2107.09067
  [hep-ph]}}.

\bibitem{Tapadar:2021kgw}
A.~Tapadar, S.~Ganguly, and S.~Roy, ``{Non-adiabatic evolution of dark sector
  in the presence of $U(1)_{L_{\mu} - L_{\tau}}$ gauge symmetry},''
  \href{http://dx.doi.org/10.1088/1475-7516/2022/05/019}{{\em JCAP} {\bfseries
  05} no.~05, (2022) 019}, \href{http://arxiv.org/abs/2109.13609}{{\ttfamily
  arXiv:2109.13609 [hep-ph]}}.

\bibitem{Heeck:2022znj}
J.~Heeck and A.~Thapa, ``{Explaining lepton-flavor non-universality and
  self-interacting dark matter with $L_{\mu} - L_{\tau}$},''
  \href{http://dx.doi.org/10.1140/epjc/s10052-022-10437-3}{{\em Eur. Phys. J.
  C} {\bfseries 82} no.~5, (2022) 480},
  \href{http://arxiv.org/abs/2202.08854}{{\ttfamily arXiv:2202.08854
  [hep-ph]}}.

\bibitem{Nagao:2022osm}
K.~I. Nagao, T.~Nomura, H.~Okada, and T.~Shimomura, ``{Neutrinophilic
  dark-matter annihilation in a model with $U(1)_{L_{\mu} - L_{\tau}} \times
  U(1)_H$ gauge symmetry},''
  \href{http://dx.doi.org/10.1103/PhysRevD.108.055032}{{\em Phys. Rev. D}
  {\bfseries 108} no.~5, (2023) 055032},
  \href{http://arxiv.org/abs/2212.14528}{{\ttfamily arXiv:2212.14528
  [hep-ph]}}.

\bibitem{KA:2023dyz}
S.~K.~A., A.~Das, G.~Lambiase, T.~Nomura, and Y.~Orikasa, ``{Probing chiral and
  flavored $Z^\prime$ from cosmic bursts through neutrino interactions},''
  \href{http://arxiv.org/abs/2308.14483}{{\ttfamily arXiv:2308.14483
  [hep-ph]}}.

\bibitem{Figueroa:2024tmn}
P.~Figueroa, G.~Herrera, and F.~Ochoa, ``{Direct detection of light dark matter
  charged under a $L_{\mu} - L_{\tau}$ symmetry},''
  \href{http://arxiv.org/abs/2404.03090}{{\ttfamily arXiv:2404.03090
  [hep-ph]}}.

\bibitem{Asai:2020qax}
K.~Asai, K.~Hamaguchi, N.~Nagata, and S.-Y. Tseng, ``{Leptogenesis in the
  minimal gauged U(1)$_{L_\mu-L_\tau}$ model and the sign of the cosmological
  baryon asymmetry},''
  \href{http://dx.doi.org/10.1088/1475-7516/2020/11/013}{{\em JCAP} {\bfseries
  11} (2020) 013}, \href{http://arxiv.org/abs/2005.01039}{{\ttfamily
  arXiv:2005.01039 [hep-ph]}}.

\bibitem{Borah:2021mri}
D.~Borah, A.~Dasgupta, and D.~Mahanta, ``{TeV scale resonant leptogenesis with
  L{\ensuremath{\mu}}-L{\ensuremath{\tau}} gauge symmetry in light of the muon
  g-2},'' \href{http://dx.doi.org/10.1103/PhysRevD.104.075006}{{\em Phys. Rev.
  D} {\bfseries 104} no.~7, (2021) 075006},
  \href{http://arxiv.org/abs/2106.14410}{{\ttfamily arXiv:2106.14410
  [hep-ph]}}.

\bibitem{Eijima:2023yiw}
S.~Eijima, M.~Ibe, and K.~Murai, ``{Muon g {\ensuremath{-}} 2 and non-thermal
  leptogenesis in $ \textrm{U}{(1)}_{L_{\mu }-{L}_{\tau }} $ model},''
  \href{http://dx.doi.org/10.1007/JHEP05(2023)010}{{\em JHEP} {\bfseries 05}
  (2023) 010}, \href{http://arxiv.org/abs/2303.09751}{{\ttfamily
  arXiv:2303.09751 [hep-ph]}}.

\bibitem{Granelli:2023egb}
A.~Granelli, K.~Hamaguchi, N.~Nagata, M.~E. Ramirez-Quezada, and J.~Wada,
  ``{Thermal leptogenesis in the minimal gauged $ \textrm{U}{(1)}_{L_{\mu
  }-{L}_{\tau }} $ model},''
  \href{http://dx.doi.org/10.1007/JHEP09(2023)079}{{\em JHEP} {\bfseries 09}
  (2023) 079}, \href{http://arxiv.org/abs/2305.18100}{{\ttfamily
  arXiv:2305.18100 [hep-ph]}}.

\bibitem{Wada:2024cbe}
J.~Wada, ``{Majoron-driven leptogenesis in gauged
  U(1)L{\ensuremath{\mu}}-L{\ensuremath{\tau}} model},''
  \href{http://dx.doi.org/10.1103/PhysRevD.110.103510}{{\em Phys. Rev. D}
  {\bfseries 110} no.~10, (2024) 103510},
  \href{http://arxiv.org/abs/2404.10283}{{\ttfamily arXiv:2404.10283
  [hep-ph]}}.

\bibitem{Asai:2018ocx}
K.~Asai, K.~Hamaguchi, N.~Nagata, S.-Y. Tseng, and K.~Tsumura, ``{Minimal
  Gauged U(1)$_{L_\alpha - L_\beta}$ Models Driven into a Corner},''
  \href{http://dx.doi.org/10.1103/PhysRevD.99.055029}{{\em Phys. Rev. D}
  {\bfseries 99} no.~5, (2019) 055029},
  \href{http://arxiv.org/abs/1811.07571}{{\ttfamily arXiv:1811.07571
  [hep-ph]}}.

\bibitem{Ibe:2025rwk}
M.~Ibe, S.~Shirai, and K.~Watanabe, ``{Global neutrino constraints on the
  minimal U(1)L{\ensuremath{\mu}}-L{\ensuremath{\tau}} model},''
  \href{http://dx.doi.org/10.1103/PhysRevD.111.095034}{{\em Phys. Rev. D}
  {\bfseries 111} no.~9, (2025) 095034},
  \href{http://arxiv.org/abs/2503.01399}{{\ttfamily arXiv:2503.01399
  [hep-ph]}}.

\bibitem{Escudero:2019gzq}
M.~Escudero, D.~Hooper, G.~Krnjaic, and M.~Pierre, ``{Cosmology with A Very
  Light L$_{\mu}$ {\ensuremath{-}} L$_{\tau}$ Gauge Boson},''
  \href{http://dx.doi.org/10.1007/JHEP03(2019)071}{{\em JHEP} {\bfseries 03}
  (2019) 071}, \href{http://arxiv.org/abs/1901.02010}{{\ttfamily
  arXiv:1901.02010 [hep-ph]}}.

\bibitem{Araki:2021xdk}
T.~Araki, K.~Asai, K.~Honda, R.~Kasuya, J.~Sato, T.~Shimomura, and M.~J.~S.
  Yang, ``{Resolving the Hubble tension in a $U(1)_{L_\mu-L_\tau}$ model with
  the Majoron},'' \href{http://dx.doi.org/10.1093/ptep/ptab108}{{\em PTEP}
  {\bfseries 2021} no.~10, (2021) 103B05},
  \href{http://arxiv.org/abs/2103.07167}{{\ttfamily arXiv:2103.07167
  [hep-ph]}}.

\bibitem{Carpio:2021jhu}
J.~A. Carpio, K.~Murase, I.~M. Shoemaker, and Z.~Tabrizi, ``{High-energy cosmic
  neutrinos as a probe of the vector mediator scenario in light of the muon g-2
  anomaly and Hubble tension},''
  \href{http://dx.doi.org/10.1103/PhysRevD.107.103057}{{\em Phys. Rev. D}
  {\bfseries 107} no.~10, (2023) 103057},
  \href{http://arxiv.org/abs/2104.15136}{{\ttfamily arXiv:2104.15136
  [hep-ph]}}.

\bibitem{Asai:2023ajh}
K.~Asai, T.~Asano, J.~Sato, and M.~J.~S. Yang, ``{Contribution of Majoron to
  Hubble Tension in Gauged U(1)L\ensuremath{\mu} \textendash{}
  L\ensuremath{\tau} Model},''
  \href{http://dx.doi.org/10.1093/ptep/ptae094}{{\em PTEP} {\bfseries 2024}
  no.~7, (2024) 073E01}, \href{http://arxiv.org/abs/2309.01162}{{\ttfamily
  arXiv:2309.01162 [hep-ph]}}.

\bibitem{Aliberti:2025beg}
R.~Aliberti {\em et~al.}, ``{The anomalous magnetic moment of the muon in the
  Standard Model: an update},''
  \href{http://arxiv.org/abs/2505.21476}{{\ttfamily arXiv:2505.21476
  [hep-ph]}}.

\bibitem{Chakraborty:2024xxc}
D.~Chakraborty, A.~Chatterjee, A.~Kaushik, and K.~Nishiwaki, ``{Prospects of
  five-dimensional L{\ensuremath{\mu}}-L{\ensuremath{\tau}} gauge interactions
  in the light of elastic neutrino-electron scatterings: The scope of the DUNE
  near detector},'' \href{http://dx.doi.org/10.1103/PhysRevD.110.095030}{{\em
  Phys. Rev. D} {\bfseries 110} no.~9, (2024) 095030},
  \href{http://arxiv.org/abs/2407.20615}{{\ttfamily arXiv:2407.20615
  [hep-ph]}}.

\bibitem{Ponton:2012bi}
E.~Ponton, \href{http://dx.doi.org/10.1142/9789814390163_0007}{``{TASI 2011:
  Four Lectures on TeV Scale Extra Dimensions},''} in {\em {Theoretical
  Advanced Study Institute in Elementary Particle Physics}: {The Dark Secrets
  of the Terascale}}, pp.~283--374.
\newblock 2013.
\newblock \href{http://arxiv.org/abs/1207.3827}{{\ttfamily arXiv:1207.3827
  [hep-ph]}}.

\bibitem{Arkani-Hamed:1998jmv}
N.~Arkani-Hamed, S.~Dimopoulos, and G.~R. Dvali, ``{The Hierarchy problem and
  new dimensions at a millimeter},''
  \href{http://dx.doi.org/10.1016/S0370-2693(98)00466-3}{{\em Phys. Lett. B}
  {\bfseries 429} (1998) 263--272},
  \href{http://arxiv.org/abs/hep-ph/9803315}{{\ttfamily arXiv:hep-ph/9803315}}.

\bibitem{Randall:1999ee}
L.~Randall and R.~Sundrum, ``{A Large mass hierarchy from a small extra
  dimension},'' \href{http://dx.doi.org/10.1103/PhysRevLett.83.3370}{{\em Phys.
  Rev. Lett.} {\bfseries 83} (1999) 3370--3373},
  \href{http://arxiv.org/abs/hep-ph/9905221}{{\ttfamily arXiv:hep-ph/9905221}}.

\bibitem{Anchordoqui:2024ajk}
L.~A. Anchordoqui, I.~Antoniadis, and D.~L{\"u}st, ``{Landscape, Swampland, and
  Extra Dimensions},'' \href{http://dx.doi.org/10.22323/1.463.0215}{{\em PoS}
  {\bfseries CORFU2023} (2024) 215},
  \href{http://arxiv.org/abs/2405.04427}{{\ttfamily arXiv:2405.04427
  [hep-th]}}.

\bibitem{Antoniadis:1990ew}
I.~Antoniadis, ``{A Possible new dimension at a few TeV},''
  \href{http://dx.doi.org/10.1016/0370-2693(90)90617-F}{{\em Phys. Lett. B}
  {\bfseries 246} (1990) 377--384}.

\bibitem{Appelquist:2000nn}
T.~Appelquist, H.-C. Cheng, and B.~A. Dobrescu, ``{Bounds on universal extra
  dimensions},'' \href{http://dx.doi.org/10.1103/PhysRevD.64.035002}{{\em Phys.
  Rev. D} {\bfseries 64} (2001) 035002},
  \href{http://arxiv.org/abs/hep-ph/0012100}{{\ttfamily arXiv:hep-ph/0012100}}.

\bibitem{Kakuda:2013kba}
T.~Kakuda, K.~Nishiwaki, K.-y. Oda, and R.~Watanabe, ``{Universal extra
  dimensions after Higgs discovery},''
  \href{http://dx.doi.org/10.1103/PhysRevD.88.035007}{{\em Phys. Rev. D}
  {\bfseries 88} (2013) 035007},
  \href{http://arxiv.org/abs/1305.1686}{{\ttfamily arXiv:1305.1686 [hep-ph]}}.

\bibitem{Deutschmann:2017bth}
N.~Deutschmann, T.~Flacke, and J.~S. Kim, ``{Current LHC Constraints on Minimal
  Universal Extra Dimensions},''
  \href{http://dx.doi.org/10.1016/j.physletb.2017.06.004}{{\em Phys. Lett. B}
  {\bfseries 771} (2017) 515--520},
  \href{http://arxiv.org/abs/1702.00410}{{\ttfamily arXiv:1702.00410
  [hep-ph]}}.

\bibitem{Flores:2021xwx}
M.~M. Flores, J.~S. Kim, K.~Rolbiecki, and R.~R. d.~A. Bazan, ``{Updated LHC
  bounds on MUED after run~2},''
  \href{http://dx.doi.org/10.1142/S0217751X23500021}{{\em Int. J. Mod. Phys. A}
  {\bfseries 38} no.~01, (2023) 2350002},
  \href{http://arxiv.org/abs/2110.00500}{{\ttfamily arXiv:2110.00500
  [hep-ph]}}.

\bibitem{Saldanha:2012vba}
R.~N. Saldanha, {\em {Precision Measurement of the $^7$Be Solar Neutrino
  Interaction Rate in Borexino}}.
\newblock PhD thesis, Princeton U., 2012.

\bibitem{Wong:2015kgl}
H.~T.-K. Wong, ``{Taiwan EXperiment On NeutrinO {\textemdash} History and
  Prospects},'' \href{http://dx.doi.org/10.1142/S0217751X18300144}{{\em The
  Universe} {\bfseries 3} no.~4, (2015) 22--37},
  \href{http://arxiv.org/abs/1608.00306}{{\ttfamily arXiv:1608.00306
  [hep-ex]}}.

\bibitem{CHARM-II:1993phx}
{\bfseries CHARM-II} Collaboration, P.~Vilain {\em et~al.}, ``{Measurement of
  differential cross-sections for muon-neutrino electron scattering},''
  \href{http://dx.doi.org/10.1016/0370-2693(93)90408-A}{{\em Phys. Lett. B}
  {\bfseries 302} (1993) 351--355}.

\bibitem{CHARM-II:1994dzw}
{\bfseries CHARM-II} Collaboration, P.~Vilain {\em et~al.}, ``{Precision
  measurement of electroweak parameters from the scattering of muon-neutrinos
  on electrons},'' \href{http://dx.doi.org/10.1016/0370-2693(94)91421-4}{{\em
  Phys. Lett. B} {\bfseries 335} (1994) 246--252}.

\bibitem{DUNE:2016hlj}
{\bfseries DUNE} Collaboration, R.~Acciarri {\em et~al.}, ``{Long-Baseline
  Neutrino Facility (LBNF) and Deep Underground Neutrino Experiment (DUNE)}:
  {Conceptual Design Report, Volume 1: The LBNF and DUNE Projects},''
  \href{http://arxiv.org/abs/1601.05471}{{\ttfamily arXiv:1601.05471
  [physics.ins-det]}}.

\bibitem{DUNE:2020fgq}
{\bfseries DUNE} Collaboration, B.~Abi {\em et~al.}, ``{Prospects for beyond
  the Standard Model physics searches at the Deep Underground Neutrino
  Experiment},'' \href{http://dx.doi.org/10.1140/epjc/s10052-021-09007-w}{{\em
  Eur. Phys. J. C} {\bfseries 81} no.~4, (2021) 322},
  \href{http://arxiv.org/abs/2008.12769}{{\ttfamily arXiv:2008.12769
  [hep-ex]}}.

\bibitem{Holdom:1985ag}
B.~Holdom, ``{Two U(1)'s and Epsilon Charge Shifts},''
  \href{http://dx.doi.org/10.1016/0370-2693(86)91377-8}{{\em Phys. Lett. B}
  {\bfseries 166} (1986) 196--198}.

\bibitem{NA64:2024nwj}
{\bfseries NA64} Collaboration, Y.~M. Andreev {\em et~al.}, ``{Shedding light
  on dark sectors with high-energy muons at the NA64 experiment at the CERN
  SPS},'' \href{http://dx.doi.org/10.1103/PhysRevD.110.112015}{{\em Phys. Rev.
  D} {\bfseries 110} no.~11, (2024) 112015},
  \href{http://arxiv.org/abs/2409.10128}{{\ttfamily arXiv:2409.10128
  [hep-ex]}}.

\bibitem{Antoniadis:2021mqz}
I.~Antoniadis and F.~Rondeau, ``{Minimal embedding of the Standard Model into
  intersecting D-brane configurations with a bulk leptonic U(1)},''
  \href{http://dx.doi.org/10.1140/epjc/s10052-022-10660-y}{{\em Eur. Phys. J.
  C} {\bfseries 82} no.~8, (2022) 701},
  \href{http://arxiv.org/abs/2112.07587}{{\ttfamily arXiv:2112.07587
  [hep-th]}}.

\bibitem{Bjorken:2009mm}
J.~D. Bjorken, R.~Essig, P.~Schuster, and N.~Toro, ``{New Fixed-Target
  Experiments to Search for Dark Gauge Forces},''
  \href{http://dx.doi.org/10.1103/PhysRevD.80.075018}{{\em Phys. Rev. D}
  {\bfseries 80} (2009) 075018},
  \href{http://arxiv.org/abs/0906.0580}{{\ttfamily arXiv:0906.0580 [hep-ph]}}.

\bibitem{Tsai:1986tx}
Y.-S. Tsai, ``{AXION BREMSSTRAHLUNG BY AN ELECTRON BEAM},''
  \href{http://dx.doi.org/10.1103/PhysRevD.34.1326}{{\em Phys. Rev. D}
  {\bfseries 34} (1986) 1326}.

\bibitem{vonWeizsacker:1934nji}
C.~F. von Weizsacker, ``{Radiation emitted in collisions of very fast
  electrons},'' \href{http://dx.doi.org/10.1007/BF01333110}{{\em Z. Phys.}
  {\bfseries 88} (1934) 612--625}.

\bibitem{Williams:1935dka}
E.~J. Williams, ``{Correlation of certain collision problems with radiation
  theory},'' {\em Kong. Dan. Vid. Sel. Mat. Fys. Med.} {\bfseries 13N4} no.~4,
  (1935) 1--50.

\bibitem{Kim:1973he}
K.~J. Kim and Y.-S. Tsai, ``{IMPROVED WEIZSACKER-WILLIAMS METHOD AND ITS
  APPLICATION TO LEPTON AND W BOSON PAIR PRODUCTION},''
  \href{http://dx.doi.org/10.1103/PhysRevD.8.3109}{{\em Phys. Rev. D}
  {\bfseries 8} (1973) 3109}.

\bibitem{Tsai:1973py}
Y.-S. Tsai, ``{Pair Production and Bremsstrahlung of Charged Leptons},''
  \href{http://dx.doi.org/10.1103/RevModPhys.46.815}{{\em Rev. Mod. Phys.}
  {\bfseries 46} (1974) 815}. [Erratum: Rev.Mod.Phys. 49, 421--423 (1977)].

\bibitem{Liu:2016mqv}
Y.-S. Liu, D.~McKeen, and G.~A. Miller, ``{Validity of the
  Weizs{\"a}cker-Williams approximation and the analysis of beam dump
  experiments: Production of a new scalar boson},''
  \href{http://dx.doi.org/10.1103/PhysRevD.95.036010}{{\em Phys. Rev. D}
  {\bfseries 95} no.~3, (2017) 036010},
  \href{http://arxiv.org/abs/1609.06781}{{\ttfamily arXiv:1609.06781
  [hep-ph]}}.

\bibitem{Liu:2017htz}
Y.-S. Liu and G.~A. Miller, ``{Validity of the Weizs{\"a}cker-Williams
  approximation and the analysis of beam dump experiments: Production of an
  axion, a dark photon, or a new axial-vector boson},''
  \href{http://dx.doi.org/10.1103/PhysRevD.96.016004}{{\em Phys. Rev. D}
  {\bfseries 96} no.~1, (2017) 016004},
  \href{http://arxiv.org/abs/1705.01633}{{\ttfamily arXiv:1705.01633
  [hep-ph]}}.

\bibitem{Voronchikhin:2024vfu}
I.~V. Voronchikhin and D.~V. Kirpichnikov, ``{Implication of the
  Weizsacker-Williams approximation for the dark matter mediator production},''
  \href{http://dx.doi.org/10.1103/PhysRevD.111.035034}{{\em Phys. Rev. D}
  {\bfseries 111} no.~3, (2025) 035034},
  \href{http://arxiv.org/abs/2409.12748}{{\ttfamily arXiv:2409.12748
  [hep-ph]}}.

\bibitem{Gorbunov:2023jnx}
D.~Gorbunov and E.~Kriukova, ``{Dark photon production via elastic proton
  bremsstrahlung with non-zero momentum transfer},''
  \href{http://dx.doi.org/10.1007/JHEP01(2024)058}{{\em JHEP} {\bfseries 01}
  (2024) 058}, \href{http://arxiv.org/abs/2306.15800}{{\ttfamily
  arXiv:2306.15800 [hep-ph]}}.

\bibitem{Blumlein:2013cua}
J.~Bl{\"u}mlein and J.~Brunner, ``{New Exclusion Limits on Dark Gauge Forces
  from Proton Bremsstrahlung in Beam-Dump Data},''
  \href{http://dx.doi.org/10.1016/j.physletb.2014.02.029}{{\em Phys. Lett. B}
  {\bfseries 731} (2014) 320--326},
  \href{http://arxiv.org/abs/1311.3870}{{\ttfamily arXiv:1311.3870 [hep-ph]}}.

\bibitem{Foroughi-Abari:2021zbm}
S.~Foroughi-Abari and A.~Ritz, ``{Dark sector production via proton
  bremsstrahlung},'' \href{http://dx.doi.org/10.1103/PhysRevD.105.095045}{{\em
  Phys. Rev. D} {\bfseries 105} no.~9, (2022) 095045},
  \href{http://arxiv.org/abs/2108.05900}{{\ttfamily arXiv:2108.05900
  [hep-ph]}}.

\bibitem{Foroughi-Abari:2024xlj}
S.~Foroughi-Abari, P.~Reimitz, and A.~Ritz, ``{Closer look at dark vector
  splitting functions in proton bremsstrahlung},''
  \href{http://dx.doi.org/10.1103/fzlm-gsd7}{{\em Phys. Rev. D} {\bfseries 112}
  no.~1, (2025) 015030}, \href{http://arxiv.org/abs/2409.09123}{{\ttfamily
  arXiv:2409.09123 [hep-ph]}}.

\bibitem{Harlander:2020cyh}
R.~V. Harlander, S.~Y. Klein, and M.~Lipp, ``{FeynGame},''
  \href{http://dx.doi.org/10.1016/j.cpc.2020.107465}{{\em Comput. Phys.
  Commun.} {\bfseries 256} (2020) 107465},
  \href{http://arxiv.org/abs/2003.00896}{{\ttfamily arXiv:2003.00896
  [physics.ed-ph]}}.

\bibitem{Harlander:2024qbn}
R.~Harlander, S.~Y. Klein, and M.~C. Schaaf, ``{FeynGame-2.1 -- Feynman
  diagrams made easy},'' \href{http://dx.doi.org/10.22323/1.449.0657}{{\em PoS}
  {\bfseries EPS-HEP2023} (2024) 657},
  \href{http://arxiv.org/abs/2401.12778}{{\ttfamily arXiv:2401.12778
  [hep-ph]}}.

\bibitem{Bundgen:2025utt}
L.~B{\"u}ndgen, R.~V. Harlander, S.~Y. Klein, and M.~C. Schaaf, ``{FeynGame
  3.0},'' \href{http://dx.doi.org/10.1016/j.cpc.2025.109662}{{\em Comput. Phys.
  Commun.} {\bfseries 314} (2025) 109662},
  \href{http://arxiv.org/abs/2501.04651}{{\ttfamily arXiv:2501.04651
  [hep-ph]}}.

\bibitem{ParticleDataGroup:2024cfk}
{\bfseries Particle Data Group} Collaboration, S.~Navas {\em et~al.}, ``{Review
  of particle physics},''
  \href{http://dx.doi.org/10.1103/PhysRevD.110.030001}{{\em Phys. Rev. D}
  {\bfseries 110} no.~3, (2024) 030001}.

\bibitem{Beranek:2013nqa}
T.~Beranek and M.~Vanderhaeghen, ``{Study of the discovery potential for hidden
  photon emission at future electron scattering fixed target experiments},''
  \href{http://dx.doi.org/10.1103/PhysRevD.89.055006}{{\em Phys. Rev. D}
  {\bfseries 89} no.~5, (2014) 055006},
  \href{http://arxiv.org/abs/1311.5104}{{\ttfamily arXiv:1311.5104 [hep-ph]}}.

\bibitem{Mertig:1990an}
R.~Mertig, M.~Bohm, and A.~Denner, ``{FEYN CALC: Computer algebraic calculation
  of Feynman amplitudes},''
  \href{http://dx.doi.org/10.1016/0010-4655(91)90130-D}{{\em Comput. Phys.
  Commun.} {\bfseries 64} (1991) 345--359}.

\bibitem{Shtabovenko:2016sxi}
V.~Shtabovenko, R.~Mertig, and F.~Orellana, ``{New Developments in FeynCalc
  9.0},'' \href{http://dx.doi.org/10.1016/j.cpc.2016.06.008}{{\em Comput. Phys.
  Commun.} {\bfseries 207} (2016) 432--444},
  \href{http://arxiv.org/abs/1601.01167}{{\ttfamily arXiv:1601.01167
  [hep-ph]}}.

\bibitem{Shtabovenko:2020gxv}
V.~Shtabovenko, R.~Mertig, and F.~Orellana, ``{FeynCalc 9.3: New features and
  improvements},'' \href{http://dx.doi.org/10.1016/j.cpc.2020.107478}{{\em
  Comput. Phys. Commun.} {\bfseries 256} (2020) 107478},
  \href{http://arxiv.org/abs/2001.04407}{{\ttfamily arXiv:2001.04407
  [hep-ph]}}.

\bibitem{Shtabovenko:2023idz}
V.~Shtabovenko, R.~Mertig, and F.~Orellana, ``{FeynCalc 10: Do multiloop
  integrals dream of computer codes?},''
  \href{http://dx.doi.org/10.1016/j.cpc.2024.109357}{{\em Comput. Phys.
  Commun.} {\bfseries 306} (2025) 109357},
  \href{http://arxiv.org/abs/2312.14089}{{\ttfamily arXiv:2312.14089
  [hep-ph]}}.

\bibitem{Doble:1994np}
N.~Doble, L.~Gatignon, G.~von Holtey, and F.~Novoskoltsev, ``{The Upgrated muon
  beam at the SPS},''
  \href{http://dx.doi.org/10.1016/0168-9002(94)90212-7}{{\em Nucl. Instrum.
  Meth. A} {\bfseries 343} (1994) 351--362}.

\bibitem{Chen:2018vkr}
C.-Y. Chen, J.~Kozaczuk, and Y.-M. Zhong, ``{Exploring leptophilic dark matter
  with NA64-$\mu$},'' \href{http://dx.doi.org/10.1007/JHEP10(2018)154}{{\em
  JHEP} {\bfseries 10} (2018) 154},
  \href{http://arxiv.org/abs/1807.03790}{{\ttfamily arXiv:1807.03790
  [hep-ph]}}.

\bibitem{Lista:2016tva}
L.~Lista, \href{http://dx.doi.org/10.1007/978-3-319-20176-4}{{\em {Statistical
  Methods for Data Analysis in Particle Physics}}}, vol.~909.
\newblock Springer, 2016.

\bibitem{LDMX:2018cma}
{\bfseries LDMX} Collaboration, T.~{\r{A}}kesson {\em et~al.}, ``{Light Dark
  Matter eXperiment (LDMX)},''
  \href{http://arxiv.org/abs/1808.05219}{{\ttfamily arXiv:1808.05219
  [hep-ex]}}.

\bibitem{Mans:2017vej}
{\bfseries LDMX} Collaboration, J.~Mans, ``{The LDMX Experiment},''
  \href{http://dx.doi.org/10.1051/epjconf/201714201020}{{\em EPJ Web Conf.}
  {\bfseries 142} (2017) 01020}.

\bibitem{Rohlf:1994wy}
J.~W. Rohlf, {\em {Modern Physics from A to Z}}.
\newblock John Wiley and Sons, New York, 1994.

\bibitem{Davoudiasl:2024fiz}
H.~Davoudiasl, H.~Liu, R.~Marcarelli, Y.~Soreq, and S.~Trifinopoulos, ``{New
  physics at the Muon (Synchrotron) Ion Collider: MuSIC for several scales},''
  \href{http://dx.doi.org/10.1007/JHEP03(2025)046}{{\em JHEP} {\bfseries 03}
  (2025) 046}, \href{http://arxiv.org/abs/2412.13289}{{\ttfamily
  arXiv:2412.13289 [hep-ph]}}.

\bibitem{Bandyopadhyay:2022klg}
T.~Bandyopadhyay, S.~Chakraborty, and S.~Trifinopoulos, ``{Displaced searches
  for light vector bosons at Belle II},''
  \href{http://dx.doi.org/10.1007/JHEP05(2022)141}{{\em JHEP} {\bfseries 05}
  (2022) 141}, \href{http://arxiv.org/abs/2203.03280}{{\ttfamily
  arXiv:2203.03280 [hep-ph]}}.

\bibitem{CCFR:1991lpl}
{\bfseries CCFR} Collaboration, S.~R. Mishra {\em et~al.}, ``{Neutrino Tridents
  and W Z Interference},''
  \href{http://dx.doi.org/10.1103/PhysRevLett.66.3117}{{\em Phys. Rev. Lett.}
  {\bfseries 66} (1991) 3117--3120}.

\bibitem{Altmannshofer:2014pba}
W.~Altmannshofer, S.~Gori, M.~Pospelov, and I.~Yavin, ``{Neutrino Trident
  Production: A Powerful Probe of New Physics with Neutrino Beams},''
  \href{http://dx.doi.org/10.1103/PhysRevLett.113.091801}{{\em Phys. Rev.
  Lett.} {\bfseries 113} (2014) 091801},
  \href{http://arxiv.org/abs/1406.2332}{{\ttfamily arXiv:1406.2332 [hep-ph]}}.

\bibitem{Altmannshofer:2019zhy}
W.~Altmannshofer, S.~Gori, J.~Mart{\'\i}n-Albo, A.~Sousa, and M.~Wallbank,
  ``{Neutrino Tridents at DUNE},''
  \href{http://dx.doi.org/10.1103/PhysRevD.100.115029}{{\em Phys. Rev. D}
  {\bfseries 100} no.~11, (2019) 115029},
  \href{http://arxiv.org/abs/1902.06765}{{\ttfamily arXiv:1902.06765
  [hep-ph]}}.

\bibitem{Ballett:2019xoj}
P.~Ballett, M.~Hostert, S.~Pascoli, Y.~F. Perez-Gonzalez, Z.~Tabrizi, and
  R.~Zukanovich~Funchal, ``{$Z^\prime$s in neutrino scattering at DUNE},''
  \href{http://dx.doi.org/10.1103/PhysRevD.100.055012}{{\em Phys. Rev. D}
  {\bfseries 100} no.~5, (2019) 055012},
  \href{http://arxiv.org/abs/1902.08579}{{\ttfamily arXiv:1902.08579
  [hep-ph]}}.

\bibitem{Shimomura:2020tmg}
T.~Shimomura and Y.~Uesaka, ``{Kinematical distributions of coherent neutrino
  trident production in gauged $L_\mu-L_\tau$ model},''
  \href{http://dx.doi.org/10.1103/PhysRevD.103.035022}{{\em Phys. Rev. D}
  {\bfseries 103} no.~3, (2021) 035022},
  \href{http://arxiv.org/abs/2009.13773}{{\ttfamily arXiv:2009.13773
  [hep-ph]}}.

\bibitem{Chakraborty:2026ocj}
D.~Chakraborty, A.~Chatterjee, A.~Datta, A.~Kaushik, and K.~Nishiwaki,
  ``{Aspects of a Five-Dimensional $U(1)_{L_\mu- L_\tau}$ Model at Future
  Muon-Based Colliders},'' \href{http://arxiv.org/abs/2604.05041}{{\ttfamily
  arXiv:2604.05041 [hep-ph]}}.

\end{thebibliography}\endgroup

\end{document}